\documentclass{article}
\usepackage{arxiv}

\usepackage[utf8]{inputenc} % allow utf-8 input
\usepackage[T1]{fontenc}    % use 8-bit T1 fonts
\usepackage{hyperref}       % hyperlinks
\usepackage{url}  % simple URL typesetting
\usepackage{booktabs}       % professional-quality tables
\usepackage{amsfonts}       % blackboard math symbols
\usepackage{nicefrac}       % compact symbols for 1/2, etc.
\usepackage{microtype}      % microtypography
\usepackage{lipsum}

\usepackage[cmex10]{amsmath}
\usepackage[table,xcdraw,dvipsnames]{xcolor}
\usepackage[english]{babel}
\usepackage{amssymb, amsthm}
\usepackage[boxed]{algorithm2e}
\usepackage{graphicx} 
\usepackage{placeins}
\usepackage{subcaption} 
\usepackage{amsfonts}
\usepackage[inline]{enumitem}
\usepackage{exercise}
\usepackage{ragged2e}
\usepackage{caption}
\usepackage{tikz}
\usepackage{pgfplots}
    \usepgfplotslibrary{fillbetween}
\usepackage{mathtools}
\usepackage{microtype}
\usepackage{hyperref}
\usepackage[nameinlink, capitalise]{cleveref}
\usepackage{makecell}
\usepackage{rotating,tabularx}
\usepackage{tablefootnote}
\usepackage{footmisc}
\usepackage{adjustbox}
\usepackage{array}
\newcolumntype{x}[1]{>{\centering\arraybackslash\hspace{0pt}}p{#1}}
\usepackage{pifont}% http://ctan.org/pkg/pifont
\usepackage{svg,svg-extract}
\usepackage{multirow}
\usepackage{longtable}
\usepackage{xltabular}

\newcommand{\A}{\mathcal{A}}
\newcommand{\mS}{\mathcal{S}}

\newcommand{\R}{\mathbb{R}}

\newcommand{\E}{\mathbb{E}}
\usepackage{bbm}

\newcommand{\bbm}{\begin{bmatrix}}
\newcommand{\ebm}{\end{bmatrix}}

\DeclareMathOperator*{\argmax}{arg\,max}

\newcommand{\cmark}{\textcolor{Green}{\checkmark}}

\newcommand{\cgray}{\cellcolor{gray!30}}

\newlist{mylist}{enumerate*}{1}
\setlist[mylist]{label=(\arabic*)}

\newcommand{\margins}{\textcolor{purple}{R_{Margins}}}
\newcommand{\rwrho}{\textcolor{purple}{R_{\rho}}}
\newcommand{\loss}{\textcolor{purple}{R_{Loss}}}
\newcommand{\rwalpha}{\textcolor{purple}{R_{Alpha}}}

\newcommand{\Asym}{\textcolor{black}{\A_{sym}(v)}}
\newcommand{\Anzero}{\textcolor{black}{\A_{n-0}(v)}}
\newcommand{\Anone}{\textcolor{black}{\A_{n-1}(v)}}
\newcommand{\AT}{\rho_{\textrm{act}}}
\newcommand{\RT}{\rho_{\textrm{rev}}}

\newcommand{\aaacfirst}{\hyperlink{sol:a3c2019}{\textcolor{violet}{A3C-2019}}}
\newcommand{\ddqn}{\hyperlink{sol:d2qn2019}{\textcolor{violet}{DDQN-2019}}}
\newcommand{\aaacsecond}{\hyperlink{sol:a3c2020}{\textcolor{violet}{A3C-2020}}}
\newcommand{\smaac}{\hyperlink{sol:smaac}{\textcolor{violet}{SMAAC-2021}}}
\newcommand{\sas}{\hyperlink{sol:SAS}{\textcolor{violet}{SAS-2021}}}
\newcommand{\binbin}{\hyperlink{sol:binbin}{\textcolor{violet}{Binbinchen-2020}}}
\newcommand{\dddqnfirst}{\hyperlink{sol:d3qn2020}{\textcolor{violet}{D3QN-2020}}}
\newcommand{\cem}{\hyperlink{sol:cem}{\textcolor{violet}{CEM-2021}}}
\newcommand{\dddqnsecond}{\hyperlink{sol:d3qn2022}{\textcolor{violet}{D3QN-2022}}}
\newcommand{\powrl}{\hyperlink{sol:powrl}{\textcolor{violet}{PowRL-2022}}}
\newcommand{\alphazero}{\hyperlink{sol:alpha}{\textcolor{violet}{AlphaZero-2022}}}
\newcommand{\brute}{\hyperlink{sol:brute}{\textcolor{violet}{BruteForce-2022}}}
\newcommand{\hri}{\hyperlink{sol:hri}{\textcolor{violet}{HRI-EU-2022}}}
\newcommand{\curriculum}{\hyperlink{sol:cur}{\textcolor{violet}{Curriculum-2023}}}
\newcommand{\hrl}{\hyperlink{sol:hri}{\textcolor{violet}{HRL-2023}}}
\newcommand{\marl}{\hyperlink{sol:d2qn2019}{\textcolor{violet}{MARL-2023}}}
\newcommand{\ljn}{\hyperlink{sol:ljn}{\textcolor{violet}{LJNAgent-2024}}}
\newcommand{\artelys}{\hyperlink{sol:artelys}{\textcolor{violet}{Artelys-2024}}}
\newcommand{\hugo}{\hyperlink{sol:hugo}{\textcolor{violet}{HUGO-2024}}}
\newcommand{\il}{\hyperlink{sol:il}{\textcolor{violet}{IL-2024}}}
\newcommand{\bdqn}{\hyperlink{sol:bdqn}{\textcolor{violet}{BDQN-2024}}}
\newcommand{\zonal}{\hyperlink{sol:zonal}{\textcolor{violet}{Zonal-2024}}}
\newcommand{\gnnil}{\hyperlink{sol:gnn}{\textcolor{violet}{GNNIL-2025}}}
\newcommand{\CCMA}{\hyperlink{sol:ccma}{\textcolor{violet}{CCMA-2025}}}
\newcommand{\SoftIL}{\hyperlink{sol:softil}{\textcolor{violet}{SoftIL-2025}}}

\title{
Optimizing Power Grid Topologies with Reinforcement Learning: A Survey of Methods and Challenges
}

\author{
 Erica van der Sar\\
  Department of Mathematics\\
  Vrije Universiteit\\
  Amsterdam, The Netherlands \\
  \texttt{e.t.van.der.sar@vu.nl} \\
   \And
 Alessandro Zocca\\
  Department of Mathematics\\
  Vrije Universiteit\\
  Amsterdam, The Netherlands \\
  \texttt{a.zocca@vu.nl} \\
  \And
 Sandjai Bhulai\\
  Department of Mathematics\\
  Vrije Universiteit\\
  Amsterdam, The Netherlands \\
  \texttt{s.bhulai@vu.nl} \\
}

\begin{document}

\maketitle

\begin{abstract}
% Introduce the problem and explain the research goal.
% \small
Power grid operation is becoming increasingly complex due to the rising integration of renewable energy sources and the need for more adaptive control strategies. Reinforcement Learning (RL) has emerged as a promising approach to power network control (PNC), offering the potential to enhance decision-making in dynamic and uncertain environments.  The Learning To Run a Power Network (L2RPN) competitions have played a key role in accelerating research by providing standardized benchmarks and problem formulations, leading to rapid advancements in RL-based methods. This survey provides a comprehensive and structured overview of RL applications for power grid topology optimization, categorizing existing techniques, highlighting key design choices, and identifying gaps in current research. Additionally, we present a comparative numerical study evaluating the impact of commonly applied RL-based methods, offering insights into their practical effectiveness. By consolidating existing research and outlining open challenges, this survey aims to provide a foundation for future advancements in RL-driven power grid optimization.

\end{abstract}

% keywords 


\tableofcontents

%%%%%%%%%%%%%%%%%%%%%%%%%%%%%%%%%%%%%%%%%%%%%%%%%%%%%%%%%%%%%%%%%%%%%%%%%%%%%%%
%%%%%%%%%%%%%%%%%%%%%%%%%%%%%%%%%%%%%%%%%%%%%%%%%%%%%%%%%%%%%%%%%%%%%%%%%%%%%%%
\section{Introduction} \label{Sec:Intro}

\subsection{Background}\label{Subsec:Background}
Electrical power grids form the backbone of modern society, being responsible for transporting electricity from producers to consumers 24 hours a day, 365 days a year. Operating these grids is a demanding control task that requires continuous monitoring and frequent interventions by skilled experts to maintain network stability, keep power flow within the thermal limits of the equipment, and ensure voltage and frequency levels are met, as highlighted by \cite{marot2020learning}. 
% The network control rooms are staffed by operators who rely heavily on their experience and expertise to anticipate and resolve undesirable system behavior\cite{Subramanian2021}.
The increasing integration of renewable energy sources into the power grid and the rising energy demand add new layers of complexity to this task. With its unpredictable generation patterns, renewable energy makes power flows more variable and less predictable, altering how the network behaves and responds to disturbances. Due to the operational costs of simulation with the current models, operators often rely on their experience to resolve network problems in real-time. Continuing this manual approach may compromise system security or lead to significantly higher costs.
According to projections by \cite{tennet2024rapport}, energy consumption in the Netherlands is expected to grow 30\% between 2024 and 2033, with similar trends across Europe (\cite{ENTSO2023rapport}). Meanwhile, the share of controllable energy sources is expected to decrease by approximately 40\% due to the shift towards renewable energy. As a result, \textit{Transmission System Operators} (TSOs) will face greater challenges in maintaining network security and reliability, underscoring the need for new grid management strategies.

With its recent advancements and breakthroughs (\cite{wang2016dueling}, \cite{vanhasselt2016deep}, \cite{schulman2017proximal}, \cite{silver2017mastering} and \cite{haarnoja2018soft}), \textit{Reinforcement learning} (RL) has emerged as a promising approach that could assist network operators in decision-making, offering the potential to find cost-effective flexibilities that may currently go unnoticed by human operators. 
For this reason, in 2019, the French TSO RTE launched the \emph{Learning to Run a Power Network} (L2RPN) challenge introduced in \cite{marot2020learning}, encouraging researchers to use RL for \textit{power network control} (PNC). This challenge seeks to explore how RL can enhance real-time operations and assist with operational planning, ultimately supporting TSOs in navigating the growing complexities of modern power grids.
    Moreover, the authors of \cite{marot2020learning} note that RL has already demonstrated potential in a variety of power system problems, reviewed in \cite{glavic2017reinforcement,glavic2019deep}, highlighting its suitability for complex control tasks in this domain.

    \subsection{Introduction to Power Grids}\label{App:IntroPowerGrids}
    In this section, we provide a brief background on power grids to facilitate researchers from the RL community stepping into the field of PNC. For a more extensive introduction, see also \cite{kelly2020reinforcement}.

    \subsubsection{From generation to consumption}\label{subsec:Gen2Cons}
\begin{figure}[!tbh]
    \centering
    \includegraphics[width=\linewidth]{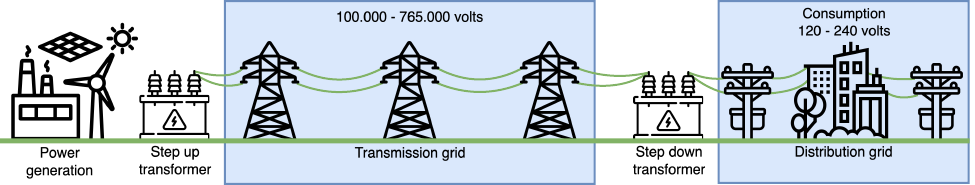}
    \caption{Schematic illustration of power distribution from generation sources to consumers.} 
    \label{Fig:Gen2Cons}
\end{figure}

    Electrical power grids are typically divided into two layers, the \textit{transmission grid}, controlled by TSOs, and the \textit{distribution grid}, see \cref{Fig:Gen2Cons}. Power is supplied to the network by generators. In the past, these generators were often located close to the consumers, the loads, reducing the transmission distances of the power. However, generation sources such as hydro and nuclear power stations or wind and solar are often situated in geographically remote areas, far from load centers. To reduce resistive losses over these long distances, electricity in the transmission grid is transported at high voltage levels, between 100.000 - 765.000 volts. The voltage in homes in Europe and North America is typically 240 and 120 volts, respectively. Therefore, step-down transformers are used to reduce the voltage for the distribution grid. From the power network’s perspective, consumption is aggregated as \textit{load}, and the grid operator’s responsibility generally ends at the step-down transformer. 
    To allow for voltage conversion using transformers, the transmission grid operates under \textit{alternating current} (AC). While AC allows for efficient voltage transformation, it also adds complexity compared to direct current (DC) systems, as power flows are governed by nonlinear equations involving both active and reactive power balances. 
    
\FloatBarrier
    
\begin{figure}[!tbh]
    \centering
    \includegraphics[width=\linewidth]{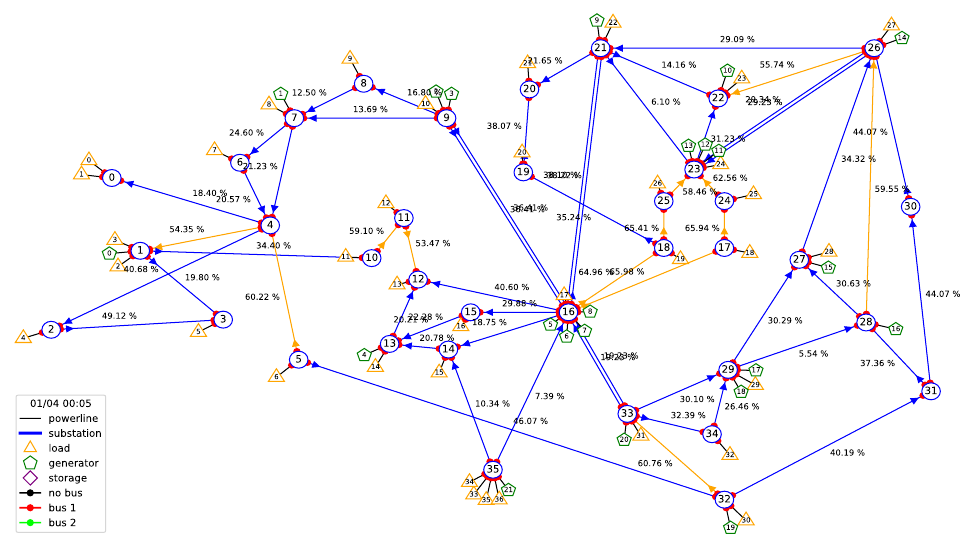}
    \caption{Example of a transmission power grid represented as a directed graph. This example is part of the environment \texttt{l2rpn\_icaps\_2021}, available in Grid2Op (see \cref{Sec:ProblemDescr}).} 
    \label{Fig:PowerGrid}
\end{figure}

    A transmission power grid can be represented as a highly interconnected directed graph, with transmission lines represented as edges and \textit{substations} as nodes, see \cref{Fig:PowerGrid} for a small example. Substations contain the transformers to step up and step down the voltages and connect the generation sources and loads to the grid.
    
% \FloatBarrier

    Human operators continuously monitor and operate the power system from the TSO control centers, intervening when necessary. Operators must always respect three key constraints, described in \cite{kelly2020reinforcement}:
    \begin{itemize}
        \item \textbf{Keep the transmission lines within their thermal limits.} Each transmission line has an upper limit to the amount of power that can flow through, determined by its materials' thermal properties. Overloading can lead to overheating, causing lines to sag dangerously close to the ground or damaging critical equipment such as transformers and cables. In such cases, automatic or manual disconnection is often preferable to irreversible damage.
        \item \textbf{Maintain the voltage within a safe range.} %In the (current) L2RPN challenge, the voltage control is handled by the environment, and agents are not required to optimize it.
        \item \textbf{Balance generation and load.} %The generation-load balance is also maintained by the environment in the (current) L2RPN setting. Nevertheless, redispatching generation can help relieve line congestion and can be considered a useful action in the RL challenge. 
    \end{itemize}
    % \paragraph{Circuit Theory} 
    Electricity flows within a power network in the way prescribed by power flow physics. In particular, this means that power cannot be directed from a specific generator to a specific consumer. If a line is disconnected, the power originally flowing on that line redistributes itself across the remaining lines, potentially overloading others.
    The job of the network operator and the challenge of L2RPN is to efficiently manage the network while avoiding overloads and cascading failures.  
    %route the power across the network while avoiding overloads and cascading failures.  

    Indirectly managing the power flows during congestion by controlling injections, such as redispatching power generation sources, is the most common and well-documented approach in the literature. 
    However, this strategy imposes constraints on producers to be flexible, which comes with additional costs, as mentioned by \cite{marot2018expert}. TSOs can alleviate congestion also by adjusting the grid topology via two types of topological actions, which, in contrast, have a negligible cost. We distinguish two types of actions: line switching and bus splitting.
    \textit{Line switching} consists of remotely disconnecting or reconnecting a transmission line. Another less explored option is \textit{bus splitting}, modifications performed within a substation (using \textit{node breakers}).

    Within each substation, there are multiple \textit{busbars} available to connect the elements. A consistent busbar connection makes a substation a singular node; differing busbar connections split the substation, changing the local network topology. \cref{Fig:Grid2OpExpl} shows an example of how a bus-splitting action changes the power flows and relieves the congested line in a simulated environment of Grid2Op (the package provided for L2RPN discussed in \cref{Sec:ProblemDescr}).

    \subsubsection{Topology Optimization}
    The challenges and solutions of L2RPN discussed in this survey paper mostly focus on topology optimization strategies, i.e., those using discrete actions such as line switching and substation reconfigurations. Continuous actions, such as redispatching, are only considered in a few solutions, often as a secondary option.
    
    Topology optimization is a promising tool for relieving congestion, as it exploits existing infrastructure without resorting to costly interventions such as redispatching, load shedding, or demand response, \cite{bacher1986network,marot2020learning}. 
    Despite this potential, topology optimization remains underutilized. Traditionally, the power systems community relies on optimization approaches to solve \textit{optimal power flow} (OPF) problems, including augmented OPF with discrete topological actions, \cite{frank2012optimal,capitanescu2011state,capitanescu2016critical}.
    However, existing optimization techniques struggle to scale up, due to the problem’s nonlinear and discrete combinatorial nature, leading to a large and complex optimization space. Including all topology configuration actions results in a nonconvex mixed-integer nonlinear programming problem, which is NP hard.
    Most of the topology optimization research focuses on line switching \cite{fisher2008optimal,hedman2009optimal,khanabadi2012optimal, fuller2012fast,hedman2011review,numan2023role}, but neglects the more complex bus-splitting actions, which could offer much more flexibility to the operator \cite{marot2018expert}. 
    While bus-splitting is explored in \cite{zhou2021substation,heidarifar2015network,park2020optimal,babaeinejadsarookolaee2023transmission}, many of these methods still exhibit one or more of the following limitations:
    \begin{itemize}
        \item they are computationally expensive,
        \item they are limited to only a small number of substations with bus-splitted options, 
        \item they rely on linear (DC) approximation, which neglects reactive power and voltage constraints, and/or 
        \item they treat the problem as a static single snapshot, whereas power network control is inherently a sequential decision problem that evolves over time \cite{viebahn2022potential}.
    \end{itemize}
    These shortcomings highlight the need for new approaches, where RL, a fast-developing field of research specifically designed for sequential decision problems, could play a pivotal role.

    \subsection{Introduction to Reinforcement Learning}\label{App:IntroRL}
    This section provides a brief background on reinforcement learning (RL) for researchers in the power systems community interested in its application to power network control (PNC). For a comprehensive overview of RL, we refer to the book by \cite{sutton2018reinforcement}.
    
    \subsubsection{Key concepts reinforcement learning}
    Reinforcement learning is a machine learning technique where an \textit{agent} (the learner and decision maker) \textbf{learns from experience} how to make sequential decisions in an environment, \textbf{guided by a reward} (feedback).

    The sequential decision problem is typically modeled as a Markov Decision Process (MDP), which consists of:
    \begin{itemize}
        \item State space $\mathcal{S}$. A state $s_t \in \mathcal{S}$ is the snapshot of the environment observed by the agent at time $t$, usually represented by a vector or matrix.
        \item Action space $\A$ (or $\A_s$ the set of available actions from state $s$). After observing the state, the agent interacts with the environment at time $t$ by taking an action (decision) $a_t \in \A$ according to its \textit{policy} $\pi$.
        \item Transition function $p:\mathcal{S} \times \mathcal{S} \times \A \to [0,1]$. The quantity $p(s_{t+1}|s_t,a_t)$ describes the probability that action $a_t$ in state $s_t$ leads to state $s_{t+1}$.
        \item Reward function $r:\mathcal{S} \times \A \times \mathcal{S} \to \R$. The quantity $r(s_t,a_t,s_{t+1})$ represents the reward received after transitioning from $s_t$ to $s_{t+1}$ via action $a_t$. In some settings, this reduces to $r(s_t,a_t)$ or $r(s_t)$.%, depending only on the state-action pair (or state).
        \item Discount factor $\gamma \in (0,1)$, which in a discounted MDP setting, determines the relative importance of future rewards. 
    \end{itemize}
    \cref{Sec:ProblemDescr} presents the specific MDP setup for the L2RPN problem.

    The RL agent interacts with the environment according to a (stochastic) policy $\pi: \mathcal{S} \times \A \to [0,1]$, where $\pi(a_t|s_t)$ defines the probability of choosing the action $a_t$ in state $s_t$. 
    The goal in RL is for the agent to learn a policy $\pi$ that maximizes the \textit{expected discounted return}:
    \begin{align*}
        V_{\pi}(s) := 
        %& \E_{\pi}\left[ r(s_0,a_0) + \gamma r(s_{1},a_{1}) +\gamma^2 r(s_{2},a_{2}) + \dots | s_0 = s \right] \\
                    & \E_{\pi}\left[ \sum_{t=0}^T \gamma^t r(s_t,a_t) ~\middle|~ s_0 = s \right].
    \end{align*}
    The function $V_\pi$ is also referred to as the \textit{state-value function} under policy $\pi$. A fundamental property of RL introduced by \cite{bellman1966dynamic} is the recursive property of the value function, which can be stated as
    \begin{align*}
        % \label{Eq:Bellman}
        V_{\pi}(s) = & \sum_{a\in\A} \pi(a|s) \sum_{s' \in \mS}p(s' | s, a) \left[ r(s,a) + \gamma V_\pi(s') \right]  \\
                    = & \E_{\substack{a\sim\pi \\ s'\sim P}}\left[  r(s,a) + \gamma V_\pi(s') \right] .
    \end{align*}
    where $s'\sim P$ is shorthand for $s' \sim p(\cdot|s,a)$.
    
    Similarly, the \textit{action-value function} (or \textit{$Q$-value} function) for policy $\pi$ is defined by 
    \begin{align*}
        Q_\pi(s,a) : =
        %& \E_\pi\left[ r(s_0,a_0) + \gamma r(s_{1},a_{1}) +\gamma^2 r(s_{2},a_{2}) + \dots | s_0 = s, a_0 = a  \right]\\
                    & \E_\pi\left[ \sum_{t=0}^T \gamma^t r(s_t,a_t) \middle| s_0 = s, a_0 = a \right] 
                    =  \E_{s' \sim P} \left[ r(s,a) + \gamma  V_\pi(s') \right] . 
    \end{align*}
    
    A policy $\pi^*$ that maximizes the expected return is an \textit{optimal policy}. The optimal state-value function defined by  $V^*(s) := \max_{\pi} V_\pi(s)$, satisfies the \textit{Bellman optimality equation}
    \begin{align*}
        V^*(s) := \max_{a\in \A} \E_{s'\sim P} \left[ r(s,a) +  \gamma V^*(s') \right].
    \end{align*}
    In terms of the $Q$-values, it can be equivalently rephrased as
    \begin{align*}
        Q^*(s,a) :%= & \E_{s'\sim P} \left[ r(s,a) + \gamma  V^*(s') \right] \\
                = &  \E_{s'\sim P} \left[ r(s,a) + \gamma  \max_{a'\in\A} Q^*(s',a') \right]. 
    \end{align*}
    % Main elements: policy $\pi$, reward signal $R_t$, value function $V(s)$, optionally model of environment.

    The general idea of RL is to collect data by interacting with the environment and use this to find an approximation of the value function $V$ or $Q$ and/or to update the policy $\pi$.

    \paragraph{Deep RL}
    The algorithms applied to the L2RPN problem usually make use of the so-called \textit{Deep RL}. In Deep RL (DRL), the value functions and policies are approximated using a neural network, which is referred to as a parameterized policy and value function, where the parameters, often denoted by $\theta$ or $\phi$, represent the weights and biases of the neural network. This function approximation enables scalability to large or continuous state/action spaces, in contrast to table-based methods like tabular $Q$-learning.
    % Deep RL
    % Function approximation
    
    \subsubsection{RL methods and terminology}\label{subsec:RL_terminology}
\begin{figure}[!tbh]
    \centering
    \includegraphics[width=0.8\linewidth]{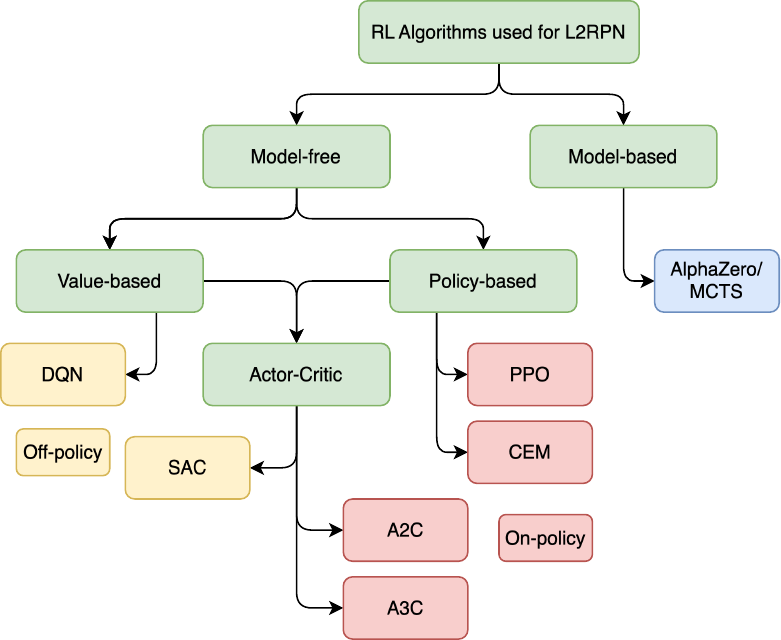}
    \caption[Caption]{An overview of the algorithms applied to L2RPN categorized by the features discussed in this section.\footnotemark} 
    \label{Fig:RL_Taxonomy}
    \end{figure}

    This survey discusses a variety of (advanced) RL algorithms applied to L2RPN. Describing each algorithm is outside the scope of this survey; here, we provide a few references and highlight the key distinctions between the algorithms discussed in this paper. To help the reader, we categorize the main algorithms used in L2RPN in \cref{Fig:RL_Taxonomy}. 

    \footnotetext{This figure is inspired by \url{https://spinningup.openai.com/en/latest/spinningup/rl_intro2.html}.}
    \FloatBarrier
    
    \paragraph{Model-based vs.~model-free}
    One of the main distinctions made in RL is between methods that require a model of the environment and methods that do not.
    
    \textit{Model-based} methods use a model of the environment (either known or learned) to simulate future outcomes and plan ahead. However, an accurate model of the environment is often unavailable or the model needs to be learned, which comes with additional challenges. The most famous example of a model-based approach is \textit{AlphaZero} by \cite{silver2017mastering}, also applied in L2RPN.
    
    \textit{Model-free} methods learn directly from experience without requiring a model. These methods tend to be easier to implement and tune, making them a more popular choice in L2RPN competitions and most other research. %The lack of a model means that one needs to sample the MDP to gather statistical knowledge about the environment. This can be done by performing actions and using them to update the state-value or state-action value function. 
    Learning is typically done using some variant of \textit{temporal-difference (TD)} updates. A basic TD method update of the value function is
    \begin{align*}
        V_{k+1}(s_t) \leftarrow V_k(s_t) +  \alpha \left[ R_{t+1} + \gamma V_k(s_{t+1}) - V_k(s_t) \right],
    \end{align*}
    where $R_{t+1}$ and $s_{t+1}$ are the observed reward and state at time $t+1$ and $\alpha\in[0,1]$ is the \textit{learning rate}. The quantity $V_{k}$ is the old estimate of the value function, while $V_{k+1}$ is the new estimate in iteration $k$. The quantity $R_{t+1} + \gamma V_k(s_{t+1})$ is the \textit{TD target}.
    
    % Model-free: $Q$-learning, PPO, Actor-Critic
    % Model-based: MCTS - Alpha zero 
    
    \paragraph{Value-based vs. policy-based methods}
    Another distinction made between RL methods is whether they are \textit{value-based}, \textit{policy-based}, or a hybrid version of the two, referred to as an \textit{actor-critic} method.

    In \textit{value-based} methods, the aim is to learn the values of the states or state-action pairs and then derive a policy from these. A breakthrough in RL was the development of $Q$-learning, an \textit{off-policy} value-based TD method. In the context of L2RPN, some approaches use a DRL version of $Q$-learning (DQN), i.e., Double-DQN (DDQN) and Dueling-Double-DQN (D3QN), by \cite{wang2016dueling} and \cite{vanhasselt2016deep}.
    
    \textit{Policy-based} methods learn a policy directly rather than searching for a value function and extracting a policy. This is particularly well-suited for dealing with large or continuous action spaces. 
    In Deep RL, they typically rely on \textit{policy gradient} methods, which optimize the parameterized policy $\pi_\theta(a|s)$ by taking the gradient ascent of $J$, the expected return of a policy $\pi_\theta$, with respect to $\theta$, where
    \begin{align*}
        J(\pi_\theta) :=  V_{\pi_\theta}(s_0) = \E_{\pi_\theta}\left[ R_\tau \right]
    \end{align*}
    and $\tau = (s_0,a_0, s_1, a_1,\dots)$ is a possible trajectory or episode following policy $\pi_\theta$. 
    Typically, policy gradient methods are \textit{on-policy} and learn \textit{stochastic} policies, meaning that the policy specifies a probability distribution over the actions. 
    Proximal Policy Optimization (PPO) by \cite{schulman2017proximal} and Cross Entropy Method (CEM) by \cite{deboer2005tutorial} are policy-based methods used in the context of L2RPN. However, CEM uses population-based search and does not rely on policy gradients.

    \textit{Actor-critic} methods combine both paradigms, using a learned value function to reduce the variance of policy gradient updates. Soft Actor-Critic (SAC) and (Asynchronous) Advantage Actor-critic (A2C/A3C) are examples of actor-critic methods applied in L2RPN.

    \paragraph{On-policy vs. off-policy}
    \textit{On-policy} methods learn from data generated by the current policy being improved. In contrast, \textit{off-policy} methods can learn from experience generated by a different behavior policy, for example $\epsilon$-greedy with respect to current $Q$-values, enabling more efficient data reuse and exploration. Using a behavior policy different from the current policy allows for applying techniques such as prioritized experience replay and different exploration strategies, as discussed in \cref{Subsub:PrioReplay,ExplStrat}.

    % Both use \textit{temporal difference learning}
    % Value based: $Q$-learning, .. - usually \textit{off-policy}
    % Policy-based:  PPO, ... - usually \textit{on-policy}
    % Actor Critic: 

\subsection{Problem Description}\label{Sec:ProblemDescr}
This section describes the platform provided for L2RPN and how the PNC problem is formulated as a Markov Decision Process in \cref{subsec:L2RPNasMDP}. Additionally, \cref{subsec:RulesL2RPN} describes the rules within the L2RPN environment.

% A challenge that aims to help network operators to control the networks and to guide the decision-making of network operators in real-time operations and in operational planning. 
% The network operator’s role is to monitor the electricity network 24 hours per day, 365 days per year. The operator must keep the network within its thermal limits, its frequency ranges and voltage ranges.
\subsubsection{Reinforcement Learning Framework for Power Grid Management} \label{subsec:L2RPNasMDP}
To facilitate the L2RPN challenge and further research on this topic, RTE provides an open-source Python package \textit{Grid2Op} (\cite{donnot2020grid2op}) that simulates realistic power grids.
The Grid2Op package enables researchers to develop a sequential decision-making RL model, termed \textit{agent}, capable of controlling the power grid in real-time. The package provides environments with different episodes, each simulating a continuous sequence of power grid states at 5-minute intervals. 

\begin{figure}[!tbh]
    \centering
    \footnotesize
    \includegraphics[width=0.6\linewidth]{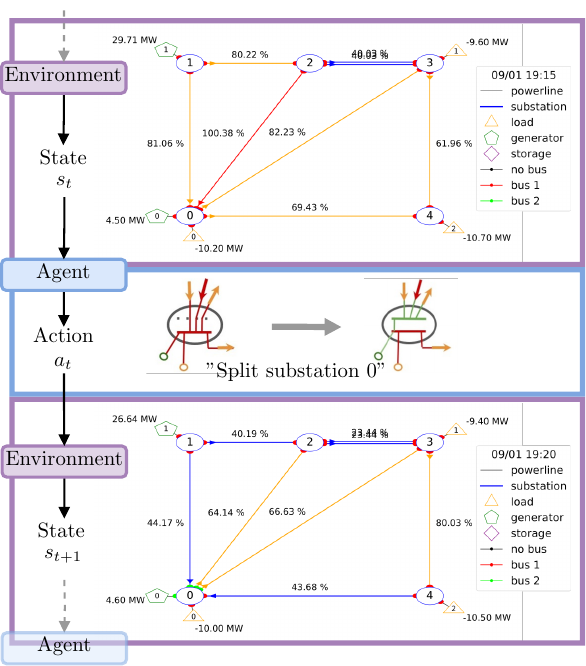}
    \caption{Example of a bus-splitting action by an agent and the effect on the environment. Some elements of substation 0 are assigned to bus 2, which results in a split of substation 0, and a consequent power flow redistribution.} 
    \label{Fig:Grid2OpExpl}
\end{figure}

Grid2Op provides a typical setup of a \textit{Markov Decision Process} (MDP) defined by a tuple $(\mS,\A, p, r)$, where at each time step $t$ an agent observes a state $s_t\in \mS$ from the environment and takes an action $a_t \in \A$. The Grid2Op environment computes and returns the next state $s_{t+1}\in \mS$ to the agent with probability $p(s_{t+1}|s_t, a_t)$, which is the unknown state transition probability, after which the agent receives an immediate reward $r(s_t, a_t) \in \R$ according to a defined reward function. This agent-environment interaction is visualized in \cref{Fig:Grid2OpExpl}. With this MDP formulation, reinforcement learning (RL) can be used to learn a (stochastic) policy $\pi(a_t|s_t)$ that optimizes the expected discounted reward $\E_\pi [\sum_{t=0}^T \gamma^t r(s_t, a_t)]$, where $\gamma \in (0,1)$ is the discount factor. 

\FloatBarrier

In Grid2Op, the states $s_t \in S$ include all grid information such as the topology of the grid, which includes the connections of the electrical elements, the power flow on the lines, power generation, and load consumption.
As mentioned in \cref{App:IntroPowerGrids}, the topology of the power grid can be represented as a graph, which we formalize as graph $G=(V,E)$, with nodes $v\in V$, as substations, and edges $e\in E$, as transmission lines, as described in \cite{lerousseau2021design}. Each substation connects various \textit{elements} such as line ends, \textit{generators} (producers), \textit{loads} (consumers) and \textit{storage units} (batteries, introduced in \cite{serre2022reinforcement}).

The total action space $\A$ consists of four types of actions:
\begin{enumerate}
    \item Bus splitting\footnote{By default, Grid2Op assigns two busbars per substation and only recently, starting from Grid2Op 1.9.9., it became possible to change this configuration and increase the number of busbars. Therefore, all competitions up to now and the solutions discussed in this paper consider only two busbars per substation.}, shown in \cref{Fig:Grid2OpExpl} and explained in \cref{subsec:Gen2Cons}; 
    \item Line switching (line reconnection/disconnection);
    \item Redispatch or curtailment of generator productions;
    \item Storage changes (if available).
\end{enumerate}
 Bus-splitting and line-switching actions are cost-effective for dynamic response to contingencies, as they indirectly redistribute line flows. These \textit{topological actions} can effectively mitigate contingencies or alleviate line congestion without requiring more expensive generator or load adjustments. 

\subsubsection{Rules of the game}\label{subsec:RulesL2RPN}
As discussed in \cref{subsec:Gen2Cons} grid operators need to keep the transmission lines within their thermal limit, maintain the voltage within a safe range and balance the generation and load. Similarly, agents in the L2RPN context must ensure that no line exceeds its thermal limit. Line loading, denoted by $\rho$, is defined by the ratio between the amount of power flow passing through and the thermal limit. When a line $e\in E$ is overloaded for too long, i.e. $\rho_e > 1.0$, it will automatically disconnect and a specific number of time steps is needed before this element can be activated again.

Voltage control, and generation-load balance are handled by the environment in the current L2RPN challenge. However, redispatching generation can help relieve line congestion and can be considered a useful action in the RL challenge. 

To mimic the physical properties of the power network, each substation and each power line has a \textit{cooldown time}, which is the time needed after activation before this element can be adjusted again. Furthermore, there is a limited number of actions that can be done by an operator or by the technology, therefore, by default, Grid2Op can handle only one substation and one line-switching action in each time step.

% Each power line has a thermal limit, which represents the maximum amount of power it can safely carry without being damaged. Line loading, represented by $\rho$, is defined by the ratio between the amount of power flow passing through and the thermal limit. When a line $e\in E$ is overloaded for too long, i.e. $\rho_e > 1.0$, it will automatically disconnect and a specific number of time steps is needed before this element can be activated again.

In the context of this RL challenge, a \textit{game-over} state will occur when an agent fails to safeguard the power grid, i.e., any network disconnect or isolation of a load or generator. The overarching objective in grid management is to control power flows in a way that prevents overloads and avoids cascading failures that could destabilize the entire network.

    The rules described above align with an $N-0$ setting, meaning no contingencies are anticipated. By adding an adversarial agent, or opponent, in the L2RPN challenge described in \cref{SubSec:Neurips2020}, the proposed RL solutions are required to act more securely.
    In real-world operations, TSOs typically adhere to the stricter $N-1$ reliability standards, ensuring that the grid remains secure even after the failure of any single grid component.

\subsection{Research motivation and contributions}\label{Sec:Motivation}
% \paragraph{Motivation} 
Since the introduction of the L2RPN challenge in 2019 in \cite{marot2020learning}, the field of RL applied to power grid control has seen remarkable growth. Several subsequent challenges have been organized (\cite{kelly2020reinforcement, marot2020l2rpn, marot2022l2rpn, serre2022reinforcement, delft2023l2rpn}) and the Grid2Op package has been extended to include larger, more complex networks. These updates introduce additional challenges such as unexpected line attacks, increased penetration of renewable energy, integration of battery storage, and sending timely alerts in case of dangerous contingencies. Each challenge has prompted innovative solutions from the research community, contributing to a rapidly evolving landscape.

This rapid development presents challenges on its own. With a growing number of approaches and techniques emerging from various research efforts, it can be difficult to navigate the landscape and select the most appropriate method.
Moreover, some solutions lack thorough documentation,  and publicly available codebases often do not clearly convey the rationale behind specific design choices. As a result, replicating or building upon existing methods becomes a cumbersome and time-consuming task.

% \section{Our Contributions}
% \paragraph{Our contributions} 
To address these challenges, this survey paper aims to provide a comprehensive and structured overview of the methods and challenges in reinforcement learning (RL) applied to grid topology control. Our aim is to guide researchers in selecting effective strategies and more easily identifying open research questions in this fast-moving field. 

In this paper, we present the following contributions. 
\begin{enumerate}
    \item First, we summarize and analyze the evolution of RL challenges and solutions, providing insights into how the field has progressed over time in \cref{sec:overview}. 
    \item Next, we review the various techniques that have been proposed to enhance the performance of RL agents in grid topology optimization in \cref{sec:techniques}.
    \item In \cref{chap:benchmark}, we discuss benchmarking practices, performance metrics, and baseline methods used in the literature to evaluate RL agents in grid topology optimization, highlighting the best practices.
    \item In \cref{sec:experiments,sec:results} we conduct comparative numerical studies to evaluate the effectiveness of selected techniques, providing empirical evidence of their impact on performance.
    \item  We provide practical guidelines and recommendations to support future research and development in the field, summarized in \cref{sec:discussion}.
    \item Lastly, \cref{sec:conclusion} concludes with a summary of our key insights and directions for future research.
\end{enumerate}

%%%%%%%%%%%%%%%%%%%%%%%%%%%%%%%%%%%%%%%%%%%%%%%%%%%%%%%%%%%%%%%%%%%%%%%%%%%%%%%
%%%%%%%%%%%%%%%%%%%%%%%%%%%%%%%%%%%%%%%%%%%%%%%%%%%%%%%%%%%%%%%%%%%%%%%%%%%%%%%
\section{Overview of Challenges and Solutions}\label{sec:overview}
This chapter provides a comprehensive overview of the L2RPN competitions, highlighting the most significant contributions to each challenge. \cite{web2022l2rpnwebpage} has been used to help create this overview. Several solutions proposed outside the L2RPN competitions are also discussed in the last part of this section.  For a broader understanding and detailed description of the challenges, we direct readers to \cite{kelly2020reinforcement} and the articles mentioned in the specific sections. 
For many of the proposed solutions, we introduce acronyms for easier reference and recognition throughout the paper. These acronyms follow the format \textcolor{violet}{AAA-YYYY}, where \textit{AAA} represents a few descriptive letters and \textit{YYYY} indicates the year. We provide an overview of all used acronyms and the paper references in \cref{Table:SolAcr} in \cref{App:Acronyms}.

% As mentioned in \cref{Sec:Motivation}, we remark here that the lack of comprehensive documentation in some of these works has sometimes made it difficult to systematically navigate them, which is why we sometimes provide less detailed information. 

%%%%%%%%%%%%%%%%%%%%%%%%%%%%%%%%%%%%% IJCN 2019 %%%%%%%%%%%%%%%%%%%%%%%%%%%%%%%%%%%%%
\subsection{Competition and solutions L2RPN 2019 IJCNN - Sandbox}
The first challenge, introduced in 2019 and described in \cite{marot2020learning}, was based on the IEEE 14-bus system. The platform used for this challenge was PyPowNet (\cite{lerousseau2021design}), which has been replaced with Grid2Op in subsequent challenges. In this challenge, the goal was to design an agent that can operate a power grid in real-time over a one-month period using only topology changes. That is, only bus-splitting and line-switching actions were used. 

This competition had two winning solutions by \cite{matavalam2022curriculum} (\hypertarget{sol:a3c2019}{\textcolor{violet}{A3C-2019}}), and \cite{lan2020ai} (\hypertarget{sol:d2qn2019}{\textcolor{violet}{DDQN-2019}}). 

\aaacfirst\ is the only solution that does not use the Grid2Op simulation function and is, therefore, purely RL-based. It uses an asynchronous-advantage-actor-critic (A3C) algorithm by \cite{mnih2016asynchronous} with a curriculum learning strategy, see \cite{bengio2009curriculum}, where the idea is that the neural networks learn more effectively when they are first trained on a simpler task. Furthermore, the authors reduce the action space by eliminating all busbar symmetry actions within a substation, since these result in an identical situation and are therefore redundant. This is something that, in subsequent solutions, is done by default and is included when collecting all substation topology actions in the Grid2Op package.

The solution \ddqn\ uses \textit{imitation learning} (IL), see \cite{kober2010imitation}, combined with a Dueling DQN algorithm, \cite{wang2016dueling}, that is trained using importance sampling or prioritized experience replay,  \cite{schaul2015prioritized}. They introduce a warning flag based on the loading of a line level. This warning flag can be seen as an \textit{activation threshold} that determines when the agent should interfere with the network to avoid further escalation. When the load on all lines is below the activation threshold, the agent is not activated, but a simple do-nothing action is given to the environment. In all subsequent solutions, an activation threshold is used. In addition, \textit{guided exploration} is used to speed up training. The final solution simulates the top $N$ actions given by the agent and takes the best action.

%%%%%%%%%%%%%%%%%%%%%%%%%%%%%%%%%%%%% WCCI 2020 %%%%%%%%%%%%%%%%%%%%%%%%%%%%%%%%%%%%%
\subsection{L2RPN 2020 WCCI Competition - A feasibility challenge}\label{Subsec:2020WCCI}
In spring 2020, the L2RPN challenge was scaled up to a larger grid of 36 substations, representing one-third of the IEEE 118-bus system. The time horizon was increased to one year, with planned maintenance outages that the agent had to account for when operating the network. The set of actions available to the agents in this competition was still limited to topology actions. Compared to the previous challenge in 2019, this challenge is notably much more difficult due to the increased action space size, which is approximately 65k. This underscores the importance of finding an effective method for managing the large action space.

The winning solution of this competition is the approach of \cite{yoon2021winning} (\hypertarget{sol:smaac}{\textcolor{violet}{SMAAC-2021}}). This solution combines the Soft Actor-Critic (SAC) algorithm \cite{haarnoja2018soft} with the use of \textit{afterstates}, which can also be referred to as the post-decision state \cite{powell2007approximate}. The afterstate refers to the state after the agent made its move, but before the environment responded. Learning the afterstate instead of the action reduces the output of the agent's neural network drastically. An intriguing element of this solution is that the SAC features a \textit{Graph Neural Network} (GNN) structure, which is well-suited due to the graphical nature of the power grid. Several other studies have followed up on this idea; for a recent overview of graph reinforcement learning methods applied to power grids, see \cite{hassouna2024graph}.
This survey discusses \cite{vandersar2023marl,dejong2024imitation,dejong2025generalizable}. 

Furthermore, the authors of \cite{yoon2021winning} note that the usage of the activation threshold yields a semi-Markov decision problem (semi-MDP) for the RL agent. Together, this leads to the Semi-MDP Afterstate Actor-Critic solution (SMAAC). In \smaac, the RL part of the agent focuses solely on the topology actions of the substations, while the line-switching actions are rule-based and follow a simple rule: reconnect a line whenever it becomes disconnected. This line-reconnection rule is a new addition compared to previous solutions, and has been adopted in most subsequent approaches.

To the best of our knowledge, the solution of the second-place winner, %team zenghsh3, 
is unfortunately not publicly available nor documented. However, we believe this solution likely served as a precursor to the winning approach of the L2RPN 2020 NeurIPS competition, based on the team's name, which matches the GitHub handle of a contributor to that solution.

The third-place winners of the 2020 WCCI competition \cite{yan2020l2rpn} (\hypertarget{sol:a3c2020}{\textcolor{violet}{A3C-2020}}) created a solution inspired by the two winners of the previous competition. \aaacsecond\ employs an A3C algorithm. They reduced the action space to 596 actions based on simulation experience and some random selection. Similarly to \ddqn, they used guided exploration during training. Two agents were trained and combined in their final submission. When one agent fails to propose a good action according to the simulation, the other agent is activated.

%%%%%%%%%%%%%%% NEURIPS 2020 %%%%%%%%%%%%%%%%%%%%%
\subsection{L2RPN 2020 NeurIPS - In a sustainable world} \label{SubSec:Neurips2020}
The L2RPN 2020 NeurIPS challenge was held in the summer of 2020. The competition design can be found in \cite{marot2020l2rpn}, and the results of this challenge are discussed in \cite{marot2021learningretrospective}. This challenge consisted of two parts; the robustness challenge and the adaptability challenge. For both parts, the agent now also has access to continuous redispatching and curtailment actions in addition to topological actions.

% \paragraph{Robustness challenge.}
The robustness challenge uses the same grid employed for the 2020 WCCI challenge. In addition, an adversarial agent is introduced that attacks certain grid lines \cite{omnes2021adversarial}.

% \paragraph{Adaptability challenge}
In the adaptability challenge, the grid size is increased to 118, thus including the whole IEEE 118-bus system. Additionally, there has been a shift in the energy production mix, with the share of renewable sources increasing from 10\% to 30\%.

The solution of \cite{zhou2021action}, Search with Action Set 
(\hypertarget{sol:SAS}{\textcolor{violet}{SAS-2021}}), won first place in both competitions. \sas\ works as follows; 
the stochastic policy $\pi_\theta(a_t|s_t)$, parameterized by $\theta$ using a neural network, first outputs a vector of probabilities with which the actions are sampled. Next, the top $K$ actions with the highest probabilities are selected to form the action set $A$. The final action is greedily selected using a risk function based on the load of the lines. This part can be compared with the guided exploration technique first used in \ddqn. To update the policy, they apply evolutionary strategies \cite{salimans2017evolution} with black-box optimization to maximize the reward. In this competition, a new rule-based strategy is introduced: The agent \sas\ reverses the topology to the reference topology, where all elements are connected to the original busbars, when the network is in a safe state\footnote{Based on their code accessed on 25-10-2024 https://github.com/PaddlePaddle/PARL/tree/develop/examples/NeurIPS2020-Learning-to-Run-a-Power-Network-Challenge.}.

\cite{binbinchen2020neurips} (\hypertarget{sol:binbin}{\textcolor{violet}{Binbinchen-2020}}) won the second-place for the robustness challenge. The authors propose a \textit{Teacher-Tutor-Junior-Senior} framework. Here, the Teacher first defines a smaller action space of 208 actions by applying a greedy agent to the entire action space and saves all used greedy actions in a reduced action space (RAS). The RAS is used by the Tutor, the tutor applies actions, again greedily, and the Junior updates a neural network using IL. Ultimately, the Senior employs this neural network to initialize the actor network for the Proximal Policy Optimization (PPO) algorithm by \cite{schulman2017proximal}. 

The second place in the adaptability challenge was won
%by the team \textit{kunjietang}. Their 
with an approach that used a straightforward expert system that ran simulations on a limited set of 200 topological actions. 

The third place for both challenges was achieved by \cite{zhihong2020neurips} (\hypertarget{sol:d3qn2020}{\textcolor{violet}{D3QN-2020}}) using a Dueling-Double-DQN algorithm, combining \cite{wang2016dueling} and \cite{vanhasselt2016deep}. Their solution was inspired by the solution \aaacsecond\ of \cite{yan2020l2rpn}, using guided exploration and combining two agents that work together with different strategies. Similarly to the \sas\ solution, they also included the rule-based strategy to revert to the reference topology when in a safe state for the robustness track. Furthermore, they focused on creating a suitable reward function to stimulate the agent to take the right actions.

\subsection{L2RPN 2021 ICAPS - Run a power network with trust}
In 2021, the L2RPN challenge was organized with the focus on building trust between an agent and the human operator. The grid used was the 36 substation grid as in the 2020 NeurIPS robustness challenge, again including events such as planned maintenance and line attacks. In addition to the previous challenge, the agent can raise an alarm when the assistance of an operator is needed due to a risk of failure.  The alarm score function is defined as follows: When a game-over occurs, but the agent raises an alarm within a decent time frame, the agent receives a positive alarm score. Otherwise, the agent will receive a negative score. The grid is divided into three areas, and if the alarm is raised in the correct area, the score will be higher. To avoid overactive agents, the alarm has an attention budget. For more details, see the description of the challenge and the results in \cite{marot2022l2rpn}.

None of the ICAPS 2021 competition winners wrote a paper to our knowledge, making it difficult to find documentation on their solutions, but the presentations of their solutions can be found at \cite{rteandepri2021present}.

The first place was achieved 
%by team xd-silly 
by \cite{wang2021l2rpnicaps}. Their solution consists of four modules: an \textit{Expert module}, an \textit{Agent module}, an \textit{Emergency module} and a \textit{Alarm module}. The expert module consists of rule-based parts that are similar to previous solutions, e.g., reconnecting disconnected lines and going back to the reference topology when the grid is in a safe state. The agent module seems to be based on the solution \sas\ (\cite{zhou2021action}), using the same action space and the neural network from this solution for the policy network that recommends actions. They added two extensions to the solution:
\begin{mylist}
    \item In case a single topology action is not able to solve the contingency, a multi-step topology action is tried.
    \item An emergency module; If topology actions cannot solve the problem, redispatch actions are tried to reduce the overflow. Moreover, if these actions are insufficient and the grid is at risk of failure in the next time step, line disconnection actions are evaluated using simulation.
\end{mylist}
Lastly, the alarm module raises an alarm when the agent cannot solve the overflow.

% * Expert module: reconnect disconnected lines. If no overflow go back to reference topology. 
% * Agent module: Policy network to recommend actions, unclear what kind of RL algorithm they used. The top actions are simulated and if the best action can solve the current issue this action is returned. Unclear which action space is used. If a single topology step is not able to solve the contingency a multi-step topology action is tried using a "Policy network wth planning", again unclear what they used exactly. If the problem can't be solved using only topology actions, redispatch actions are tried to alleviate the overflow.
% * Simple alarm module: If the agent cannot solve the overflow, an alarm is raised.

The team of \cite{martinez2021l2rpnicaps} won second place in the 2021 ICAPS competition. The problem is split into an operational and an alarm strategy. Their solution uses the action space of \binbin\ (\cite{binbinchen2020neurips}) with only 208 actions. 
The operational strategy is a simple greedy strategy: If the grid is in danger, all actions are simulated, and the action with the lowest maximum line load $\rho_{\max}$ value is chosen. The alarm strategy is a bit more advanced; they used a combination of rule-based strategies, 
%raising an alarm when the $\rho_\max$ value is above a certain value $X_i$ and the alarm budget is above a certain value $Y_i$, 
with a neural network.

The third place in the competition was secured by \cite{enlite2021mazerl}. This solution served as a precursor to their winning approach for the 2022 WCCI competition \alphazero\ (\cite{dorfer2022power}) also using the AlphaZero approach with Monte Carlo Tree Search (MCTS) sampling by \cite{silver2017mastering}, as the core RL algorithm. Notably, in this solution, they preserved the original action space to avoid losing potential beneficial actions. A benefit of their solution is that the MCTS tree already requires looking into the future and, therefore, helps identify critical situations early. This capability is then used for the alarm strategy.

\subsection{L2RPN 2022 WCCI and 2023 TU Delft - Energies of the Future and carbon neutrality}
In the challenge of 2022 introduced in \cite{serre2022reinforcement}, the energy mix is changed to obtain more realistic future scenarios to reach carbon neutrality in 2050; 40\% of the energy mix consists of renewables; wind and solar power and less than 3\% is fossil fuels. The use of renewable energy generators is privileged and the use of fossil fuels is penalized to steer agents. The results of the competition can be found in the blog post of \cite{pavao2023learning}.

As mentioned in the previous section, \cite{enlite2021mazerl} improved their solution submitted to the L2RPN 2021 ICAPS competition. Unlike their previous solution, the action space is now narrowed down to 2000 actions using a brute-force selection method. In addition to the learned topology agent, they introduce a redispatching controller that uses Cross-Entropy (CE) optimization by \cite{deboer2005tutorial}. Topology actions are preferred over redispatch actions, but when both cannot solve the congestion, they rerun the redispatch optimization on the top five proposed topology action candidates. With the improved solution \hypertarget{sol:alpha}{\textcolor{violet}{AlphaZero-2022}}, \cite{dorfer2022power} won the 2022 WCCI competition. The authors also wrote a follow-up paper on how to integrate such an RL-based agent into existing grid operation workflows, \cite{fuxjager2023reinforcement}.

The solution of \cite{alibaba2022l2rpn} (\hypertarget{sol:brute}{\textcolor{violet}{BruteForce-2022}},) obtained the second place, using a brute-force search for topology actions and the optimization agent optimCVXPY from L2RPN baselines \cite{rte2020baselines}. They used the method of the Teacher in \binbin\ to gain a reduced action space of 314 actions. The authors also experimented with RL-based agents but were unable to achieve better results, showing the difficulty of this problem. 

The third place in the 2022 WCCI competition was achieved by \cite{hri2022trial} (\hypertarget{sol:hri}{\textcolor{violet}{HRI-EU-2022}}). Their agent also did not include any learning. They randomly selected 1000 line-switching actions combined with redispatching actions, and at each time step, select the action that reduced the maximum line loads $\rho_{\max}$ the most. Note that they do not make use of substation topology actions.

The competition of \cite{serre2022reinforcement} was repeated in 2023 \cite{delft2023l2rpn}, which resulted in two new interesting results.

The solution of \cite{lajavaness2023l2rpn} (\hypertarget{sol:ljn}{\textcolor{violet}{LJN-2024}}) won the competition, using the curriculum agent (\curriculum) approach of \cite{lehna2023managing}, discussed in \cref{Sec:SolutionsOutside}, combined with the optimCVXPY baseline agent of L2RPN. They reduced the action space using three types of Teachers. 
A general teacher who finds the best action when an overflow occurs on the grid, an attacking teacher who finds the best action in case of an attack on a power line, and finally a $N-1$ teacher who tries to find the best actions to mitigate the effect of a potential attack on a power line.

The second place was for the team of \cite{artelys2024} (\hypertarget{sol:artelys}{\textcolor{violet}{Artelys-2024}}) who used a very similar approach, also building an agent based on the \curriculum\ approach of \cite{lehna2023managing}. For the continuous redispatching actions, they used convex optimization (optimCVXPY). Their approach for using redispatching actions is very much the same as in the approach of \alphazero\ (\cite{dorfer2022power}), where an additional note is made that the agent prefers actions that do not limit renewable energy sources.

\subsection{Solutions outside of competitions}\label{Sec:SolutionsOutside}
Quite some researchers proposed interesting solutions outside of competitions; the most relevant solutions to our knowledge will be chronologically discussed in this section.

\paragraph{2021} 
In the paper of \cite{Subramanian2021} (\hypertarget{sol:cem}{\textcolor{violet}{CEM-2021}}), the Cross-Entropy Method by \cite{deboer2005tutorial} is applied to a modified version of the 14-bus power grid, \texttt{case\_14\_realistic} in Grid2Op. With this simple RL approach and a smart selection of episodes, the authors were able to train an agent to successfully operate $96.5\%$ of the scenarios consisting of one week. The research considers bus-splitting actions only and proposes a reduction of the action space by excluding actions where only one element is connected to a busbar and actions where none of the connected elements is a line (these are infeasible actions and are currently excluded by default when collecting topology actions in Grid2Op). Furthermore, they show how the number of actions per substation can be computed. An interesting part of this research is the evaluation of the agent, which shows that the agent only needs very few different topologies to manage the grid. Their agent manages the grid with topology actions at substations 1, 3, and 8. 

\paragraph{2022} 
\cite{damjanovic2022deep} (\hypertarget{sol:d3qn2022}{\textcolor{violet}{D3QN-2022}}) also consider the 14-bus power grid with only bus-splitting actions. They use RL-lib to apply multiple versions of the DQN algorithm and compare their performance, noting that Double Dueling DQN with prioritized replay has the best performance. They use a highly reduced action space considering only 9 actions and a limited part of the observation space compared to other solutions. The final agent is tested on scenarios of one month and, on average, survives $82\%$ of the 8064 time steps.

% \cite{qiu2022distribution} \textcolor{red}{REPEATING WORK OF SMAAC... PERHAPS IGNORE IN THIS OVERVIEW?}

The topological agent in \cite{henka2022power} minimizes line overflow by solving a mixed integer linear programming (MILP) based on linear constraints from DC approximation. \cite{henka2022power} propose a three-step pipeline to define a grid segmentation used for a multi-agent approach (a decentralized control technique with topological closed-loop controllers). First, they create an influence graph using \textit{line outage distribution factor} (LODF). Second, a clustering of this influence graph is made. Lastly, clusters are selected using quality criteria.

% \cite{henka2022power} Using \textit{line outage distribution factor} (LODF) to find segmentations and MILP to solve the DC problem on case 118. I wonder why it is even computationally feasible to do this using MILP and somehow it is not able to survive so the MILP is not optimal? 
\paragraph{2023} 
The work of \cite{chauhan2023powrl} (\hypertarget{sol:powrl}{\textcolor{violet}{PowRL-2022}}) considers the 2020 NeurIPS robustness challenge, which concerns the 36-bus power grid and an attacking opponent. Similarly to \cite{Subramanian2021}, they show a formula to compute the number of feasible topological bus-splitting actions, without the one-element constraint. The RL agent considers bus-splitting actions only. Before training, the action space is reduced to 240 actions using brute-force simulation. The authors describe how they combine heuristics with deep RL. Here, the heuristics are the rule-based parts of the method, which are similar to previously proposed methods: do not activate the RL agent when there is a safe state, reconnect lines when possible, and revert to the default network topology when the contingency ends. The authors introduce one rule that is different from previous solutions: Disconnect lines when there is a sustained period of overflow to avoid permanent damage. The RL agent is trained using PPO with prioritized replay.

The solution \binbin\ is extended in the paper of \cite{lehna2023managing} (\hypertarget{sol:cur}{\textcolor{violet}{Curriciculum-2023}}). This work focuses on the robustness track of the 2020 NeurIPS challenge. They extend the action space of \binbin\ by introducing an $N-1$ \textit{Teacher} and an $N-1$ \textit{Tutor} for imitation learning of the \textit{Junior} agent. The $N-1$ algorithm used for the Teacher and the Tutor selects the best $N-1$ secure action, rather than defaulting to a greedy action that assumes all lines remain connected. This is achieved by simulating the combination of each topological action with the disconnection of each line from a predefined subset, and selecting the best $N-1$ secure actions based on the resulting $\rho_{\max}$ values.
% by combining all topological actions with the disconnection of a subset of lines that can be disconnected and using simulation to see the resulting $\rho_{\max}$ values. 
This enhancement increased the considered action space of \binbin\ from 208 to 508 actions. In addition, they added a topology reversion improvement, as used in solutions \sas,% (\cite{zhou2021action}), 
\dddqnfirst\ %(\cite{zhihong2020neurips}) 
and \powrl.% (\cite{chauhan2023powrl})
Lastly, the authors made some important code improvements; for details, see \cite{lehna2023managing}. This leads to an improved score of $49.12$ where \binbin\ got $46.89$ according to \cite{marot2021learningretrospective}.

A novel approach using hierarchical reinforcement learning is introduced in \cite{manczak2023hierarchical} (\hypertarget{sol:hrl}{\textcolor{violet}{HRL-2023}}). The problem considered is the modified 14-bus network (as in \cite{Subramanian2021}) with and without an adversarial agent. In the hierarchical structure, the authors consider three levels. At the highest level, the agent needs to decide whether an action is needed. This is a rule-based agent that uses the activation threshold as in previous solutions. At the intermediate level, a new agent is introduced that determines where an action is needed. This agent is RL-based, the experiments include both SAC as well as PPO algorithms. At the lowest level, the agent picks the specific bus-splitting action. Experiments are performed with both greedy and RL-based algorithms at the lowest level. The intermediate level forces the lowest level agent to act at the correct substation by applying an action mask.

The authors of \cite{vandersar2023marl} (\hypertarget{sol:marl}{\textcolor{violet}{MARL-2023}}) play with a similar idea, but extend the RL-based lowest level with substation-specific agents in contrast to the method of \hrl\ where the lowest level consists of one agent that is steered by the intermediate level agent using an action mask. This concept results in a multi-agent reinforcement learning method (MARL). The authors do experiments using the SACD or PPO algorithm with dependent or independent multi-agents. In this research, the intermediate or mid-level agent is purely rule-based. The concept is applied to the 5-bus power grid, leaving the research applied to larger grids to be explored.  

\paragraph{2024} 
\cite{hu2024towards} introduce a hierarchical multiobjective Markov decision problem (HMO-MDP) to address more fairness between the power plants. This is done by introducing a suitable reward function and is tested using (Advantage Actor-Critic) A2C and PPO.

In \cite{lehna2024hugo} (\hypertarget{sol:hugo}{\textcolor{violet}{HUGO-2024}}), the authors propose new enhancements to their previous \curriculum\ solution. In this approach, the greedy Tutor is used to learn \textit{Target Topologies} (TTs), which are selected based on their robustness. A target topology is a topology that was found to be safe after a contingency. Here, safe means that the agent is not triggered by the activation threshold to adjust the topology of the network due to a high $\rho_{\max}$ value. The most frequent TTs are saved and used Greedily in the final agent when the maximum load on the network lines is above a certain activation threshold. This activation threshold is slightly lower than usual because often multiple actions are needed to reach the final TT. The authors tested the proposed method on the 118-bus network of the WCCI 2022 challenge. 
They improved on their previous solution, which had a mean survival time of $1160$, to a mean survival time of $1436$. 

In the study conducted by \cite{dejong2024imitation} (\hypertarget{sol:il}{\textcolor{violet}{IL-2024}}) IL is evaluated on the 14-bus network with and without planned and unplanned outages. An agent is trained using data from two expert agents, a greedy and an $N-1$ expert, for details, see \cite{dejong2024imitation}. 
These expert agents can be regarded as the Tutor in Teacher-Tutor-Junior-Senior framework used in \binbin\ and \curriculum . The agent trained using imitation learning can be compared to the Junior.

An interesting topic that is addressed in the paper of \cite{wang2024alleviating} (\hypertarget{sol:bdqn}{\textcolor{violet}{BDQN-2024}}) is the imbalance of states and actions in the PNC problem. They note that this leads to degradation in the agents' performance and propose a Balance DQN algorithm to tackle this issue. This algorithm is built upon the ApeX-DQN framework of \cite{horgan2018distributed} and uses $k$-means-based predefined curriculum learning (KPCL) in combination with Option-DQN, inspired by the option framework in \cite{sutton1999between}, to tackle the imbalance issue.

In \cite{boguslawskiemulation} (\hypertarget{sol:zonal}{\textcolor{violet}{Zonal-2024}}) the authors address a slightly different approach to the L2RPN problem. Their agent acts as a zonal controller that receives a plan from a planner. The RL-based part of the agent focuses on the continuous actions; curtailment and storage charging plans. A heuristic expert agent manages the topological actions and determines whether an intervention from the RL agent is necessary in emergency situations. The RL-agent, trained with PPO, receives both grid information and the planner's target plan as state input, aiming to adhere to the plan as closely as possible.

\cite{losapio2024state} propose a state and action space factorization technique to distribute the PNC task across multiple agents. Unlike previous multi-agent approaches that rely on the network’s graph structure, their method uses a data-driven decomposition of the problem. While this technique has not yet been integrated into an RL-based solution, it shows potential for PNC applications.

\paragraph{2025} 
The authors of \cite{dejong2025generalizable} build on their previous work on IL in \cite{dejong2024imitation}, investigating the use of graph neural networks (GNNs) versus fully connected neural networks (FCNNs) for grid topology control (\hypertarget{sol:gnn}{\textcolor{violet}{GNNIL-2025}}). They propose two GNN variants, referred to as heterogeneous and homogeneous, based on different graph representations. Their results show that both GNN types generalize better than FCNNs, with the heterogeneous GNN achieving the best overall performance.

\cite{demol2025centrally} propose a centrally coordinated multi-agent architecture (\hypertarget{sol:ccma}{\textcolor{violet}{CCMA-2025}}) where multiple regional agents suggest actions, and a central agent selects the final action for the grid topology configuration.
The authors explore several variants, differing in whether the agents follow rule-based, RL-based, or greedy strategies. Among these, the RL-Greedy approach — where multiple greedy agents propose actions and an RL-based central agent makes the final decision — performs particularly well. Their experiments focus on case 14, both with and without an opponent, and suggest potential for action space factorization in larger cases.

A recent study on IL for PNC, \cite{hassouna2025learning}, introduces a \textit{soft-label} IL approach (\hypertarget{sol:SoftIL}{\textcolor{violet}{SoftIL-2025}}). Instead of directly learning from greedy-based action selection, as in previous methods, all actions are labeled based on the line loads during the simulations, providing a richer learning signal for the agent.
They evaluate this method using FCNN and GNN architectures on the 118-bus case (WCCI 2022 L2RPN). The soft-labeled GNN, in particular, achieves strong performance, outperforming current state-of-the-art RL agents.

% \textcolor{red}{
% TO DO: \cite{ji2024power}. -> REPLICA OF SMAAC\\ 

%%%%%%%%%%%%%%%%%%%%%%%%%%%%%%%%%%%%%%%%%%%%%%%%%%%%%%%%%%%%%%%%%%%%%%%%%%%%%%%
%%%%%%%%%%%%%%%%%%%%%%%%%%%%%%%%%%%%%%%%%%%%%%%%%%%%%%%%%%%%%%%%%%%%%%%%%%%%%%%
\section{Design choices for RL in Power Grid Control} \label{sec:techniques}
This chapter explores the key design choices involved in applying RL to PNC, with a focus on the implementation of various techniques and how they differ across solutions. \cref{Table:Techniques} provides an overview of the techniques used in a selection of the proposed methods described in \cref{sec:overview}. Due to incomplete documentation, not all solutions are represented in this or subsequent tables. 

This section is divided into two distinct sections. \cref{sec:RLTechniques} covers RL-specific design choices, such as action and state space definitions, reward function shaping, and other related matters. \cref{sec:RuleBasedTechn} focuses on heuristic, rule-based decisions, such as activation thresholds and line reconnection strategies.

\renewcommand{\arraystretch}{1.5}
\begin{table}[h]
\begin{adjustbox}{center}
\resizebox{1.2\textwidth}{!}
{
\tabcolsep=3pt
\begin{tabular}{|l|l|r|c|r|l|c|c|c|r|c|c|}
% \tagpdfsetup{table/header-rows=1}
\hline
{\textbf{Solution Acronym}} & \textbf{Competition} & \multicolumn{1}{l|}{\textbf{Size Grid}} & \makecell[tl]{\textbf{With}\\ \textbf{Opponent}} & \multicolumn{1}{l|}{\textbf{Rank}} & \makecell[tl]{\textbf{Algorithm}\\ \textbf{used}}  & \makecell[tl]{\textbf{Imitiation}\\ \textbf{learning}} & \makecell[tl]{\textbf{Prioritized}\\ \textbf{replay}} & \makecell[tl]{\textbf{Simulation}\\ \textbf{used}} & \makecell[tl]{\textbf{Activation}\\ \textbf{threshold}} & \makecell[tl]{\textbf{Reconnect}\\ \textbf{lines}}  & \makecell[tl]{\textbf{Revert}\\ \textbf{topology}} \\ 
\hline
\hline
\aaacfirst & IJCNN 2019 & 14 %(PyPowNet)  
&    & 1/2  & A3C  &    &   &   & 0.6  \tablefootnote{Based on code.\label{footnote:CodeBased}}   &     &  \\ 
\hline
\ddqn & IJCNN 2019 & 14 %(PyPowNet)  
&    & 1/2  & DDQN & \cmark  & \cmark & \cmark & 0.885   &    & \\ 
\hline 
\hline
\smaac   & WCCI 2020  & 14, 36 &   & 1& SAC  &   &  &  & 0.9  & \cmark &    \\ 
\hline
\aaacsecond & WCCI 2020  & 36    &   & 3& A3C  &   &  & \cmark & 0.8   &    & \\ 
\hline
\hline
\sas & NeurIPS 2020  & 36, 118& \cmark  & 1& ES    &  &  & \cmark & 1.0 \tablefootnote{\label{footnote:ActThresh} This activation threshold is compared with the $\rho_{\max}(t+1)$  after a simulated DN-action. So unlike other solutions where the current $\rho_{\max}(t)$ is checked.}     & \cmark   & \cmark\\ 
\hline
\binbin & NeurIPS 2020  & 36    & \cmark  & 2& PPO  & \cmark  &   & \cmark & \makecell[tr]{1.0 (Teacher)\\ 0.925 (Tutor)\\ 0.9 (Senior)\\ 0.999 (final agent)}  & \cmark &   \\ 
\hline
\dddqnfirst & NeurIPS 2020  & 36, 118& \cmark  & 3& D3QN &   & \cmark & \cmark & 1.0 \footref{footnote:CodeBased} & \cmark   & \cmark\\ 
\hline
\hline
\alphazero  & WCCI 2022  & 118   & \cmark  & 1& MCTS + CE &    &   & \cmark & 0.98 \tablefootnote{Although not explicitly documented, this value was used in their ICAPS 2021 solution, suggesting it was applied again here.}  & \cmark   & \cmark\\ 
\hline
\brute & WCCI 2022  & 118   & \cmark  & 2& \makecell[tl]{Greedy +\\ OptimCVXPY}    &    &   & \cmark & 0.95  & \cmark   & \cmark\\ 
\hline
\hri  & WCCI 2022  & 118   & \cmark  & 3& Greedy  &    &   & \cmark & 1.0 \footref{footnote:ActThresh} & \cmark &     \\ 
\hline
\hline
\ljn & WCCI 2023  & 118   & \cmark  & 1 & \makecell[tl]{PPO +\\ OptimCVXPY}  & \cmark  &   & \cmark & \makecell[tr]{ $[0.6;1.99]$ (Teachers)\\ 0.99 (final agent)}  & \cmark  & \cmark\\ 
\hline
\artelys & WCCI 2023  & 118   & \cmark  & 2& \makecell[tl]{PPO +\\ OptimCVXPY}  & \cmark  &   & \cmark & 0.98      & \cmark   & \cmark\\ 
\hline
\hline
\cem  & -& 14 %(realistic)        
&   &  & CEM      &  &  &  & 0.95  & \cmark &   \\ 
\hline
\dddqnsecond & -& 14 %(sanbox)  
&   &   & D3QN &    &  \cmark &   & 0.95  &     &  \\ 
\hline
\powrl   & -& 36    & \cmark  &  & PPO  &    &  \cmark & \cmark & 0.95 \footref{footnote:ActThresh} \tablefootnote{The specific activation threshold is not documented, but given that the reward function uses a safety threshold of 0.95, likely, this value was also applied as the activation threshold.} & \cmark   & \cmark\\ 
% \hline
\hline
\curriculum & - & 36    & \cmark  &  & PPO  & \cmark  &   & \cmark & \makecell[tr]{0.925 (Teacher)\\ 0.9 (Tutor)\\ 0.9 (Senior)\\ 0.95 (final agent) \footref{footnote:CodeBased}}  & \cmark   & \cmark\\ 
\hline
\hrl & - & 14 %(realistic)  
& \cmark  &  & PPO / SAC  &    &   & \cmark & 0.95  & \cmark &    \\ 
\hline
\marl & - & 5     &    &  & \makecell[tl]{MA-PPO /\\MA-SACD}&    &   &   & 0.9  & \cmark &     \\ 
% \hline
\hline
\hugo & - & 118   & \cmark  &  & PPO  & \cmark  &   & \cmark & \makecell[tr]{0.85 (TopoAgent)\\ 0.95 (final Senior)}  & \cmark   & \cmark \\ 
\hline
\il  & - & 14    & \cmark  &  & IL& \cmark  &   & \cmark & 0.97   &  & \cmark  \\ 
\hline
HMO-2024 & - & 14, 118& \cmark  &  & A2C/PPO        &    &   &   & N.A.\tablefootnote{Not Available. It is either not documented or not used.}  & N.A. & N.A. \\ \hline
\bdqn   & - & 14, 36    &    &  & ApeX-DQN   &    & \cmark &   & N.A. & N.A. & N.A. \\ \hline
\zonal  & - & 118       & \cmark    &  & PPO  &         &     &           & 0.9  & N.A.   & N.A.   \\ \hline
\gnnil  & - & 14        & \cmark    &  & IL   & \cmark  &     & \cmark    & 0.97 & \cmark &        \\ \hline
\CCMA   & - & 5, 14     & \cmark    &  & PPO  &         &     & \cmark    & 0.95 & \cmark & \cmark \\ \hline
\SoftIL & - & 118       & \cmark    &  & IL   & \cmark  &     & \cmark    & 0.9  & \cmark & \cmark \\ \hline
\end{tabular}
}
\end{adjustbox}
\caption{Overview of discussed solution methods.}
\label{Table:Techniques}
\end{table}
\FloatBarrier

\subsection{General RL Techniques}\label{sec:RLTechniques}

\subsubsection{Algorithm}\label{subsub:Algorithm}
As shown in \cref{Table:Techniques}, the most frequently used RL algorithm for PNC is PPO, featured in 11 of the 26 solutions reviewed in this overview, significantly more often than any other algorithm. Studies comparing PPO with other algorithms (\cite{manczak2023hierarchical, vandersar2023marl, hu2024towards}), such as SAC and A2C, consistently show that PPO performs better in PNC. 
Note, however, that the agent's performance depends not only on the choice of algorithm but also on other implementation details, which are discussed in the following paragraphs.
Additionally, various versions of DQN have been applied to the PNC problem. In \cite{damjanovic2022deep} (\dddqnsecond), the authors demonstrate that the prioritized D3QN, an advanced variant of DQN, significantly speeds up the training. 
To get an idea of the differences between these methods, we refer the reader to \cref{subsec:RL_terminology}.

% PPO is a state-of-the-art policy optimization algorithm that improves traditional policy gradient methods by introducing a clipped surrogate objective function. This ensures that policy updates remain close to the original policy, effectively balancing the trade-off between exploration and exploitation. 

% Papers where PPO is applied to learn to control the power network using topology actions in the "Learning to Run a Power Network" challenge:
% \cite{manczak2023hierarchical}, 

% In combination with the Teacher-Tutor-Junior-Senior framework (Curriculum agent):
% \cite{binbinchen2020neurips} 
% \cite{lehna2023managing} 
% \cite{chauhan2023powrl}
% \cite{lajavaness2023l2rpn} 
% \cite{artelys2024}

\subsubsection{Action space}\label{Subsub:ActSpace}
As outlined in \cref{Sec:Intro}, the actions available to operate the power network include bus splitting, line switching, generator re-dispatch or curtailment and, optionally, storage changes (introduced in \cite{serre2022reinforcement}). Topological actions -- bus splitting and line switching -- serve as a low-cost option to alleviate thermal overloads on the network, as highlighted in \cite{kelly2020reinforcement}. However, the large, non-linear, and combinatorial nature of the action space makes optimal control of the grid topology beyond current state-of-the-art capabilities (\cite{marot2020learning, marot2020l2rpn}), leaving these topological actions as a significant, yet underutilized, source of operational flexibility in power networks. 

The inability of existing methods to navigate such a vast action space has motivated the L2RPN competitions (see \cite{marot2020learning, kelly2020reinforcement, marot2020l2rpn}), where the primary goal is to use RL, to explore the full range of switching options to control power flows. As a result, topological actions are the main focus of agents used in L2RPN. \cref{Table:ActSpaces} provides an overview of the number of feasible bus-splitting and line-switching actions per case. It shows how the number of bus-splitting actions increases drastically when going from case 14 to case 36. 

Note that Grid2Op considers two types of actions for bus splitting and line switching; \textit{set} actions and \textit{change} actions. A \textit{set bus}  (or \textit{set status}) action specifies the configuration (or status) that the substation (or line) gets, regardless of its current state. This means that if a substation (or line) is already in the specified configuration (or status), executing the action leaves the configuration unchanged. In contrast, a \textit{change bus} (or change status) action defines which elements of a substation (or which lines) to switch to the other bus (or status). This results in an action that depends on the current configuration (status) of the substation (line). Executing the change action will always affect the topology (unless the affected element is in a cool-down state). It is easier to consider only the set actions since the feasibility of a change action depends on the current state of the affected element(s). Therefore, the action spaces discussed in the following refer only to set actions.

To describe the size of the action spaces, we first introduce some useful notations. As mentioned earlier, the power grid can be represented by a graph $G=(V,E)$, where $V$ is the set of substations (nodes) and $E$ is the set of transmission lines (edges). The size of substation $v\in V$ is denoted as $|v|$ and is equal to the total number of elements connected to the substation, that is $|v| = g(v) +l(v) + b(v) + e(v)$, where $g(v)$ is the number of generators, $l(v)$ the number of loads, $b(v)$ the number of storage (batteries), and $e(v)$ the number of lines connected to substation $v$. A line (edge) $e\in E$ that, when switched off, directly disconnects part of the grid is referred to as a \textit{bridge} (or cut-edge), denoted as $b\in B\subset E$. The total number of lines and bridges in the grid is denoted by $|E|$ and $|B|$, respectively.

In most Grid2Op environments and L2RPN challenges, the environment allows for only one new substation configuration at each time step. Therefore, the formula for the total number of feasible bus-splitting actions $|\A_{bus}|$ is simply the sum of all possible configurations per substation, that is
\begin{align}\label{Eq:SizeActionSpace}
    |\A_{bus}| &= \sum_{v \in V} |\A(v)|,
\end{align}
where $|\A(v)|$ is the number of topology actions that can be performed on substation $v$.

Computing all possible configurations of a substation $v$ with two busbars simply leads to $|\A(v)| = 2^{|v|}$ options. However, symmetric actions and actions that directly result in a game over by disconnecting a non-line element, i.e., load, generator or battery, from the rest of the grid can be excluded. This simple reduction leads to $|\A(v)|=|\Asym|$, which can be computed as 
\begin{align}\label{Eq:SubActionSpace}
   |\Asym| &= 2^{|v| -1} - (2^{g(v) + l(v)+b(v)}-1),
\end{align}
also presented by \cite{chauhan2023powrl}. Note that this computation also includes the actions of substations where only one configuration is possible due to connectivity constraints. %This is the case for all substations that are not part of a cycle, since adjusting the configuration will always disconnect one or more elements from the grid. 
Therefore, the result of this computation is slightly higher compared to what is presented in \cref{Table:ActSpaces}, which excludes these actions.

The number of feasible line-switching actions is given by 
\begin{align}\label{Eq:ActionLines}
    |\mathcal{A}_{line}| = |E| - |B|.
\end{align}
$|\mathcal{A}_{line}|$ excludes all line-switching actions involving bridges, as they would isolate parts of the network.

\begin{table}[h]
\begin{adjustbox}{center}
% \resizebox{1.2\textwidth}{!}
% {
\begin{tabular}{|l|r|r|r|r|r|r|r|}
\hline
\textbf{Environment} & \multicolumn{1}{l|}{\textbf{\# Subs}} & \multicolumn{1}{l|}{\textbf{\# Lines}} & \multicolumn{1}{l|}{\textbf{\# Gens}} & \multicolumn{1}{l|}{\textbf{\# Loads}} & \multicolumn{1}{l|}{\textbf{\# Bridges}} & \multicolumn{1}{l|}{\textbf{$|\mathcal{A}_{bus}|$}} & \multicolumn{1}{l|}{\textbf{$|\mathcal{A}_{line}|$}} \\ \hline
case 14 realistic     & 14 & 20  & 5        & 11  & 1          & 150             & 19  \\ \hline
case 14 sandbox       & 14 & 20  & 6        & 11  & 1          & 178             & 19  \\ \hline
case 36 all versions     & 36 & 59  & 22       & 37  & 1          & 66810           & 58  \\ \hline
% case 36 WCCI 2020     & 36 & 59  & 22       & 37  & 1          & 66810           & 58  \\ \hline
% case 36 2020 Neurips  & 36 & 59  & 22       & 37  & 1          & 66810           & 58  \\ \hline
% case 36 2021 Icaps    & 36 & 59  & 22       & 37  & 1          & 66810           & 58  \\ \hline
case 118 2020 NeurIPS & 118& 186 & 62       & 99  & 8          & 72107           & 178 \\ \hline
case 118 2022 WCCI\tablefootnote{This environment also includes 7 storage units.}    & 118& 186 & 62       & 91  & 8          & 72957           & 178 \\ \hline
\end{tabular}
% }
\end{adjustbox}
\caption{Overview of action spaces per environment.\tablefootnote{Some papers report a different number of bus-splitting actions. This might be related to different Grid2Op versions. The numbers reported here refer to the version adopted in this paper, that is Grid2Op 1.9.3.}}
\label{Table:ActSpaces}
\end{table}
\FloatBarrier

Due to the large number of feasible actions and the even greater number of possible network topology states, a majority of the proposed solutions incorporate some form of action space reduction. \cref{Table:Case14ActReduction,Table:Case36ActReduction,Table:Case118ActReduction} provide an overview of the actions considered (both by the RL agent and the rule-based heuristics), the reduced action space sizes, and the reduction strategies used in the solutions discussed in \cref{sec:overview}. The action space sizes refer specifically to those controlled by the RL agent, excluding actions handled by the heuristics. For most solutions, the RL agent focuses on handling the discrete topology actions, in some cases with, but in most cases without, line-switching actions. 
Combining line-switching and bus-splitting actions is possible, but this naturally yields an even larger action space of size $|\mathcal{A}_{tot}| = |\mathcal{A}_{bus}| \times |\mathcal{A}_{line}|$, which might unnecessarily complicate the problem. Only \ddqn, one of the solutions proposed for the first challenge, tried to combine both action spaces in its solution. In all other solutions, the agents stick to only one action per time step, either a bus-splitting or a line-switching action.

\begin{table}[h]
\begin{adjustbox}{center}
\resizebox{1.2\textwidth}{!}
{
\begin{tabular}{|l|p{0.15\linewidth}|p{0.12\linewidth}|p{0.12\linewidth}|p{0.12\linewidth}|p{0.3\linewidth}|p{0.15\linewidth}|}
\hline
\textbf{Solution Acronym} & \textbf{Actions for RL agent} & \textbf{Original Size Action Space} & \textbf{Reduced Size Action Space} & \textbf{\% of actions included} & \textbf{Reduction method} & \textbf{Actions for rule-based heuristics} \\ \hline
\aaacfirst    & Topology actions           & $312+20 = 332$ \tablefootnote{This action space includes 312 bus-splitting and 20 line-switching actions.}    & $156+20 = 176$     & 50.0\%& \textit{Symmetry reduction}.          & N.A. \\ \hline
\ddqn   & Combined topology actions: bus splitting and line switching. & $156 \times 20 = 3120 $   & 155 + 19 + 76 = 251 \tablefootnote{This action space includes 155 bus-splitting, 19 line-switching, 76 bus-splitting and line-switching combinations and 1 do-nothing action(s). }   & 8.0\% & N.A. & N.A. \\ \hline
\cem  & Bus splitting     & 150     & 112    & 74.7\%& \textit{($N-0$)-reduction}.     & Line switching. \\ \hline
\dddqnsecond   & Bus splitting     & 178     & 9      & 5.1\% & \textit{($N-1$)-reduction}, after which a subset of actions is selected that lead to "desirable power network topologies". Not documented what this entails. & N.A. \\ \hline
\hrl    & Bus splitting     & 150     & 106    & 70.7\%& \textit{($N-0$)-reduction}. Next, the actions for substations where only one configuration is possible are removed, and one explicit do-nothing action is added. & Line switching. \\ \hline
\il     & Bus splitting     & 150     & 112    & 74.7\%& \textit{($N-0$)-reduction}. & N.A.           \\ \hline
\gnnil  & Bus splitting     & 150     & 112    & 74.7\%& \textit{($N-0$)-reduction}. & N.A.           \\ \hline
\CCMA   & Bus splitting     & 178     & 73     & 41.0\%& \textit{($N-1$)-reduction}. & line switching \\ \hline
\end{tabular}
}
\end{adjustbox}
\caption{Action space reductions for case 14, the PyPowNet, realistic, and sandbox version. The first two solutions use case 14 in PyPowNet, which is not mentioned in \cref{Table:ActSpaces}.}
\label{Table:Case14ActReduction}
\end{table}
\FloatBarrier

In \cref{Table:Case36ActReduction,Table:Case118ActReduction} we omitted the ``Original Size Action Space'' column, as these solutions primarily focus on bus-splitting actions, $\A_{bus}$, for which the action space sizes are already detailed in \cref{Table:ActSpaces}. As can be seen in \cref{Table:Case36ActReduction}, the exceptions to this are \aaacsecond\ and \dddqnfirst.

% For most solutions, the RL agent focuses on handling the discrete topology actions, in some cases with, but in most cases without, line-switching actions. Only in the solution \dddqnfirst of \cite{zhihong2020neurips} applied to case 36 with an adversarial agent, the RL agent also uses continuous redispatching actions. 
% Due to the focus on bus splitting actions and since this part of the action space increases exponentially, \cref{Table:Case36ActReduction,Table:Case118ActReduction} consider the reduction of these topological bus splitting actions only. For the original action space size of the bus splitting actions, one can refer to \cref{Table:ActSpaces}.

Based on the current solutions, we distinguish seven methods applied to reduce the bus-splitting action space, each of which is explained in detail below.
\begin{enumerate}
    \item \textbf{\textit{Symmetry reduction}}: 
    This method includes all \textit{feasible} actions while excluding symmetric ones. Although it is now the default approach, it was first introduced during the 2019 L2RPN challenge. The formula for computing the size of this action space is given in \cref{Eq:SizeActionSpace}.
    \item \textbf{\textit{($N-0$)-reduction}} introduced by \cite{Subramanian2021} (solution \cem ):  
    This approach only includes actions where at least two elements (or zero) are connected to a busbar, with at least one of these being a line. According to experts, this strategy yields more stable actions. The number of actions for a substation $v\in V$ now reduces to $|\A(v)|= |\Anzero|$, where
    \begin{align}\label{Eq:N-0ActionSpace}
    |\Anzero| &= |\Asym| - (e(v)- \delta_{|v|,2} \cdot \delta_{e(v),2} - \delta_{e(v),1}) ,
    \end{align}
    % , \quad \text{with}  \\
    % \delta_{|v|,2} &= \begin{cases}
    %     1 \quad \text{ for }  |v| = 2,\\
    %     0 \quad \text{ otherwise}
    % \end{cases}  \quad \text{(the Kronecker delta)},
    % \end{align}
    with $\delta_{m,n}$ the Kronecker delta function, that is $\delta_{m,n}=1$ if $m=n$ and $0$ otherwise. \footnote{\cref{Eq:N-0ActionSpace} is a slightly improved version of the formula presented by \cite{Subramanian2021}.}

    \item \textbf{\textit{($N-1$)-reduction}} used in \dddqnsecond\ and \CCMA : 
    This method only includes actions where at least two elements (or zero) are connected to a busbar of which at least two elements are lines. The resulting reduced action space is more robust to potential line failures or planned maintenance. Using this strategy, \cite{demol2025centrally} showed that the number $|\A(v)| = |\Anone|$ of feasible substation configurations in substation $v$ is equal to 
    \begin{align}\label{Eq:N-1ActionSpace}
        |\Anone| &= |\Asym| -  2^{g(v) + l(v) + b(v)} \cdot (e(v) - \delta_{e(v),2}- \delta_{e(v),1}).
    \end{align}
    
    \item \textbf{\textit{Greedy reduction}}:
    This method narrows the action space to include only the highest-performing actions identified by brute-force search. Extensive simulations are run with a greedy agent that selects actions based on the $\rho_{\max}$ value or, in some cases, based on the reward function. The most frequently selected actions across multiple scenarios are then retained. For robustness, the scenarios can include an opponent, as in \binbin. Another method to select more robust actions used in \curriculum\ and \ljn, is with an $N-1$ Teacher, as described in \cref{Sec:SolutionsOutside}.
    
    \item \textbf{\textit{Expert reduction}}:
    The action space can be reduced according to domain knowledge or experience, as in \aaacfirst, \dddqnfirst,\ and \hugo. % \cite{yan2020l2rpn,zhihong2020neurips, lehna2024hugo}. 
    Unfortunately, it is difficult to check what this entails. The experience could also be based on greedy simulations.
    \item \textbf{\textit{Mask reduction}} introduced by \cite{yoon2021winning}: This approach restricts to actions from substations larger than a specified size $M$. Let $V_{>M}:= \{v \in V: |v| > M \}$. The total size of the action space is now defined by
    \begin{align}
        |\A_{bus}| = \sum_{v\in V_{>M}} |\A(v)|.
    \end{align}
    Note that, as mentioned in \cref{Subsec:2020WCCI}, in the solution \smaac\ the output of the neural network was mainly reduced by learning the afterstates instead of the specific actions. The size of the output of the neural network, denoted by $|\mathcal{N}_{output}|$ scales therefore linearly, as shown in \cref{Eq:NNoutputGNN}.
    \begin{align} \label{Eq:NNoutputGNN}
        |\mathcal{N}_{output}| = \sum_{v\in V_{>M}} |v|-1.
    \end{align}
    \item \textbf{\textit{Stable Topo reduction}}: This method selects actions that result in a stable topology. For example, in \hugo, the target topologies (TTs) are identified by recording all the topologies that brought the network in a \textit{safe state}; i.e., where $\rho_{\max} < $ \textit{activation threshold}. The 500 most frequently used TTs are saved for the final agent.
    Another example can be found in \bdqn, where each topology action is tested and considered successful if it remains stable for at least four time steps. After an extensive number of simulations, actions with a high success rate are kept, while others are filtered out based on a set threshold.
\end{enumerate}

\begin{table}[h]
\begin{adjustbox}{center}
\resizebox{1.2\textwidth}{!}
{
\begin{tabular}
{|l|p{0.15\linewidth}|p{0.12\linewidth}|p{0.12\linewidth}|p{0.35\linewidth}|p{0.15\linewidth}|}
\hline
\textbf{Solution Acronym} & \textbf{Actions for RL agent} & \textbf{Reduced Size Action Space} & \textbf{\% of actions included} & \textbf{Reduction method} & \textbf{Actions for rule-based heuristics} \\ \hline
\smaac  & Bus splitting        & 66674    & 99.8\%& \textit{Mask reduction} with $N=5$.  & Line switching.           \\ \hline
\aaacsecond   & Topology actions & 300      & 0.4\% & 200 based on \textit{experience}, 100 randomly selected. In addition, the action space contains 295 line-switching actions. & Line switching.           \\ \hline
\sas   & Bus splitting        & 732 \tablefootnote{Two action spaces are used, $\mathcal{A}_{normal}$, which is used by default and $\mathcal{A}_{exception}$, which is used only when specific lines are disconnected. Here, $|\mathcal{A}_{normal}|=500$ and $|\mathcal{A}_{exception}|=232$.} & 1.0\% & N.A.       & Line switching and redispatching.          \\ \hline
\binbin     & Bus splitting        & 208 \tablefootnote{\label{footnote:ChangeActions}This action space includes \textit{change} bus actions. This is different from \textit{set} bus actions which most solutions use. See https://grid2op.readthedocs.io/en/latest/.}  & 0.3\% & \textit{Greedy reduction.}   & Line switching when line is disconnected.             \\ \hline
\dddqnfirst   & Topology actions or redispatch actions    & 786 \footref{footnote:ChangeActions} \tablefootnote{In addition, the action space contains 58 line switching, 40 redispatching and one explicit do-nothing action.  } & 1.2\% & \textit{Expert reduction}.    & Line switching when line is disconnected.           \\ \hline
\powrl  & Bus splitting        & 240 & 0.4\% & \textit{Greedy reduction}.       & Line switching.            \\ \hline
\curriculum           & Bus splitting        & 508 \footref{footnote:ChangeActions} & 0.8\% & \textit{Greedy reduction}.   & Line switching.     \\ \hline
\bdqn   & N.A., but only bus splitting actions seem to be used & 1080& 1.6\% & \textit{Stable Topo reduction}.  & N.A.   \\ \hline
\end{tabular}
}
\end{adjustbox}
\caption{Action space reductions for case 36.}
\label{Table:Case36ActReduction}
\end{table}
\FloatBarrier

Although the ($N-0$)- and ($N-1$)-reduction methods may be more sensible, they are only applied to the smaller case 14 instances. Applying the ($N-1$)-reduction to case 36 would reduce the action space to $65985$, see \cite{demol2025centrally}, which is still nearly $99\%$ of the action space. Hence, for the larger cases, such as 36 and 118, more drastic reduction strategies are typically employed to deal with the exponential growth of the action space. For case 36 and case 118, the percentage of selected bus-splitting actions generally falls within $[0.3 - 1.6]$ and $[0.4 - 2.8]$, respectively, excluding the exceptional cases \hri\ and \smaac. 

The challenge in effective action space reduction lies in balancing two critical factors: ensuring that the reduced space still encompasses all the necessary actions to mitigate potential contingencies and avoiding an overly large action space that hinders the RL agent from converging.   
Most current solutions employ a greedy reduction method, which quickly selects the most promising actions. However, this approach may exclude actions that, while not optimal on their own, could prove valuable when combined with others. For instance, an action might appear less rewarding in the short term, but when considered in conjunction with other actions, it could yield a higher long-term benefit.

% the difficulty of knowing whether the reduced action space still encompasses all the necessary actions to mitigate each contingency. An excessive discard of actions may lead to a suboptimal action space that does not contain enough actions to address the contingencies the power network is facing. On the other hand, keeping too many actions can make it too difficult for the DRL agent to converge. Most of the currently proposed solutions use a greedy reduction method, which is an easy way to quickly select the good actions. However, it should be noted that this could exclude actions that are good when combined with other actions, but are not the most effective action by itself. To be more concrete, the direct reward of an action might be less than the best action, but when looking at the long-term reward, this action, perhaps combined with others, could be better. 

\begin{table}[h]
\begin{adjustbox}{center}
\resizebox{1.2\textwidth}{!}
{
\begin{tabular}
{|l|p{0.15\linewidth}|p{0.12\linewidth}|p{0.12\linewidth}|p{0.35\linewidth}|p{0.15\linewidth}|}
\hline
\textbf{Solution Acronym} & \textbf{Actions for RL agent} & \textbf{Reduced size action space} & \textbf{\% of actions included} & \textbf{Reduction method} & \textbf{Actions for rule-based heuristics} \\ \hline
\sas      & Bus splitting          & 1000       & 1.0\%   & N.A.            & Line switching.          \\ \hline
\dddqnfirst     & Topology actions       & 978       & 1.4\%   &  \textit{Expert reduction}\footref{footnote:ChangeActions}. In addition, the action space contains 185 line switching and one do-nothing action.      & Line switching.          \\ \hline
\alphazero            & Bus splitting          & 2000       & 2.7\%   &  \textit{Greedy reduction}.         & Line switching and redispatching. \\ \hline
\brute          &            & 314        & 0.4\%   & \textit{Greedy reduction}.           & Bus splitting, line switching and redispatching.     \\ \hline
\hri   &            & N.A.          & N.A.   & Randomly draw 1000 line switch combined with redispatch actions each time the grid is in danger.    & Line switching combined with redispatch.          \\ \hline
\ljn & Bus splitting          & 1516       & 2.1\%   & \textit{Greedy reduction}.      & Line switching and redispatching.           \\ \hline
\artelys  & Bus splitting          & 500        & 0.7\%   & \textit{Greedy reduction}.    & Line switching and redispatching.           \\ \hline
\hugo     & Bus splitting          & 2030       & 2.8\%   & Apply the action space of \alphazero\ with an additional 30 expert actions selected by RTE (\textit{Expert reduction}). \textit{Stable Topo Reduction} for the target topologies (TTs). & Line switching.          \\ \hline
\SoftIL     & Bus splitting          & 2030       & 2.8\%   & Apply the action space of \hugo. & Line switching.          \\ \hline
\end{tabular}
}
\end{adjustbox}
\caption{Action space reductions for case 118.}
\label{Table:Case118ActReduction}
\end{table}
\FloatBarrier

% What to include? How to present it to the DL NN? Feature matrix or not. Observation including history
\subsubsection{Observation space}\label{subsub:ObsSpace}
The observation or state space of the power network includes all relevant information about the physical network needed for PNC, described in \cref{Subsec:Background}. It should be noted that while extensive information may provide a comprehensive view of the system, it may also introduce unnecessary complexity, thereby posing a challenge for a Deep-RL (DRL) agent to focus on the most crucial aspects. Conversely, an overly reduced observation space might result in suboptimal or uninformed decisions due to missing key information. 

Based on the approaches evaluated in \cref{Table:ObsSpaces}, it can be noted that most authors incorporate key variables such as active power of generators, loads, and lines ($gen\_p$, $load\_p$, $p\_{or}$, $p\_{ex}$), topology configuration ($topo\_vect$) and consistently across all solutions, the line loading ($\rho$) as features representing the state $S_t$ observed by the RL agent. Notably, the solutions \ddqn, \alphazero\  and \zonal\ utilize the entire observation space without applying any reductions in state representation referred to as the \texttt{CompleteObservation}\footnote{https://grid2op.readthedocs.io/en/latest/observation.html}.

The solution \smaac\ introduces an additional feature \textit{danger}, a binary vector that flags lines $e\in E$ when the loading $\rho_e$ exceeds a predefined danger threshold $\rho_{danger}=0.9$. Moreover, this solution enriches the state $S_t$ by incorporating a history of previous states, $S_t = [s_{t-n}, \dots , s_t]$, enabling the agent to discern trends, such as rising or falling feature values. 
Like other solutions that utilize a GNN architecture — such as \marl, \gnnil, and \SoftIL\ — \smaac\ structures feature values into a \textit{feature matrix}, alongside an \textit{adjacency matrix} that represents the current topology configuration of the graph.

In \CCMA, the coordination agent receives, in addition to the information on the grid, the proposed actions and/or action values from the regional agents.

The most heavily reduced state space appears in \dddqnsecond, where only four features are retained: the voltages of the lines ($v\_or$, $v\_ex$), the currents or line flows ($a\_or$, $a\_ex$), the line loads ($\rho$), and the topology configuration.

It should be noted that not all authors explicitly documented which parts of the observation space were included in the state space provided to the RL agent. In such cases, we extracted this information from publicly available code repositories, if available. Solutions for which the exact observation space could not be determined have been excluded from \cref{Table:ObsSpaces}.

None of the reviewed solutions explicitly applies systematic feature selection techniques to reduce the state space. Instead, most authors rely on domain knowledge and intuition to select the most relevant features. While this approach has been effective in previous work, it raises the question of whether more principled feature selection methods could further improve performance.

\begin{sidewaystable}[h]
\begin{adjustbox}{center}
\resizebox{1.05\textwidth}{!}
{
\scriptsize
\tabcolsep=3pt
\begin{tabular}{l|c|x{0.06\textwidth}|ccc|ccc|ccc|x{0.07\textwidth}|x{0.05\textwidth}|x{0.05\textwidth}|x{0.05\textwidth}|x{0.07\textwidth}|cc|c|p{0.08\textwidth}}
             &    &               & \multicolumn{3}{c|}{\textbf{Active power}}             & \multicolumn{3}{c|}{\textbf{Reactive power}}           & \multicolumn{3}{c|}{\textbf{Voltages}}    &             &              &   &         &       & \multicolumn{2}{c|}{\textbf{Cooldown}} &         &                   \\
\textbf{Solution Acronym} & \makecell[l]{\texttt{Complete}\\ \texttt{Observation}} & \textbf{Time stamp} & \makecell[c]{\textbf{Loads}\\($load\_p$)} & \makecell[c]{\textbf{Gens}\\($gen\_p$)} & \makecell[c]{\textbf{Lines}\\($p\_ex, p\_or$)} & \makecell[c]{\textbf{Loads}\\($load\_qp$)} & \makecell[c]{\textbf{Gens}\\($gen\_q$)} & \makecell[c]{\textbf{Lines}\\($q\_ex, q\_or$)} & \makecell[c]{\textbf{Loads}\\($load\_v$)} & \makecell[c]{\textbf{Gens}\\($gen\_v$)} & \makecell[c]{\textbf{Lines}\\($v\_ex, v\_or$)} & \textbf{Line flows ($a\_{or}, a\_{ex}$)} & \textbf{Line loads ($\rho$)} & \textbf{Time step overflow} & \textbf{Line status} & \textbf{Topo config} ($topo\_vect$) & \textbf{line}        & \textbf{sub}        & \textbf{Maintenance} & \textbf{Extra}                 \\ \hline
\aaacfirst          &    &               & \cmark & \cmark      &   & \cmark&        &   &   & \cmark     &   & \cmark          & \cmark           &   & \cmark      & \cmark    &         & \cmark     &         &                   \\
\ddqn         & \cmark  &               &   &        &   &   &        &   &   &        &   &             &              &   &         &       &         &        &         &                   \\
\aaacsecond \footnote{\label{footnote:codebasednotdoc}Based on code, not (clearly) documented}       &    &               & \cmark& \cmark     & \cmark& \cmark& \cmark     & \cmark& \cmark& \cmark     & \cmark& \cmark          & \cmark           &   & \cmark      & \cmark    &         &        &         &                   \\
\smaac \footref{footnote:codebasednotdoc}     &    &               & \cmark& \cmark     & \cmark&   &        &   &   &        &   &             & \cmark           & \cmark             &         & \cmark    &         &        & \cmark      & Danger and history \\
\dddqnfirst \footref{footnote:codebasednotdoc}      &    &               & \cmark& \cmark     & \cmark&   &        &   &   &        &   & \cmark          & \cmark           & \cmark             & \cmark      & \cmark    & \cmark      & \cmark     & \cmark      &                   \\
\binbin   &    & \cmark            & \cmark& \cmark     & \cmark& \cmark& \cmark     & \cmark& \cmark& \cmark     & \cmark& \cmark          & \cmark           & \cmark             & \cmark      & \cmark    & \cmark      & \cmark     & \cmark      &                   \\
\sas \footref{footnote:codebasednotdoc}          &    & \cmark            &   &        &   & \cmark& \cmark     &   & \cmark& \cmark     &   &             & \cmark           &   &         &       &         &        &         &                   \\
\cem        &    &               & \cmark& \cmark     & \cmark& \cmark& \cmark     & \cmark& \cmark& \cmark     & \cmark& \cmark          & \cmark           & \cmark             & \cmark      & \cmark    &         &        &         &                   \\
\dddqnsecond         &    &               &   &        &   &   &        &   &   &        & \cmark& \cmark          & \cmark           &   &         & \cmark    &         &        &         &                   \\
\powrl        &    & \cmark            & \cmark& \cmark     & \cmark& \cmark& \cmark     & \cmark& \cmark& \cmark     & \cmark& \cmark          & \cmark           & \cmark             & \cmark      & \cmark    & \cmark      & \cmark     & \cmark      &                   \\
\alphazero   & \cmark &               &   &        &   &   &        &   &   &        &   &             &              &   &         &       &         &        &         &                   \\
\curriculum \footref{footnote:codebasednotdoc}   &    & \cmark            & \cmark& \cmark     & \cmark& \cmark& \cmark     & \cmark& \cmark& \cmark     & \cmark& \cmark          & \cmark           & \cmark             & \cmark      & \cmark    & \cmark      & \cmark     & \cmark      &                   \\
\hrl          &    &               & \cmark& \cmark     & \cmark&   &        &   &   &        &   &             & \cmark           & \cmark             &         & \cmark    &         &        & \cmark      &                   \\
\marl        &    &               & \cmark& \cmark     & \cmark&   &        &   &   &        &   &             & \cmark           & \cmark             &         & \cmark    &         &        & \cmark      & Danger and history \\
\ljn \footref{footnote:codebasednotdoc}     &    &               &   &        &   &   &        &   &   &        &   &             & \cmark           &   &         &       &         &        &         &                   \\
\il           &    &               & \cmark& \cmark     & \cmark& \cmark& \cmark     & \cmark& \cmark& \cmark     & \cmark& \cmark          & \cmark           &   &         &       &         &        &         & Thermal limit                  \\
\bdqn         &    &               & \cmark& \cmark     & \cmark& \cmark& \cmark     & \cmark& \cmark& \cmark     & \cmark& \cmark          & \cmark           &   & \cmark      & \cmark    &         &        &         &      \\            
\zonal        & \cmark &               &   &        &   &   &        &   &   &        &   &             &              &   &         &       &         &        &         &                   \\  
\gnnil        &    &               & \cmark& \cmark     & \cmark& \cmark& \cmark     & \cmark& \cmark& \cmark     & \cmark& \cmark          & \cmark           &   &         &       &         &        &         & Thermal limit                  \\     
\CCMA         &    &               & \cmark& \cmark     & \cmark&   &        &   &   &        &   &             & \cmark           & \cmark             &         & \cmark    &         &        &      & Actions and action values.     \\
\end{tabular}
}
\end{adjustbox}
\caption{Selected observation features for the RL agent per solution. For documentation on each feature, see \cite{Grid2Op}.
}
\label{Table:ObsSpaces}
\end{sidewaystable}
\FloatBarrier

% , such as the current date and time, the current vector of the topology, the active and passive power productions and consumptions, the current flow on the power lines, etc. All values that are available in the Grid2Op environment will be referred to as the \texttt{CompleteObservation}\footnote{https://grid2op.readthedocs.io/en/latest/observation.html}.
% This observation also includes the number of lines or the maximum length of the episode. These are values that do not change throughout the episode and, therefore, are less valuable to an RL agent who is trying to learn in which situation which action would be most interesting. Nevertheless, in the past, authors have chosen to include the entire CompleteObservation as state space for their agent. On the other side, different authors made different choices in which values to include or exclude and some authors adjusted the representation of the values in a way that it can also be interpreted for Graph Neural Networks (GNN).

\subsubsection{Normalization} \label{SubSub:Norm}
Once the features from the observation space have been selected, it is important to transform the data to train the model efficiently.
\textit{Normalization} entails rescaling the input features to ensure that the values fall within a standard range, such as $[-1,1]$. This step is essential in machine learning, as neural networks tend to train faster and more effectively when features are on a similar scale. Although most Deep RL solutions for PNC apply some form of normalization, this aspect is often under-reported in the documentation and sometimes entirely overlooked in implementations. 
% Moreover, normalization is not always applied consistently or correctly. Some approaches normalize only a single feature. In some cases, batch normalization \cite{ioffe2015batch} has been used as a preprocessing step by applying it as the first layer of a neural network. However, batch normalization is designed to standardize activations within hidden layers based on mini-batch statistics during training, rather than to normalize raw input features.

Selecting the appropriate normalization technique can be a complex task in its own right; see \cite{ioffe2015batch}, \cite{van2016learning}, \cite{klambauer2017self}, \cite{bhatt2019crossnorm} and \cite{bjorck2021towards}.
This paragraph highlights two common techniques used in L2RPN that are effective and accessible for scaling features when training a neural network. 

A widely used normalization technique, applied by \cite{yan2020l2rpn}, \cite{lehna2023managing} and \cite{manczak2023hierarchical} is \textit{min-max} (or \textit{linear}) \textit{scaling}. This method is effective when the data is relatively uniformly distributed and does not contain any extreme outliers. The formula for computing the normalized value $X_{normalized}$ of an observed feature value $X$ using the min-max scaling is described in \cref{Eq:MinMaxNorm}. 
\begin{align}\label{Eq:MinMaxNorm}
    X_{normalized} = \frac{X - X_{\min}}{X_{\max}-X_{\min}}.
\end{align}
The method guarantees $X_{normalized}\in [0,1]$. For the computation, approximate lower and upper bounds ($X_{\min}$ and $X_{\max}$) are required, which can be obtained by running simulations, for example, using a do-nothing agent. For some features, such as generator production, these bounds are predefined and can be directly derived from Grid2Op.

Another common approach, used by \cite{yoon2021winning}, \cite{enlite2021mazerl}, \cite{wang2021l2rpnicaps}, \cite{martinez2021l2rpnicaps}, \cite{dorfer2022power} and \cite{vandersar2023marl}, is \textit{Z-score normalization} (or \textit{standardization}). This technique is more appropriate when the data is roughly Gaussian distributed. In this case, the normalized value is obtained by:
\begin{align}\label{Eq:ZscoreNorm}
    X_{normalized} = \frac{X - \mu(X)}{\sigma(X)},
\end{align}
where $\mu(X)$ and $\sigma(X)$ represent the empirical mean and standard deviations, which can be estimated through simulations. This method results in Z-normalized data with a mean of $0$ and a standard deviation of $1$. 

While both techniques make assumptions about the underlying data distribution, real-world data often does not strictly conform to these assumptions. Exploring alternative normalization methods could offer improvements in the training process.

% We made remarks about how the original data should look to apply the two types of normalization described above. This is in both cases not necessarily the case, so it might be interesting to investigate other normalization techniques and see how this could improve the training process. 

\subsubsection{Reward}\label{Subsub:Reward}
Selecting an appropriate reward function is essential for a reinforcement algorithm to effectively learn the desired behavior, as emphasized by \cite{eschmann2021reward}. This section provides an overview of the reward functions proposed for the various L2RPN challenges. As mentioned in \cref{App:IntroRL}, reward can be a function of the previous state, the action and the new state. All rewards discussed in this section are based only on the state $s_t$ observed at time $t$, and simply denoted by $r(t)$.

As shown in \cref{Table:Rewards}, the reward function typically involves two components: the reward during \textit{normal operations} and the reward at the end of an episode. The latter usually incorporates a negative reward upon a game over and sometimes also includes a positive reward upon the agent's successful completion of the episode. During normal operations, the reward function should steer the agent in order to avoid a game over. 
Having a negative or positive reward at the end of an episode encourages the agent to maintain the stability of the network. However, relying solely on terminal rewards can delay learning, as it may take many episodes for the value functions to converge and reflect long-term rewards.
Providing meaningful rewards during normal operations offers more immediate feedback, helping the agent differentiate between beneficial and detrimental actions. In the following, we describe various reward functions that have been used to guide agent behavior during \textit{normal operation}.

\cite{marot2020learning} introduce a commonly used reward function referred to as $\margins$, detailed in \cref{eq:MarginRw}. This reward has been applied (directly or in a scaled fashion) in several solutions, including \ddqn, \aaacsecond, \cem, \hrl, and \CCMA\ as shown in \cref{Table:Rewards}. $\margins$ is one of the default rewards available in the Grid2Op package. This reward is designed to encourage a balanced power flow across the transmission lines by maximizing the margins between the power flow used and the \textit{thermal limits} (maximum capacity) of the lines. Let $F_e(t)$ denote the power flow on transmission line $e \in E$ at time step $t$ (in Amperes), and $L_e$ represent its thermal limit. At time step $t$, the margin $M_e(t)$ is computed as
\begin{align*}
    M_e(t) &= \max\left\{0, 1 - \frac{F_e(t)}{L_e} \right\}, \quad \forall e \in E,
\end{align*}
and the reward as
\begin{align} \label{eq:MarginRw}
    \margins(t) = \sum_{e\in E} \left (1 - (1- M_e(t))^2 \right).      
\end{align}
This reward function encourages keeping the power flows on each line below their thermal limits by rewarding larger safety margins.

Another frequently used reward function is $\rwrho$, a simple linear function based on the maximum line load $\rho_{\max}(t) := \max\{\rho_e(t): e \in E\}$, defined as:
\begin{align*}%\label{eq:RwRho}
        \rwrho(t) = 
        \begin{cases}
            2 - \rho_{\max}(t), &\text{ if } \rho_{\max}(t) < 0.95,\\
            2 - 2\rho_{\max}(t), &\text{ otherwise.}
        \end{cases} 
\end{align*}
This reward function was first introduced in \binbin\ and later applied in \powrl, \curriculum,\ and \ljn. As illustrated in \cref{Table:Rewards}, it is often combined with a high negative penalty for game over events or for perorming illegal actions\footnote{Illegal actions are actions that are currently unavailable, for example, due to maintenance of a substation or power line.} and a strong positive reward when the agent successfully completes the episode without failure.

The solution \smaac\ uses a reward function based on energy losses rather than transmission line loading, namely
\begin{align*}
     \loss(t) =  
        \frac{\sum_{i\in \text{ loads }} p_i(t) }{\sum_{i\in \text{gens}} p_i(t)} - 0.9 ,
\end{align*}
where $p_i$ is the active power production or consumption of a load or generator $i$, respectively. Notably, this reward function departs from the other line-load-based reward functions by focusing on minimizing energy losses, indirectly encouraging lower line loads since overloaded lines tend to increase energy losses. Despite being fundamentally different, this reward function has shown good performance in \cite{yoon2021winning}. 

Lastly, we discuss the reward designed by \cite{dorfer2022power} and used in \alphazero. Their reward is based on the cumulative sum over all overloaded lines.%, defined as $L_{overflow}:= \{l \in L:  \rho_l>1\}$. 
The authors introduce a function $u$ as in \cref{eq:RwAlphaZero_U} and utilize it to calculate the reward as:
\begin{align*} %\label{eq:RwAlphaZero}
    \rwalpha(t) = e^{-u(t) - 0.5 \cdot n_{\textrm{offline}}},   
\end{align*}
where
\begin{align} \label{eq:RwAlphaZero_U}
    u(t) = 
        \begin{cases}
            \max \{ \rho_{\max}(t)-0.5, 0\}, & \text{ if } \rho_{\max} \leq 1.0,\\
            \sum_{l \in L: \rho_l>1} \rho_l, & \text{ otherwise}.
        \end{cases} 
\end{align}
This reward function considers the number of disconnected lines, $n_{\textrm{offline}}$, to decrease the maximum achievable reward at that time step.

The solutions \brute\ and \hri\ rely on greedy simulations to determine the next action. The selection of greedy actions can be guided by the reward $r(t+1)$, but these approaches base their decision on the maximum line load $\rho_{\max}(t+1)$. Specifically, the greedy agent selects the action that minimizes $\rho_{\max}(t+1)$. \brute\ and \hri\ are therefore excluded from \cref{Table:Rewards}.

Since \il, \gnnil\ and \SoftIL\ do not use an RL agent and thus do not require a reward function to guide their behavior toward a stable state, these approaches are also excluded from \cref{Table:Rewards}.

Lastly, we note that not all authors documented the reward function used. In this case, the reward function was derived from the code if it was publicly available. Solutions for which we were unable to determine the reward function are excluded from \cref{Table:Rewards}.

% \begin{table}[]
% \begin{adjustbox}{center}
% \resizebox{1.1\textwidth}{!}
% {
{
\footnotesize
\tabcolsep=3pt
\begin{longtable}{|l|p{0.82\textwidth}|}
\hline
\textbf{Solution Acronym} & \textbf{Reward function} \\ 
\endhead
\hline
        \multicolumn{2}{r@{}}{Table continues on the next page\ldots}\\
\endfoot 
\endlastfoot  
\hline
\aaacfirst &    
$
r(t) = \begin{cases}
    -100,  & \text{if game over}, \\
    \sum_{l\in L} R_l(t), & \text{otherwise},
\end{cases}
$
where 
$
R_l(t) = \begin{cases}
    0.95-\frac{F_l(t)}{L_l}, & \text{ if } F_l(t) \leq 0.95\cdot L_l,\\
    \alpha \cdot (0.95-\frac{F_l(t)}{L_l}), & \text{ otherwise.}
\end{cases}
$
\\ \hline
\ddqn &  $
r(t)=\begin{cases}
    -1, & \text{ if game over},\\
    \frac{1}{n} \cdot  \margins(t), & \text{ otherwise}.
\end{cases}
$        \\ 
\hline
\aaacsecond& 
$
r(t)= \begin{cases}
    -200, & \text{ if game over},\\
    \margins(t) + R_{violated}(t), & \text{ otherwise},\tablefootnote{The documentation in the slides and the code do not agree. This equation shows how it is implemented. According to their documentation they cut it off at a minimum of $-10$ to avoid ``fear''.}
\end{cases}
\quad $ 
where 
$ \quad
 R_{violated}(t) = 
    -\sum_{l \in L} \beta_l \cdot ReLu(\frac{F_l(t)}{L_l}-1) 
\quad$ with $\quad \beta_l = 50, \quad  \forall l\in L$.
\\ \hline
\smaac &   
$
r(t) = 
    \begin{cases}
        -1, & \text{ if action results in game over},\\
        -0.5, & \text{ if action is illegal},\\
        \loss(t), & \text{ otherwise}.
    \end{cases}
$       \\ \hline
\dddqnfirst        & 
$r(t) = w_1 \cdot R_{Sandbox}(t) + w_2 \cdot R_{CloseToOverflow}(t) + w_3 \cdot R_{Distance}(t)  + w_4 \cdot R_{LineCapa}(t) $, where  
\begin{itemize}[leftmargin=20pt]
    \item $R_{Sandbox}(t)$ based on costs for redispatch and loss, 
    \item $R_{CloseToOverflow}(t) $  based on the overflow of lines; less overflow gives higher reward,  
    \item  $R_{Distance}(t) $ based on the distance from the reference topology; smaller distance gives higher reward, 
    \item $R_{LineCapa}(t)$ based on the lines capacity usage; more available capacity gives a higher reward, 
    \item and $w_1 = 30$,  $w_2=200$, $w_3 = 20$, and $w_4 =3.0$.\tablefootnote{See \url{https://github.com/lujasone/NeurIPS_2020_L2RPN_Comp_An_Approach/blob/fdb01cc42253e948562453d2fb3249fe1ded37e0/Adaptability_Track/MyRewards.py} for more information on the exact reward functions.}
\end{itemize}          
\\ \hline
\binbin &  
$
r(t) = \begin{cases}
    -300, & \text{ if action results in game over},\\
    -10, & \text{ if action is illegal},\\
    500, & \text{ if finished complete episode},\\
    \rwrho(t), & \text{ otherwise}.
\end{cases}
$
\\ \hline
\sas &     No reward function used for training. They estimate the value function (which should be the expected discounted future reward of a state) with the current risks of the lines defined as 
$ R_{isk}(t) = \max \left \{ \frac{F_l(t)}{L_l} \right \}, \forall l \in L$.
\\ 
\hline
\cem &     
$r(t) = \begin{cases}
    0, & \text{ if action results in game over},\\
    \margins(t), & \text{ otherwise}.
\end{cases} $       \\ \hline
\dddqnsecond &     
$
r(t)=\begin{cases}
    1, & \text{ if survived time step } t,\\
    0, & \text{ if game over}.
\end{cases}
$ \\ \hline
\powrl &   
$
r(t) = \begin{cases}
    -300, & \text{ if action results in game over},\\
    500, & \text{ if finished complete episode},\\
    \rwrho(t), & \text{ otherwise}.
\end{cases}
$ 
\\ \hline
\alphazero &  
        $
        r(t) = \rwalpha(t).
        $
\\ \hline
% BruteForce-2022 \cite{alibaba2022l2rpn} & No reward function used. Note: Their greedy agent makes decisions purely based on the simulated $\rho_{\max}(t+1) $ values of the next time step. \\ 
% \hline
% HRI-EU \cite{hri2022trial} & No reward function used. Note: Their greedy agent makes decisions purely based on the simulated $\rho_{\max}(t+1) $ values of the next time step.  \\ \hline
\curriculum &
$
r(t) = \begin{cases}
    -10, & \text{ if action illegal or results in game over},\\
    \rwrho(t), & \text{ otherwise}.
\end{cases}
$  
\\ \hline
\hrl & 
$r(t)=\begin{cases}
    \frac{1}{n} \cdot \margins(t), & \text{ if survived time step } t,\\
    -0.5, & \text{if game over}.
\end{cases}
$   \\ \hline
\marl &    
$
r(t) = 
    \begin{cases}
        -1, & \text{ if action results in game over},\\
        -0.5, & \text{ if action is illegal},\\
        \loss, & \text{ otherwise}.
    \end{cases}
$ 
\\ \hline
% \artelys & Not documented and no public code base available. \\ \hline
\ljn &  
$
r(t) = \begin{cases}
    -10, & \text{ if action illegal or results in game over},\\
    \rwrho(t) + R_{action}(t), & \text{ otherwise},
\end{cases}
$\newline where 
$
R_{action}(t) = \begin{cases}
    4.0, & \text{ if do-nothing action is used,}\\
    0.0, & \text{ otherwise}.
\end{cases}
$
\newline The greedy search uses a simple linear reward based on the maximum $\rho$ value: 
$R_{Greedy} = 2 - \rho_{\max}$. 
 \\ \hline
% HUGO-2024 \cite{lehna2024hugo} & \makecell[tl]{Not documented and no public code base available. We can assume that the authors probably used the same reward function as \\ in their previous paper, Curriculum-2023 \cite{lehna2023managing}.}\\ \hline
% IL-2024 \cite{dejong2024imitation} &  No reward function needed for the imitation learning method used in this paper. \\ \hline
% HMO-2024 &  Too complicated...\\ \hline
\bdqn &
$r(t) = \begin{cases}
    -20, & \text{if action results in game over},\\
    0.7 R_e(t) + 0.3 R_c(t), & \text{otherwise},
\end{cases}$ 
where $R_e(t) = \frac{\sum_{i\in gens} c_i\cdot (P_i^{\max} - P_i(t))}{\sum_{i\in gens} c_i \cdot P_i^{\max} }$, with $c_i$ the costs of generator $i\in gens$ and $P_i^{\max}$ and $P_i(t)$ the maximum and given active power of generator $i\in gens$, respectively, and 
$R_c(t) = 1 - \frac{1}{|O|} \sum_{i \in O} \frac{I_i(t)}{I_i^{\max}} $,  with $O\subseteq E$ the set of connected transmission lines and $I_i^{\max}$ and $I_i(t)$ the maximum and the given current of line $i\in O$, respectively.  
\\ \hline
% \zonal &
% TOO VAGUE
% \\ \hline
\CCMA & 
$
r(t)=\begin{cases}
    \frac{1}{n} \cdot \margins(t), & \text{ if survived time step } t,\\
    -0.5, & \text{if game over}.
\end{cases}
$
\\ \hline
\caption{Overview of the reward functions, $r(t)$, used by the solutions discussed.}
\label{Table:Rewards}
\end{longtable}
% }
% \end{adjustbox}

% \end{table}
\FloatBarrier
}

\subsubsection{Imitation learning}\label{Subsub:Imitation}
As shown in \cref{Table:Techniques}, eight out of twenty-two solutions employ \textit{Imitation Learning} (IL), see \cite{kober2010imitation,schaal1999imitation,osa2018algorithmic}, a machine learning technique in which a task is learned by observing an expert. In IL, the agent tries to mimic the behavior of an expert by predicting the actions $a(t)\in A$ the expert would choose in state $s(t)\in S$. This differs from RL, where the agent learns through exploration, receiving rewards to guide it towards a policy that maximizes long-term cumulative reward. Although RL is effective in many scenarios, learning an optimal policy can be difficult, for example, when the rewards are sparse (e.g., given only at the end of an episode), or the reward function is undefined, which is the case in power grid maintenance. As discussed earlier, different reward functions have been designed to provide the agent with the right feedback. However, manually designing a reward function that satisfies the desired behavior can be extremely complicated. IL could be a good solution to address this issue.
Moreover, even when a proper reward function is available, IL can speed up the learning process by reducing the number of samples required for RL.

IL has two main classes: \textit{behavioral cloning} (BC), described in \cite{pomerleau1988alvinn, bain1995framework}, and \textit{inverse reinforcement learning} (IRL), introduced in \cite{russell1998learning}. In BC, data trajectories $(\tau^*)$ of the expert agent are collected, consisting of state-action pairs $(s_0,a_0^*),(s_1,a_1^*),\dots$, and supervised learning is used to learn the policy $\pi_\theta$, parameterized by $\theta$ using a neural network, that minimizes the loss $\mathcal{L}\left(a^*,\pi_\theta(s)\right)$. In contrast, IRL tries to determine the reward function $r_\theta(s,a)$ that the expert agent is optimizing, based on observed demonstrations, rather than directly learning the policy. 

The IL-based solutions proposed for L2RPN%, including \ddqn, \binbin, \curriculum, \ljn, \artelys, \hugo, \il, \gnnil\ and \SoftIL\  %\cite{lan2020ai,binbinchen2020neurips,lehna2023managing,lajavaness2023l2rpn,artelys2024,lehna2024hugo,dejong2024imitation}, 
all involve BC. In \binbin, a greedy expert (referred to as Tutor) selects the action that minimizes the maximum line load $\rho_{\max}(t+1)$, while the agent (Junior) attempts to replicate this policy. In \curriculum, the experts' strategy is extended with an $N-1$ secure action selection, as described in \cref{Sec:SolutionsOutside}. This extended experts' strategy has also been used in \ljn\, \artelys\, \hugo, \il, \gnnil\ and \SoftIL.
Five of these papers use IL as a warm start to initialize RL-based agents, similar to the approach used in the AlphaGo algorithm \cite{silver2016mastering}. 
In contrast, \il, \gnnil\ and \SoftIL\ explore IL as the primary method without subsequently deploying an RL algorithm. Nevertheless, these experiments show promising results, being able to improve on the expert Greedy baseline while reducing the computation time needed.

% It is good to note that an agent trained using IL, will generally only match, not surpass, the performance of the expert. The main advantage of IL in such cases is computational efficiency, reducing the effort required to identify suitable actions compared to using a greedy expert who needs to simulate all options. To exceed the expert's performance, the agent would need to be enhanced, for example, through RL.

As pointed out by \cite{dejong2024imitation}, BC can result in a \textit{distribution shift}: small deviations in the agents' actions from the experts can lead to unseen scenarios for which the behavior is undefined. Consequently, the agent makes more mistakes, which could lead to catastrophic failures. To mitigate this issue, an improved version of BC can be used, where training alternates between the agents' policy and the expert \cite{ross2010efficient,ross2011reduction}.

% Debatable if \cite{zhou2021action} is also a form of imitation learning. They call it an evolutionary algorithm, but basically, the algorithm is learning to behave more and more Greedily with respect to their estimated ``future" reward function. 

\subsubsection{Prioritized replay}\label{Subsub:PrioReplay}
In its simplest form, RL algorithms discard experience after one update. \textit{Experience replay}, a technique designed for \textit{off-policy} algorithms (see \cref{subsec:RL_terminology}), addresses two key issues: the risk of strongly correlated updates, and the tendency to forget useful but rare experiences immediately. Transitions, $(s_t, a_t,r_t,s_{t+1})$, experienced by the agent, are stored in a replay buffer and sampled uniformly at random to be revisited.
In \textit{prioritized experience replay} (PER) \cite{schaul2015prioritized}, the idea is to replay transitions that are important more frequently and, in this way, make more efficient and effective use of the stored experiences. The priority is typically determined by the temporal difference (TD) error, which measures the difference between the predicted Q-value of a state-action pair and the target Q-value based on the current policy. Essentially, the TD error indicates how much the new experience impacts the learning process. As mentioned in \cite{schaul2015prioritized}, this prioritization might introduce a bias, which is corrected through importance sampling.

Several solutions in \cref{Table:Techniques} that employ DQN-based algorithms, \ddqn, \dddqnfirst\ and \dddqnsecond (\cite{lan2020ai,zhihong2020neurips,damjanovic2022deep}), make use of PER. In \cite{damjanovic2022deep}, the authors demonstrate the performance improvement when using PER for D3QN applied to the L2RPN problem. 

The authors of \powrl (\cite{chauhan2023powrl}) also applied PER in their solution, but combined with PPO. This is noteworthy since PPO is an on-policy algorithm, and as previously mentioned, experience replay is primarily designed for off-policy methods. The use of PER in this context shows potential benefits outside its conventional use.

\subsubsection{Exploration strategies}\label{ExplStrat}
A DRL agent needs exploration to learn more about the environment and discover potentially good actions. A common exploration strategy used for off-policy algorithms (see \cref{subsec:RL_terminology}) is \textit{Epsilon-Greedy}. Here, the agent selects a random action with probability $\epsilon$ and with probability $1-\epsilon$ the action best evaluated, $a^* = \argmax_{a \in A} \{Q(s_t, a)\}$, according to the current estimated $Q$-values.

Due to the large action space it can be difficult for the agent to converge to a good policy. \ddqn\ and \dddqnfirst, therefore, introduce a \textit{guided exploration} method. In this approach, the agent selects the top $N$ estimated $Q$-values for the current state $Q(s_t,a)$ and simulates the performance of each action to pick the action with the highest reward. This leads to a more stable training process and significantly increases training efficiency. 
It should be noted that this approach uses a greedy action selection based on simulations and, therefore, might overlook actions that do not directly yield a high reward but do result in a new state $s_{t+1}$ for which the cumulative discounted future reward is profitable.

\subsubsection{Curriculum learning or prioritized training}\label{Curriculum}
It can be noted that PNC is a very complex task. To understand and learn the concepts of a complex task, humans and animals often start with smaller, easier sub-problems and gradually move to the more complex ones. \textit{Curriculum learning}, introduced by \cite{bengio2009curriculum}, essentially copies this idea as a training strategy in the field of machine learning.

\aaacfirst (\cite{matavalam2022curriculum}) apply this technique by defining three difficulty levels and training the agent from easy to hard. The difficulty levels are defined by adjusting the Grid2Op\footnote{The research of \cite{matavalam2022curriculum} has been applied on the predecessor of Grid2Op, PyPowNet. We discuss the equivalent parameters here.} configuration parameters: \texttt{SOFT\_OVERFLOW\_THRESHOLD} ($SOT$), \texttt{NB\_TIMESTEP\_OVERFLOW\_ALLOWED} ($TOA$\footnote{In \cite{matavalam2022curriculum} referred to as consecutive overload limit, COL.}) and \texttt{HARD\_OVERFLOW\_THRESHOLD} ($HOT$). 
When a transmission line $e \in E $ has a soft overflow, i.e., $\rho_e(t)> SOT$, for more than $TOA$ consecutive time steps, the line will be disconnected. In case $\rho_e(t)> HOT$, the line $e$ immediately disconnects. In level-1 (the easiest) all parameters, $SOT$, $TOA$, and $HOT$ are set in such a way that lines would never disconnect, level-3 applies the default parameters of the environment, which are $SOT = 1.0$, $TOA = 3$ time steps and $HOT = 1.5$ and level-2 is set in such a way that the strictness of the environment gradually increases. The authors show that adding this curriculum training strategy to the A3C algorithm applied to PNC enhances the training progress and improves the results of the final agent when tested on the evaluation episodes.

The authors of \bdqn (\cite{wang2024alleviating}) introduce K-means based predefined curriculum learning. The initial states that can be used as a start during the training process are grouped into clusters using K-means. One of the clusters is then sampled based on a calculated sampling weight, which is defined by the cluster's difficulty and visiting frequency. With this sampling weight, the authors try to balance learning in a curriculum strategy, from easy to hard, while ensuring state variety.  

Another solution that orders the data given to the DRL agent is \smaac (\cite{yoon2021winning}). In this approach, the training algorithm gives priority to more difficult episodes. Contrary to what curriculum learning suggests. The difficulty of an episode is determined by how well the do-nothing agent performed and how well the DRL agent has done so far. This has the advantage that the agent enters situations where it could learn more frequently, since difficult scenarios require more interference from the agent, whereas too easy scenarios might barely exceed the thermal limit of the transmission lines.

\subsubsection{State and action space factorization}\label{Subsub:ProblemSeg}
One of the key challenges of PNC is the large state and action space which is known as the \textit{curse of dimensionality}. To deal with this problem, several authors propose to decompose the PNC problem into multiple subtasks, referred to as \textit{factorization} (or decomposition) of the (state and) action space.

Four of the evaluated papers, \cite{yoon2021winning,manczak2023hierarchical,vandersar2023marl,demol2025centrally}, suggest a hierarchical setting where one agent is responsible for determining the area, in this case, the substation where an action is needed, and another agent decides the specific action, i.e., the substation configuration, executed.  In \smaac (\cite{yoon2021winning}) and \marl (\cite{vandersar2023marl}), the agent that determines the actionable substation is a rule-based agent, and the agent that defines the specific substation configuration is one or multiple RL-based agent(s). On the other hand, the solution of \hrl (\cite{manczak2023hierarchical}) proposes an RL-based agent to select the substation together with a Greedy or RL-based agent for the substation configurations. 
The solution \CCMA\ (\cite{demol2025centrally}) reverses the order of the agents. The regional agents first propose an action, after which one central coordinating agent decides in which area the proposed action will be executed. 

In \bdqn\ (\cite{wang2024alleviating}), the authors integrate the option model inspired by \cite{sutton1999between}. In this approach, the problem is split into two subtasks based on the action frequencies. One of the neural networks of the agent is specialized on high-frequency actions and the other on low-frequency actions. 

A limitation of the previously mentioned approaches is that an agent $i$ with a certain subtask, i.e., that only controls a subset $\A_i \subset \A$, still observes the entire grid, and therefore their state space is not reduced, as pointed out by \cite{losapio2024state}. Creating a good segmentation of the state and action for PNC is a difficult task, as the highly interconnected structure of the grid and the nature of power systems can cause local actions to have an effect on a very distant portion of the grid, as shown by \cite{hines2016cascading, marot2018guided} with  \textit{influence graphs}.

In \cite{henka2022power}, the authors use an influence graph based on the \textit{Line Outage Distribution Factors}, a metric that measures the effect of disconnecting a transmission line $i$ to the power flow of transmission line $j$, to define a segmentation of the power grid. The resulting segmentation is then used to define the areas of the topological agents, which show promising performance when compared with one central agent, both MILP-based. 

\cite{losapio2024state} propose a state and action factorization method that can also be used outside the PNC context, without the use of influence graphs or domain-specific knowledge. The factorization that was found using this method has not yet been applied to an RL agent for PNC at the time of writing, so this is still open for future research.

% \cite{yoon2021winning,manczak2023hierarchical,vandersar2023marl,henka2022power,wang2024alleviating,losapio2024state}

% \subsection{(PPO) Hyperparameters}\label{subsub:Hyperpar}

% \subsection{Neural Network}\label{subsub:NN}
% GNN vs FCNN. How does the size of the NN change when going from smaller cases to larger cases?

%%%%%%%%%%%%%%%%%%%%%%%%%%%%%%%%%%%%%%%%%%%%%%%%%%%%%%%%%%%%%%%%%%%%%%%%%%%%%%
\subsection{Rule Based Techniques} \label{sec:RuleBasedTechn}
This section discusses different rule-based techniques used to enhance the agents' performance. The rule-based techniques considered are the activation threshold, reconnection of lines, and reverting to the reference topology. These expert rules are based on what is known to work well in the power grid system.
As seen in \cref{Table:Techniques}, all solutions use expert rules to enhance agents' performance.    

\subsubsection{Activation threshold}\label{subsub:ActThresh}
When the power grid is operating safely and there are no contingencies, it is best not to interfere (too often) as this will result in more instability. To accommodate for this, \ddqn\ (\cite{lan2020ai}) introduced a warning flag which we refer to as \textit{activation threshold} (AT), denoted $\rho_{\textrm{act}}$. The agent is activated only when one of the lines exceeds the AT, i.e., $\rho_{\max}(t) > \AT$. As shown in \cite{lan2020ai}, this significantly improves the training process, as the agent does not have to learn what to do during a period when the system is in a safe state. \smaac\ (\cite{yoon2021winning}) note that the MDP now becomes a semi-MDP setting for the RL-agent \cite{sutton1999between}. 

From \cref{Table:Techniques}, it can be seen that the values used for the AT range between $0.6-1.0$ and there is no golden rule. On average, the AT used for the final agent is $\AT \approx 0.93$ and from \cref{Table:ActivationThresh} it can be seen that on average a higher AT is used in the larger cases. Some agents use a lower value for $\AT$ during the training process. This can be argued by the fact that the agent otherwise might take a long time to train due to its infrequent activation. By having a lower AT, data is collected more frequently, and more training updates can take place. However, having a low AT harms the training process, as the agent needs to learn to do nothing when the network is in a safe state.

\begin{table}[h]
\centering
\begin{tabular}{l|rr}
\textbf{Case} & \multicolumn{1}{l}{\textbf{Mean AT } $\AT$} & \multicolumn{1}{l}{\textbf{Median AT $\AT$}} \\ \hline
All cases     & 0.93                            & 0.95                                      \\
Case 14       & 0.90                            & 0.95                                      \\
Case 36       & 0.94                            & 0.95                                      \\
Case 118      & 0.96                            & 0.98                    
\end{tabular}
\caption{Overview of mean and median activation thresholds $\AT$ used for the RL agents.}
\label{Table:ActivationThresh}
\end{table}

The AT is not always compared to the current $\rho_{\max}(t)$, instead some approaches estimate $\rho_{\max}(t+1)$ using a simulation of a DN-action and activate the agent when $\rho_{\max}(t+1) > \AT$.                                                               

\subsubsection{Line reconnections and disconnections} \label{subsub:LineReconnect}
As described in \cref{Subsec:Background}, when a line is overloaded for too long, it will automatically disconnect and cannot be reconnected for the next 10 time steps, defined in Grid2Op by \texttt{NB\_TIMESTEP\_RECONNECTION}. Other reasons a line could be disconnected are due to planned maintenance, introduced in the WCCI 2020 challenge, see \cref{Subsec:2020WCCI}, opponent attacks, which are part of the NeurIPS Robustness Track in 2021, see \cref{SubSec:Neurips2020}, and later challenges, or simply due to a line switching action of the agent. After such a disconnection the line cannot be reconnected until the cooldown time of the line has passed, defined in Grid2Op by \texttt{NB\_TIMESTEP\_COOLDOWN\_LINE}.

When a line is disconnected, it makes sense to reconnect it as soon as possible, as the network is usually more stable and robust when most lines are connected. Using this assumption, it makes sense to implement this straightforward rule to enhance the agent instead of letting the RL algorithm of the agent learn when to reconnect lines.  From \cref{Table:Techniques} it can be seen that, starting from the WCCI 2020 competition, most solutions use this enhancement.

In \powrl\ (\cite{chauhan2023powrl}), an additional rule for line switching is introduced: When a line is overloaded for a consecutive number of time steps, the line will be manually disconnected to avoid permanent damage.

\subsubsection{Revert to reference topology}\label{subsub:RevertTopo}
This expert rule is based on the assumption that the grid is most stable in its \textit{reference topology}, i.e., the topology where all elements are connected to the same busbar in each substation \cite{lehna2023managing}. Based on this assumption, some solutions add a heuristic that proposes actions to revert to the reference topology whenever the grid is in a safe state. To define a safe state, a \textit{revert threshold} (RT), denoted $\RT$, is usually introduced. When $\rho_{\max}(t) < \RT$ the network is assumed to be safe. The actions to revert to the reference topology are often first simulated and then greedily selected based on $\rho_{\max}(t+1)$. The code of \alphazero\ reveals that the authors also introduced a minimum period between each reversion, which prevents the grid from going back and forth between topologies and, therefore, provides a more stable grid. 
The approach \CCMA\ uses a different additional condition for topology reversion. The revert to reference topology rule is only implemented between 03:00AM and 06:00AM, as the grid tends to be more stable at night. 

The solutions \il\ and \gnnil\ do not use an RT, but instead, the scenarios are split into individual days. The authors highlight two benefits of this approach: (1) shortening the scenarios increases the amount of usable data for learning, as fewer data can be used from scenarios where the agent fails quickly, and (2) this mimics topology reversal, which is assumed to be beneficial for operating power grids, as stated earlier.

The last column of \cref{Table:Techniques} shows which solutions revert to the reference topology in their implementation. The mean value used as RT is $0.89$, see \cref{Table:RevertThresh}.

% \begin{table}[h]
% \centering
% \small
% \begin{tabular}{l|r}
%                 & \multicolumn{1}{l}{\textbf{Revert threshold (RT)}} \\ \hline
% \sas            & $0.95$\tablefootnote{\label{footnote:codebased}Based on code.}        \\
% \dddqnfirst     & $0.95$     \footref{footnote:codebased}             \\
% \alphazero  & $0.9$ \footref{footnote:codebased}                  \\
% \brute      & $0.85$ \footref{footnote:codebased}                 \\
% \ljn        & $0.9$                                               \\
% \powrl      & $0.95$                                              \\
% \curriculum & $0.8$                                               \\
% \CCMA & $0.9$                                               \\
% \SoftIL & $0.8$                                      
% \end{tabular}
% \caption{Overview of values chosen as revert threshold $\RT$ in different solutions.}
% \label{Table:RevertThresh}
% \end{table}

\begin{table}[h]
\centering
% \footnotesize
\begin{tabular}{r|l}
\textbf{Revert threshold ($\RT$)} & \textbf{Solutions} \\ \hline
0.95    & \sas\tablefootnote{\label{footnote:codebased}Based on code.}, \dddqnfirst\footref{footnote:codebased}, \powrl     \\
0.9     & \alphazero\footref{footnote:codebased}, \ljn, \CCMA   \\
0.85    & \brute\footref{footnote:codebased}    \\
0.8     & \curriculum, \SoftIL
\end{tabular}
\caption{Overview of values chosen as revert threshold $\RT$ in different solutions.}
\label{Table:RevertThresh}
\end{table}

% IDEA: make a schematic overview of how the rule-based parts of the agents can be used in combination with the RL agent.

% \cite{chauhan2023powrl} \cite{lehna2023managing} \cite{lajavaness2023l2rpn}

 % 
%%%%%%%%%%%%%%%%%%%%%%%%%%%%%%%%%%%%%%%%%%%%%%%%%%%%%%%%%%%%%%%%%%%%%%%%%%%%%%%
%%%%%%%%%%%%%%%%%%%%%%%%%%%%%%%%%%%%%%%%%%%%%%%%%%%%%%%%%%%%%%%%%%%%%%%%%%%%%%%
\section{Benchmarks, performance metrics and baselines}\label{chap:benchmark}
This chapter discusses the evaluation framework used for assessing reinforcement learning (RL) agents in the context of power network control (PNC), by covering three main components: 
benchmarks (\cref{Sec:Benchmarks}), performance metrics (\cref{sec:PerformanceMetrics}), and lastly baseline agents (\cref{Sec:Baselines}).

%%%%%%%%%%%%%%%%%%%%%%%%%%%%%%%%%%%%%%%%%%%%%%%%%%%%%%%%%%%%%%%%%%%%%%%%%%%%%%%
\subsection{Benchmarks}\label{Sec:Benchmarks}
Grid2Op provides a variety of benchmarks that can be used for the training and evaluation of an (RL) agent for PNC. See \cref{Table:Environments} for an overview of all environments that we refer to in this paper. Starting from a toy size, case 5, with only five substations, the difficulty of the benchmark environments increases to a real-size grid of 118 substations, case 118. Furthermore, additional stochasticity, more sustainable energy sources, and an adversarial agent are added to some environments to increase the difficulty. Each of the environment datasets consists of episodic scenarios with 5-minute intervals. 

The different environment datasets that can be used are more elaborately discussed in \cref{sec:overview}, and more information can be found in \cite{Grid2Op}.

\begin{table}[h]
\begin{adjustbox}{center}
% \scriptsize \tabcolsep=3pt
\begin{tabular}{ll|cccc}
\textbf{Case}         & \textbf{Grid2Op name}        & \textbf{Maintenance} & \textbf{Opponent} & \textbf{Redispatch} & \textbf{Storage units} \\ \hline
Case 5 example        & \textit{rte\_case5\_example         } &                      &                   &                  &                       \\ 
Case 14 realistic     & \textit{rte\_case14\_realistic      } &                      &                   &    \cmark        &                       \\ 
Case 14 sandbox       & \textit{l2rpn\_case14\_sandbox      } &                      &                   &    \cmark        &                       \\
Case 36 WCCI 2020     & \textit{l2rpn\_wcci\_2020           } &       \cmark         &                   &    \cmark        &                       \\
Case 36 NeurIPS 2020  & \textit{l2rpn\_neurips\_2020\_track1} &       \cmark         &      \cmark       &    \cmark        &                       \\
Case 118 NeurIPS 2020 & \textit{l2rpn\_neurips\_2020\_track2} &       \cmark         &                   &    \cmark        &                       \\
Case 36 ICAPS 2021    & \textit{l2rpn\_icaps\_2021          } &       \cmark         &      \cmark       &    \cmark        &                       \\
Case 118 WCCI 2022    & \textit{l2rpn\_wcci\_2022           } &       \cmark         &      \cmark       &    \cmark        & \cmark                \\
Case 118 IDF 2023     & \textit{l2rpn\_idf\_2023            } &       \cmark         &      \cmark       &    \cmark        & \cmark               
\end{tabular}
\end{adjustbox}
\caption{Overview of the available benchmark environments in Grid2Op, \cite{Grid2Op}.}
\label{Table:Environments}
\end{table}

%%%%%%%%%%%%%%%%%%%%%%%%%%%%%%%%%%%%%%%%%%%%%%%%%%%%%%%%%%%%%%%%%%%%%%%%%%%%%%%

\subsection{Performance metrics} \label{sec:PerformanceMetrics}
\paragraph{Training curve} A good measure of the performance of an agent is the \textit{episode return}, i.e., the accumulated reward.
To measure the performance of a novel RL algorithm, researchers often compare \textit{training curves} of the algorithm with other state-of-the-art RL algorithms. The training curve is usually a plot with the mean episode return on the y-axis and the number of time steps played in the environment on the x-axis. The training curve provides a good visualization of how agent performance improves as the RL algorithm has more interactions with the environment.

These training curves are frequently used in the research conducted on RL for PNC. However, it is difficult to compare the results of one paper with the results of the other since not everyone uses the same reward function. Therefore, it is necessary to regenerate the results of other agents for a fair comparison. To use a more general metric, one could use the mean episode \textit{duration} (or mean survival time) instead of the return in the training curve, as in \cite{dorfer2022power}.

Another aspect to consider for the training curve is that an RL-agent that uses an AT does not interact with the environment at each time step, see \cref{sec:RuleBasedTechn}. Therefore, it is more sensible to use the number of \textit{environment interactions} of the RL-agent on the x-axis instead of the number of time steps played in the environment as done in \cite{yoon2021winning,manczak2023hierarchical,vandersar2023marl}. Alternatively, the number of agent updates can be put on the x-axis as in \cite{zhou2021action}, but this quantity critically depends on the batch size chosen for the updates. 

\paragraph{Metrics for the final agent} 
To evaluate the performance of the final (trained) agent for the L2RPN challenge, \cite{marot2020l2rpn} introduced a \textit{score function}. The resulting scores range between $[-100,100]$ where $-100$ corresponds to an immediate blackout, $0$ corresponds to the performance of a do-nothing baseline agent, $80$ corresponds to safely finishing the scenario, and $100$ corresponds to an oracle agent that also optimizes all energy losses. The operational cost, the costs of energy losses, and the costs of a blackout are used to define the overall L2RPN score. 

Another frequently used metric to assess the final performance of the agent is the average number of time steps survived, as in \cite{lehna2023managing,dorfer2022power,manczak2023hierarchical}. In this case, it can be good to also include the redispatch and curtailment costs, as in \cite{dorfer2022power}, since this metric does not account for the costs of agents' actions. 

Some papers also report the mean episode reward, but since the reward function is different per proposed solution, this metric is less informative.
Other examples of statistics that, for example, are included in \cite{marot2020learning,yoon2021winning,Subramanian2021,lehna2023managing,manczak2023hierarchical} are the number of (un)solved scenarios, 
the topology depth,
the (number of) unique topologies or actions, and
the (number of) substations changed. Where the \textit{topology depth} defines the distance from the reference topology, i.e., how many substations are not in the reference topology.

Lastly, an interesting factor in the performance of the agent is the computation time needed. For example, a greedy agent performs quite well on the PNC problem, as shown by the solutions \brute\ and \hri. However, a greedy agent that has to iterate over all possible options requires a lot of computation time. \cite{dorfer2022power,lehna2024hugo} therefore include computation time as a performance measure in their evaluation.

%%%%%%%%%%%%%%%%%%%%%%%%%%%%%%%%%%%%%%%%%%%%%%%%%%%%%%%%%%%%%%%%%%%%%%%%%%%%%%%
\subsection{Baselines}\label{Sec:Baselines}
To aid a good start and help researchers evaluate their agents, the authors of \cite{marot2020l2rpn} provide a dedicated package with baselines for the L2RPN competition\footnote{These baselines are available at \url{https://github.com/Grid2op/l2rpn-baselines} and documentation is available at \cite{rte2020baselines}}. The package includes a \textit{Do-Nothing Agent}, which is a frequently used baseline and is also directly accessible in the Grid2Op package. As the name suggests, the Do-Nothing Agent does not change anything in the network and gets a corresponding L2RPN score of $0$, as described above. This is a simple baseline to confirm that your agent learned the intended behavior.
The same package also includes an \textit{Expert Agent} described by \cite{marot2018expert}, which is used as a baseline by \cite{lehna2023managing}. This agent ranks the topology actions with the information of an \textit{overload distribution graph} based on LODF, and greedily selects the best action after simulation of the top actions according to this rank.

It is also good practice to benchmark the results of an agent with some current SOTA solutions, as done in \cite{yoon2021winning, chauhan2023powrl}. Most of the proposed solutions are publicly accessible via the baseline package. However, it might be time-consuming and difficult to replicate the results of previous solutions even when the code is publicly available due to the changes to Grid2Op or other packages required by the proposed solution. 
Another option is to compare the results with earlier, different, or simpler versions of the agent to show how the proposed enhancements of the agent affect the final results, as in \cite{lan2020ai,damjanovic2022deep,lehna2023managing, manczak2023hierarchical,vandersar2023marl,dorfer2022power,wang2024alleviating}.

A baseline agent that only uses continuous actions is \textit{OptimCVXPY}. This agent is included in a couple of proposed solutions, as can be seen in \cref{Table:Techniques}. It is interesting to compare the saved costs of redispatch and curtailment actions and the computation time of a proposed agent with this agent.
\cite{serre2022reinforcement} introduces another baseline agent that uses only continuous actions, trained using PPO.

Lastly, the \textit{Greedy Agent} is an interesting baseline agent to discuss. In this baseline, used in \cite{dorfer2022power,manczak2023hierarchical,dejong2024imitation}, the agent selects the best action based on the reward $r(t+1)$ or the maximum line load $\rho_{\max}(t+1)$ simply by simulating the network at the next time step $t+1$. Similar to the RL-based agents, the Greedy Agent is activated to perform an action only when the activation threshold is exceeded. As stated earlier, this agent gives good results, and an RL agent that can get close to or improve this baseline is usually quite good. However, the Greedy Agent does not consider the potential future rewards and, therefore, omits actions that do not directly give the best reward but can result in beneficial states after execution. Additionally, simulating all possible actions is computationally expensive. Therefore, a reduction of the action space is needed when applying this to cases 36 and 118, as done in \cite{dorfer2022power}. 

%%%%%%%%%%%%%%%%%%%%%%%%%%%%%%%%%%%%%%%%%%%%%%%%%%%%%%%%%
%%%%%%%%%%%%%%%%%%%% NEW CHAPTER %%%%%%%%%%%%%%%%%%%%%%%%
%%%%%%%%%%%%%%%%%%%%%%%%%%%%%%%%%%%%%%%%%%%%%%%%%%%%%%%%%
\section{Experimental setup}\label{sec:experiments}
This chapter presents our experimental setup for our numerical studies. We evaluate a selection of modeling choices that have been discussed in \cref{sec:techniques}, by analyzing their impact on the performance of the (RL-based) agent.

First, \cref{subsec:ExpSetup} describes the general experimental setup, including the benchmark environment, performance metrics, and baselines used. The subsequent sections explain the setup for each individual evaluated modeling choice:
\begin{itemize}
    \item Action space in \cref{Subsec:ActExp},
    \item Observation space in \cref{Subsec:ObsExp},
    \item Reward function in \cref{Subsec:RwExp},
    \item Curriculum training in \cref{Subsec:ExpCurriculumTraining},
    \item Activation threshold in \cref{Subsec:RuleExp},
    \item Line reconnections and disconnections in \cref{Subsec:RuleExp}, and
    \item Revert to reference topology in \cref{Subsec:RuleExp}.
\end{itemize}

The results of all experiments are presented in \cref{sec:results}. Each experiment adjusts one modeling choice, while the other parts remain as described for the baseline presented in the following. Finally, we create a `Rainbow' agent along the same ideas and principles of \cite{hessel2018rainbow}, combining our best findings to see how much we can improve the baseline and if the combined configurations will help each other.

% This chapter presents our experimental setup and provides guidelines for applying the techniques discussed in \cref{sec:techniques}. We evaluate different modeling choices on the following topics through multiple experiments to analyze and compare their impact on performance.
% \begin{itemize}
%     \item Action space
%     \item Observation space
%     \item Reward function
%     \item Curriculum training
%     \item Activation threshold
%     \item Line reconnections and disconnections
%     \item Revert to reference topology
% \end{itemize}

% \cref{Subsec:ActExp,Subsec:ObsExp,Subsec:RwExp,Subsec:ExpCurriculumTraining,Subsec:RuleExp} explain the experimental setup for each modeling choice.
% In the experiments, we focus on a single PPO agent, the RL algorithm most frequently applied on this topic, see \cref{Table:Techniques}. 
% Since PPO is an on-policy method, techniques designed for off-policy algorithms (e.g., prioritized replay and exploration strategies) are not included. The comparison of the techniques in our experiments can still be relevant when using other algorithms such as DQN or SAC.
% Furthermore, \textit{imitation learning} and \textit{factorization} are also interesting research areas. However, exploring these in depth would require a dedicated study and falls outside the scope of this work. 

\subsection{General setup}\label{subsec:ExpSetup}
Our experiments focus on a single PPO agent, the RL algorithm most frequently applied in the L2RPN challenge, see \cref{Table:Techniques}. 
Since PPO is an on-policy method, techniques designed for off-policy algorithms (e.g., prioritized replay and exploration strategies) are not included in the evaluation studies. The comparison of the techniques in our experiments can still be relevant when using other algorithms such as DQN or SAC.
Furthermore, \textit{imitation learning} and \textit{factorization} are also interesting research areas. However, exploring these in depth would require a dedicated study and is outside the scope of this work. 

All our experiments use the open-source library RLlib from \cite{liang2018rllib}. The PPO algorithm in this library is slightly adjusted to facilitate Semi-MDP. The entire setup of our experiments is included in our RL4PNC package\footnote{\url{https://github.com/EricavanderSar/rl4pnc-survey}}.
Before training the PPO algorithm, all episodes are split into train ($80\%$), test ($10\%$), and validation scenarios ($10\%$) to avoid overfitting.
Validation scenarios are used to compare the final trained agents with the other agents (Do Nothing, Greedy, or other PPO agents).

\paragraph{Benchmarks} All experiments are tested on the benchmark case 14 sandbox with and without an added opponent, where the opponent is configured as in \cite{manczak2023hierarchical}. This test case allows for extensive experimentation while maintaining manageable computational costs. Furthermore, introducing an opponent provides a more realistic assessment, as larger environments typically involve either an opponent or maintenance constraints.

We train the PPO agents on both environments. We will refer to the agent trained on the environment without an opponent as PPO and the agent trained on an environment with an opponent as PPO$^*$.

% Some tests are validated on case 36 (To decide which one), however, due to the long computation time for the larger cases, we decided to keep these tests to a limit. 

\paragraph{(Performance) metrics} To understand how techniques affect the RL-based agent's training process,  we plot a training curve with the mean survival time (the solid lines) and the corresponding standard deviation (shaded area) against the number of environment interactions, averaged over five different model seeds. The curves are smoothed by using a moving average of 1500 environment interactions.

The final trained agents and the Greedy and Do-Nothing agents are validated on the validation scenarios. The results are presented in tables in which we compare the following metrics.
\begin{itemize}
    \item Percentage completed episodes,
    \item percentage time steps survived,
    \item percentage time steps overloaded,
    % \item average redispatch and curtailment costs, Not relevant when I don't include these actions...
    \item agent execution time: mean computation time needed when the agent is activated, i.e., when $\rho_{max}>\AT$,
    \item maximum topology depth,
    \item number of unique actions (substation configurations) used,
    \item number of unique lines in danger,
    \item number of unique substations changed,
    \item specific substations changed.
    % \item number of unique topologies,
\end{itemize} 
Most of these metrics have been examined in previous papers and are introduced in \cref{sec:PerformanceMetrics}. Here we introduce two new metrics:
\begin{mylist}
    \item percentage of time steps overloaded, tracking how often the grid operates in danger, which helps us compare agents with similar survival percentages, and
    \item number of unique lines in danger, tracking how many different lines have been overloaded. 
\end{mylist}

Each PPO agent is trained with five different seeds, and all are evaluated on the validation scenarios. Metrics are recorded for each agent and the average is reported.

\paragraph{Baselines} The baselines used for validation are 
\begin{mylist}
    \item the Do-Nothing Agent, 
    \item a Greedy Agent w.r.t. $\rho_{\max}(t+1)$, and 
    \item a single PPO agent as in \hrl~(\cite{manczak2023hierarchical})
\end{mylist} 
which is configured as described below.
\begin{itemize}
    \item The action space is $\A_{n-0}$, described in \cref{Subsub:ActSpace},
    \item the observation space is as described for \hrl\ in \cref{Table:ObsSpaces}. Without the maintenance attribute, since this is not relevant for case 14,
    \item the reward function is based on $\margins$, as described for \hrl\ in \cref{Table:Rewards},
    \item the AT is $\AT = 0.95$, see \cref{Table:Techniques},
    \item lines are reconnected when disconnected,
    \item revert to reference topology is not applied,
    \item normalization is applied as in \cref{Eq:MinMaxNorm},
    \item the hyperparameters are defined as in \cref{Table:Params},
    \item the neural network consists of three \textit{fully connected layers} (FCNN) of $[256 \times 256 \times 256]$, with ReLu activation and, 
    \item Train for $100.000$ environment interactions. % for the environment without an opponent. Train for $500.000$  with opponent.
\end{itemize}
This configuration is used as a baseline because \cite{manczak2023hierarchical} demonstrated good performance with a purely PPO-based agent, trained without simulation-based steering.\footnote{It is worth mentioning that the experiments in \cite{manczak2023hierarchical} used case 14 realistic instead of case 14 sandbox. Case 14 realistic is currently outdated, and therefore the case 14 sandbox has been used in this paper. The results may differ due to this.}

\begin{table}[h]
\centering
% \footnotesize  
% \tabcolsep=3pt
\begin{tabular}{l|rr} %p{0.35\linewidth} p{0.30\linewidth}}
% \hline
\textbf{Hyperparameter}     & \textbf{Value without opponent (PPO) }     & 
{\textbf{Value with opponent (PPO$^{*}$)}} \\ \hline
Discount ($\gamma$)         & $0.99$      & $0.99$      \\
Learning-rate               & $0.0001$    & $0.0001$    \\ %$1\times 10^{-4}$  $1$e-$4$        \\
VF coeff. ($c_1$)           & $0.5$       & $0.5$       \\
Entropy coeff. ($c_2$)      & $0.0$       & $0.01$      \\
Clipping param. ($\epsilon$)& $0.3$       & $0.3$       \\
GAE param. ($\lambda$)      & $0.95$      & $0.95$      \\ 
SGD iterations              & $5$         & $15$        \\
Minibatch size              & $256$       & $256$       \\
Batch size                  & $1024$      & $1024$      \\
% \hline
\end{tabular}
\caption{Hyperparameters used for experiments training a PPO agent on case 14.}
\label{Table:Params}
\end{table}
\FloatBarrier

Where applicable, the baseline Greedy agent uses the same configurations as the PPO agent, allowing for a fair comparison. More specifically, the action space is defined as  $\A_{n-0}$, the AT $\AT = 0.95$, and lines are reconnected whenever disconnected.

% \begin{table}[h]
% \centering
% \begin{tabular}{l|r}
% % \hline
% \textbf{Hyperparameter}     & \textbf{Value }   \\ \hline
% Discount ($\gamma$)         & $0.99$         \\
% Learning-rate               & $0.0001$       \\ %$1\times 10^{-4}$  $1$e-$4$        \\
% VF coeff. ($c_1$)           & $0.5$          \\
% Entropy coeff. ($c_2$)      & $0.01$          \\
% Clipping param. ($\epsilon$)& $0.3$          \\
% GAE param. ($\lambda$)      & $0.95$         \\ 
% SGD iterations              & $15$            \\
% Minibatch size              & $256$          \\
% Batch size                  & $1024$         \\
% % \hline
% \end{tabular}
% \caption{Hyperparameters used for experiments training a PPO agent on case 14.}
% \label{Table:Params}
% \end{table}
% \FloatBarrier

The chosen baselines allow us to assess how different modeling choices impact PPO performance relative to (1) a passive strategy (Do Nothing), (2) a heuristic-based method (Greedy), and (3) an established PPO approach from prior work.

Although we maintain consistent hyperparameters across experiments to control computational costs, we note that optimal values may vary depending on the modeling choices, such as action and observation space, which impact the output and input of the neural network used. Future work could explore adaptive tuning to further enhance performance.

We conducted additional experiments with PPO agents trained without an opponent with the hyperparameters described in the third column “Value with opponent (PPO$^*$)” of \cref{Table:Params}. A summary of these results is presented in \cref{App:ResPPO_WrongHyperParam} to demonstrate the importance of hyperparameter tuning.

% Throughout all experiments, we keep the hyperparameters consistent, as described in \cref{Table:Params}, to manage computational costs effectively while conducting diverse experiments. However, we note that when the action space changes, the output of the neural network and the number of actions that the agent needs to learn changes. Therefore, it can be beneficial to tune the parameters to achieve better results. The hyperparameters that worked well for the baseline, might not work as well with a different choice of action space. The same holds for the observation space, where we change the input of the network and other design choices we experimented with. 

\subsection{Action space experiments} \label{Subsec:ActExp}
As mentioned in \cref{Subsub:ActSpace}, most approaches reduce the action space to manage complexity in larger networks. The baseline Greedy and PPO agents use the $\A_{n-0}$ action space, which retains approximately 75\% of the original $\A_{sym}$ action space (\cref{Table:Case14ActReduction}). A more restrictive reduction, $\A_{n-1}$, results in an action space of 40\% of $\A_{sym}$. However, in the larger cases, 36 and 118, the action spaces are often reduced to around 1\%--2\%, most often using a Greedy reduction method, as described in \cref{Subsub:ActSpace}. Therefore, in the experiments, we will also consider the smallest action space for case 14, containing roughly 5\% of $\A_{sym}$, proposed by \dddqnsecond, which we denote by $\A_{d3qn}$.

To assess the number of actions required for effective network management, experiments are conducted with both the Greedy and PPO agent using four different action spaces: 
\begin{itemize}
    \item $\A_{sym}$: 178 actions,
    \item $\A_{n-0}$: 133 actions,
    \item $\A_{n-1}$: 73 actions and
    \item $\A_{d3qn}$: 9 actions.
\end{itemize}

\subsection{Observation space experiments}\label{Subsec:ObsExp}
These experiments examine how different observation spaces impact the training and final performance of the RL agent. We compare the following observation spaces:
\begin{itemize}
    \item \textbf{Baseline}: \textit{active power, line loads ($\rho$), time-step overflow, and topology configuration}. 
    \item \textbf{Complete}: \textit{time stamp, active power, reactive power, voltages, voltage angles ($\theta$), line flows, line loads, time-step overflow, line status, and topology configuration}.
    \item \textbf{\dddqnsecond} by \cite{damjanovic2022deep}: \textit{voltages of lines only ($v\_or, v\_ex$), line flows, line loads, and topology configuration}. This is interesting because this was the smallest observation space in the solutions discussed.
    \item \textbf{Line loads}: Including only the line loads $\rho$, which is the common feature across all proposed solutions.
    \item \textbf{Danger}: The baseline observation space with an additional \textit{danger} parameter as in \smaac~\cite{yoon2021winning}.
    \item \textbf{History}: Extends the previous observation space with six historical time steps, as in \smaac. Since an FCNN is used, there is no need to structure the features in a feature matrix, which would be beneficial for a GNN architecture. 
\end{itemize}
For an overview of the observation spaces considered in the experiments and their sizes, see \cref{Table:SizeObsSpaceExp} in \cref{App:ObsSpaces}. 
The features \textit{cooldown time} and \textit{maintenance} are omitted as they are irrelevant for case 14.

Continuous features are normalized in all experiments, as detailed in \cref{Eq:MinMaxNorm}. Since timestamp transformation is not explicitly discussed in prior solutions, we apply a cosine transformation, which better captures cyclic patterns than min-max normalization. The time of day (\texttt{time\_of\_day}) and the day of the year ($\texttt{day\_of\_year}$) are handled as separate cycles in view of the well-known daily and seasonal patterns in power usage; see \cref{Eq:t_norm,Eq:y_norm}.

\begin{align}
    \texttt{time\_of\_day} &= \cos(2\pi \cdot \texttt{minute\_of\_day} / 60\cdot24)   \label{Eq:t_norm} \\
    \texttt{day\_of\_year} &= \cos(2\pi \cdot \texttt{day\_of\_year} / 365)   \label{Eq:y_norm}
\end{align}

\subsection{Reward function experiments}\label{Subsec:RwExp}
It is essential to choose a suitable reward function to help the RL algorithm learn the right behavior, as mentioned in \cref{Subsub:Reward}. \cref{Table:Rewards} shows how each proposed solution uses a (sometimes slightly) different reward function. In the experiments, the following reward functions are compared for the training process of a PPO agent:
\begin{itemize}
    \item \textbf{Baseline}: The reward from \hrl\ using the $\margins$ function.
    \item \textbf{Base + Bonus/Penalty}: The baseline reward, with a bonus of $500$ for episode completion and a penalty of $-300$ in case of game over.
    \item \textbf{Constant}: A constant reward, as used by \dddqnsecond.
    \item \textbf{Binbinchen}: The reward used in \binbin\ using the $\rwrho$ function.
    \item \textbf{SMAAC}: The reward used in \smaac\ using the $\loss$ function.
    \item \textbf{AlphaZero}: The reward used in \alphazero\ using the $\rwalpha$ function.
\end{itemize}

\subsection{Curriculum training}\label{Subsec:ExpCurriculumTraining}
Different from previous design choices, curriculum training is an optional enhancement of the RL algorithm. There are many possibilities for how to apply curriculum training to the PNC problem. We compare the baseline PPO agent with a curriculum-trained PPO agent using the method proposed by \cite{matavalam2022curriculum}, described in \cref{Curriculum}. The configuration parameters for the three levels, $SOT$, $TOA$, $HOT$, discussed in \cref{Curriculum}, are shown in \cref{Table:CurrTrainingSetup}, along with an additional parameter $NOD$:NO\_OVERFLOW\_DISCONNECTION. With this setup, level 1 corresponds to a level where there is no line limit enforcement, level 2 disconnects after 15 time steps of high overload of 2, and level 3 corresponds to the default environment behavior\footnote{The exact setup slightly deviates from \cite{matavalam2022curriculum} as the Grid2Op environment changed over time.}.

\begin{table}[h]
\centering
% \small
\begin{tabular}{r|llll}
\textbf{Level} & \textbf{$NOD$} & \textbf{$SOT$} & \textbf{$TOA$} & \textbf{$HOT$} \\ \hline
1              & \cmark       & 99           & 99           & 999          \\
2              &              & 2            & 15           & 99           \\
3              &              & 1            & 2            & 2           
\end{tabular}
\caption{Setup curriculum levels.}
\label{Table:CurrTrainingSetup}
\end{table}
\FloatBarrier

We transition from level 1 to level 2 after $1/5$ of the total training time, and from 2 to 3 (the final level) after $7/15$ of the training time, following \cite{matavalam2022curriculum}. %As we train for 100.000 time steps, level 2 starts after $\frac{1}{5}\cdot 100.000 = 20.000$ and level 3 starts after $\frac{7}{15}\cdot 100.000 = 46.667$ time steps.

\subsection{Rule-based experiments}\label{Subsec:RuleExp}
For each rule-based decision described in \cref{sec:RuleBasedTechn}, we test different application strategies: applying the rule during both training and the final validation of the agent, applying the rule only at one stage, or using a more strict version in one phase in which the rule is applied more frequently. This allows us to examine how the rule timing and intensity affect the agent’s performance.

\paragraph{Activation threshold}
As mentioned in \cref{subsub:ActThresh}, there is no golden rule to choose the activation threshold $\AT$. This motivates us to examine its impact by testing different values of $\AT$. We tested different values $\AT \in \{0.8, 0.9, 0.95, 0.99\}$, evaluating both the training and validation performance in eight combinations of the training and final agent thresholds.

\paragraph{Line switching: Reconnection and disconnection}
The PPO and Greedy baselines include a rule that reconnects the lines if their reconnection improves the current action, tested using simulation. We measure the frequency of these reconnections during training and compare performance with and without this rule.
Additionally, we test a rule that disconnects overloaded lines after multiple consecutive time steps. This rule is motivated by the default reconnection delay of 10 steps\footnote{As defined by the Grid2Op parameter: \texttt{NB\_TIMESTEP\_RECONNECTION}.}, which can make early disconnection beneficial.
The rule is implemented similarly to reconnection, meaning that only beneficial disconnections are executed. 
We test incorporating this rule in combination with the line reconnection rule. Testing the line disconnection by itself is not considered, as permanent disconnections would be detrimental.

\paragraph{Revert to reference topology}
The `revert to reference topology' rule restores substations to their reference topology (i.e., all elements connected to the same bus) when the network is in a `safe state'. Each reversion action is simulated, after which the best one is executed if it improves on the current action. Safe states are defined by different revert thresholds (RTs), which we compare in training and validation experiments: $\RT \in \{0.8, 0.9, 0.95\}$. If no RT is applied, the agent follows the baseline behavior.

%%%%%%%%%%%%%%%%%%%%%%%%%%%%%%%%%%%%%%%%%%%%%%%%%%%%%%%%%%%%%%%%%%%%%%%%%%%%%%%
%%%%%%%%%%%%%%%%%%%%%%%%%%%%%%%%%%%%%%%%%%%%%%%%%%%%%%%%%%%%%%%%%%%%%%%%%%%%%%%
\section{Results, recommendations and guidelines}\label{sec:results}
This chapter describes the results for the experiments described in \cref{sec:experiments}. The results will be presented in the same order as described in the experimental setups in \cref{sec:experiments}. First, we describe the results of the Greedy, Do-Nothing and PPO baselines in \cref{Sec:ResBaselines}. The results of the numerical experiments on each of the modeling choices follow in the subsequent sections \cref{Sec:ResAct,Sec:ResObs,Sec:ResRw,Sec:ResCurr,Sec:ResAT,Sec:Res_LineSwitch,Sec:ResRT}.
Lastly, \cref{Sec:ResRainbow} presents the setup and the results of the Rainbow agent, configured based on the results of the previous experiments.

% The best values for “Completed episodes”, “Steps survived” and “Steps overloaded” in the tables with validation results are indicated in bold.
% Since the `Substations Changed' metric was consistently equal to {[}1, 2, 3, 4, 5, 8, 12{]} across all experiments in \cref{Subsec:ObsExp,Subsec:RwExp,Subsec:ExpCurriculumTraining,Subsec:RuleExp}, it has been omitted from the tables in those sections.

\subsection{Baselines}\label{Sec:ResBaselines}

The Greedy agent consistently outperforms the other baselines in maintaining grid safety, as the results in \cref{Table:ResBaselineNoOpp,Table:ResBaselineOpp} demonstrate, where
the best values for “Completed episodes”, “Steps survived” and “Steps overloaded” are highlighted in bold, following the convention used throughout the paper.

\begin{figure}[!tbh]
    \centering
    \includegraphics[width=0.8\linewidth]{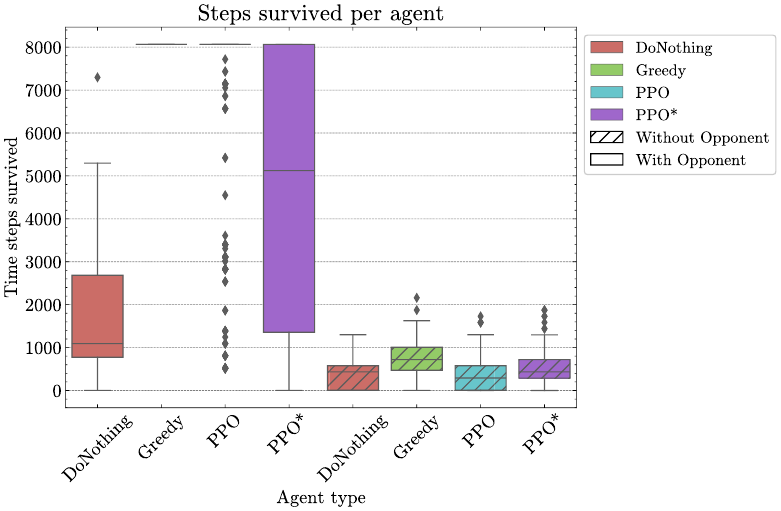}
    \caption{Boxplot of the time steps survived for the baseline agents applied to the validation episodes.} 
    \label{Fig:BaselinesBoxplot}
\end{figure}

As expected, the Greedy agent requires significantly more execution time than the RL-based agents and, of course, the Do-Nothing agent. The PPO baseline did not complete all episodes but survived 87\% of time steps on average -- significantly improving on the Do-Nothing agent, which confirms it learned meaningful behavior -- while using fewer than half the actions required by the Greedy agent (see 'Unique actions' column). PPO$^*$ performed worse than PPO in the environment without an opponent but slightly better in the environment with an opponent. \cref{Fig:BaselinesBoxplot} provides a visual comparison of the validation results, further illustrating that environments with an opponent pose a significantly greater challenge.

\begin{table}[h]
\begin{adjustbox}{center}
\resizebox{1.2\textwidth}{!}
{
\scriptsize
\tabcolsep=3pt
\begin{tabular}{|l|p{0.1\linewidth}|p{0.1\linewidth}|p{0.1\linewidth}|p{0.1\linewidth}|p{0.1\linewidth}|p{0.08\linewidth}|p{0.08\linewidth}|p{0.1\linewidth}|p{0.1\linewidth}|}
\hline
\textbf{Agent}            & \textbf{Completed episodes} & \textbf{Steps survived} & \textbf{Steps overloaded} & \textbf{Agent execution time {[}ms{]}} & \textbf{Maximum topology depth} & \textbf{Unique actions} & \textbf{Unique lines in danger} & \textbf{Unique subs changed} & \textbf{Substations changed} \\ \hline
Do Nothing & 0.0\%            & 20.90\%            & 0.185\%          & 0.063   & 1 & 1    & 4    & 0   & {[}{]}                       \\ \hline
Greedy     & \textbf{100.0\%} & \textbf{100.00 \%} & 0.014\%          & 369.669 & 5 & 65   & 12   & 5   & {[}1, 3, 4, 5, 8{]}          \\ \hline
PPO        & 78.6\%           & 87.31\%            & \textbf{0.008\%} & 1.9668  & 5 & 27.8 & 11.8 & 6   & {[}1, 2, 3, 4, 5, 8, 12{]} \\ \hline
PPO$^{*}$  & 35.4\%           & 57.89\%            & 0.041\%          & 1.9658  & 4.6 & 80.6 & 17   & 6.8& {[}1, 2, 3, 4, 5, 8, 12{]} \\ \hline
\end{tabular}
}
\end{adjustbox}
\caption{Validation results of baseline agents, described in \cref{subsec:ExpSetup}, applied to the environment case 14 sandbox without an opponent.}
\label{Table:ResBaselineNoOpp}
\end{table}
\FloatBarrier

\begin{table}[h]
\begin{adjustbox}{center}
\resizebox{1.2\textwidth}{!}
{
\scriptsize
\tabcolsep=3pt
\begin{tabular}{|l|p{0.1\linewidth}|p{0.1\linewidth}|p{0.1\linewidth}|p{0.1\linewidth}|p{0.1\linewidth}|p{0.08\linewidth}|p{0.08\linewidth}|p{0.1\linewidth}|p{0.1\linewidth}|}
\hline
\textbf{Agent}            & \textbf{Completed episodes} & \textbf{Steps survived} & \textbf{Steps overloaded} & \textbf{Agent execution time {[}ms{]}} & \textbf{Maximum topology depth} & \textbf{Unique actions} & \textbf{Unique lines in danger} & \textbf{Unique subs changed} & \textbf{Substations changed} \\ \hline
Do Nothing  & 0.0\%       & 1.920\%         & 4.137\%           & 0.062          & 1           & 1             & 12            & 0           & {[}{]}                       \\ \hline
Greedy      & 0.0\%       & \textbf{9.030\%}& 1.217\%           & 365.313        & 5           & 88            & 14            & 5           & {[}1, 3, 4, 5, 8{]}          \\ \hline
PPO         & 0.0\%       & 4.690\%         & \textbf{0.638\%}  & 2.1618  & 4 & 12.8 & 12.2 & 5.4 & {[}1, 2, 3, 4, 5, 8, 12{]}     \\ \hline
PPO$^*$     & 0.0\%       & 5.740\%         & 0.683\%           & 2.2398  & 4.6 & 91   & 15   & 7   & {[}1, 2, 3, 4, 5, 8, 12{]} \\ \hline
\end{tabular}
}
\end{adjustbox}
\caption{Validation results of baseline agents, described in \cref{subsec:ExpSetup}, applied to the environment case 14 sandbox \textit{with} an opponent.}
\label{Table:ResBaselineOpp}
\end{table}
\FloatBarrier
\FloatBarrier

\begin{figure}[!tbh]
\begin{adjustbox}{center}
    \resizebox{1.2\textwidth}{!}
    {
    \includegraphics[width=0.49\linewidth]{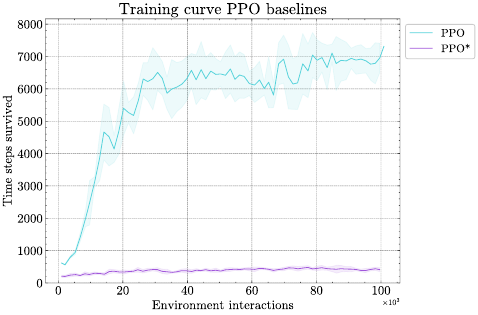}
    \includegraphics[width=0.49\linewidth]{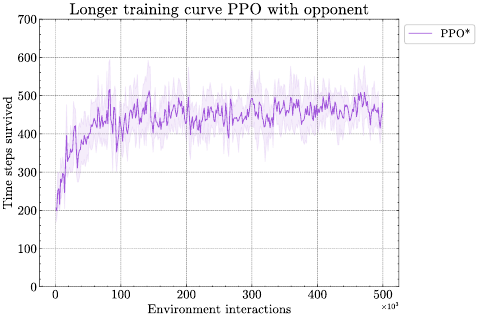}
    }
\end{adjustbox}
    \caption{Training curves of the PPO agent trained without opponent (PPO) and the PPO agent trained with opponent (PPO*).} 
    \label{Fig:BaselinesTrainCurve}
\end{figure}

\cref{Fig:BaselinesTrainCurve} illustrates the training progress of the PPO agent, showing that it converges at approximately 7000 time steps survived ($\sim$87\%), consistent with the validation results in \cref{Table:ResBaselineNoOpp}. The training curve suggests that PPO$^*$ does not learn effectively. Training was extended to 500.000 interactions (right figure in \cref{Fig:BaselinesTrainCurve}), but performance plateaued after 100.000 interactions. Therefore, our subsequent experiments stop after 100.000 interactions when training with an opponent.

\FloatBarrier

\subsection{Action spaces}\label{Sec:ResAct}
\begin{figure}[!tbh]
\begin{adjustbox}{center}
    \resizebox{1.2\textwidth}{!}
    {
    \begin{subfigure}[b]{0.45\textwidth}
        \centering
        \includegraphics[scale=1.0]{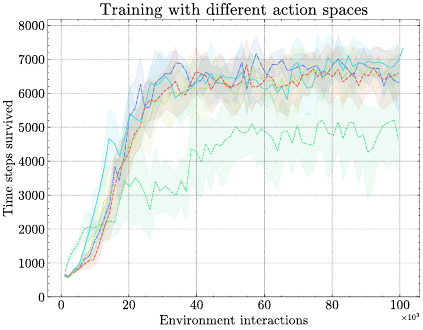}
        \caption{Without an opponent (PPO).}
        \label{Subfig:Act}
    \end{subfigure}
     \hfill
     \begin{subfigure}[b]{0.55\textwidth}
        \centering
        \includegraphics[scale=1.0]{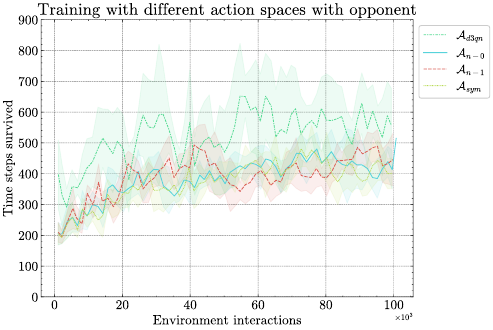}
        \caption{With an opponent (PPO$^*$).}
        \label{Subfig:ActOpp}
    \end{subfigure}
    }
\end{adjustbox}
    \caption{Training curves of PPO agents trained with different action spaces.} 
    \label{Fig:ActSpacesTrainCurve}
\end{figure}

The choice in action can have a significant impact on the performance of the (PPO) agents.  
\cref{Fig:ActSpacesTrainCurve} illustrates substantial differences in the training progress when training the PPO agent with different action spaces. In an environment without an opponent, \cref{Subfig:Act}, the agent trained with the action space $\A_{d3qn}$ yields the worst performance, whereas in an environment with an opponent, \cref{Subfig:ActOpp}, it achieves the best result.   
This suggests that expert-curated actions in $\A_{d3qn}$, which emphasize robustness, are particularly beneficial against line attacks. The results in \cref{Table:ResActSpaceOpp} support this observation, showing that $\A_{d3qn}$ consistently outperforms other action spaces across all agents (Greedy, PPO, and PPO$^*$).

Despite eliminating less robust actions, the training progress of the agent with the $\A_{n-1}$ action space does not exceed that of the agents trained with the $\A_{n-0}$ and $\A_{sym}$ action spaces. 

\FloatBarrier

Examining the validation results of the environment without an opponent reported in \cref{Table:ResActSp}, we observe that both the Greedy and PPO agents exhibit slightly reduced performance when the baseline action space $\A_{n-0}$ is reduced to $\A_{n-1}$. However, this reduction comes with a substantial decrease in execution time for the Greedy agent. The agent with  $\A_{d3qn}$, the smallest action space containing only 5\% of the available actions, survives only half of the time steps. Both the Greedy agent and PPO agents trained with $\A_{n-0}$ and $\A_{n-1}$ act only in substations 1, 3, 4, 5, and 8, suggesting that a reduced action space incorporating these substations could be effective. Furthermore, the Greedy agent with the action space $A_{sym}$ exhibits performance comparable to that of $A_{n-0}$, while the agent with the action space $A_{n-0}$ employs significantly fewer distinct configurations.

\cref{Table:ResPPOWrongHyperpar} in \cref{App:ResPPO_WrongHyperParam} shows that training the PPO agent using different hyperparameters can give substantially different results. Specifically, the reduced action space $\A_{n-1}$ achieves an overall better score in \cref{Table:ResPPOWrongHyperpar} compared to \cref{Table:ResActSp}. This fact highlights the importance of hyperparameter tuning when changing the number of actions an RL agent needs to learn.

\begin{table}[h]
\begin{adjustbox}{center}
\resizebox{1.2\textwidth}{!}
{
\scriptsize
\tabcolsep=3pt
\begin{tabular}{|l|p{0.07\linewidth}|p{0.1\linewidth}|p{0.1\linewidth}|p{0.1\linewidth}|p{0.1\linewidth}|p{0.1\linewidth}|p{0.08\linewidth}|p{0.08\linewidth}|p{0.1\linewidth}|p{0.1\linewidth}|}
\hline
\textbf{Agent}           & \textbf{action space} & \textbf{completed episodes} & \textbf{steps survived} & \textbf{steps overloaded} & \textbf{agent execution time {[}ms{]}} & \textbf{maximum topology depth} & \textbf{unique actions} & \textbf{unique lines in danger} & \textbf{unqique subs changed} & \textbf{substations changed}                    \\ \hline
\multirow{4}{*}{Greedy}  & $\A_{sym}$ & \textbf{100.0\%} & \textbf{100.00\%} & 0.017\%          & 487.991 & 6   & 90   & 14   & 10   & {[}0, 1, 2, 3, 4, 5, 8, 9, 11, 12{]}                    \\ \cline{2-11} 
      & \cgray $\A_{n-0}$ & \cgray \textbf{100.0\%} & \cgray \textbf{100.00\%} & \cgray 0.014\% & \cgray 369.669 & \cgray 5   & \cgray 65   & \cgray 12   & \cgray 5    & \cgray {[}1, 3, 4, 5, 8{]}                                 \\ \cline{2-11} 
                         & $\A_{n-1}$ &  98.0\%          &  99.74\%          & 0.022\%          & 210.361 & 5   & 52   & 16   & 5    & {[}1, 3, 4, 5, 8{]}                                 \\ \cline{2-11} 
                         & $\A_{d3qn}$&  19.0\%          &  52.05\%          & 0.088\%          & 31.585  & 3   & 8    & 4    & 3    & {[}1, 3, 4{]}                                     \\ \hline\hline
\multirow{4}{*}{PPO}     & $\A_{sym}$ & 74.6\%           & 85.68\%           & 0.011\%          & 2.2462  & 5.4 & 43   & 12.4 & 9.4  & {[}0, 1, 2, 3, 4, 5, 6, 8, 9, 10, 11, 12, 13{]} \\ \cline{2-11}
      & \cgray $\A_{n-0}$ & \cgray 78.6\% & \cgray 87.31\% & \cgray \textbf{0.008\%} & \cgray 1.9668  & \cgray 5   & \cgray 27.8 & \cgray 11.8 & \cgray 6    & \cgray {[}1, 2, 3, 4, 5, 8, 12{]}                      \\ \cline{2-11}
                         & $\A_{n-1}$ & 53.6\%           & 77.87\%           & 0.030\%          & 2.2202  & 4   & 35.2 & 15.8 & 4.8  & {[}1, 3, 4, 5, 8{]}                             \\ \cline{2-11} 
                         & $\A_{d3qn}$& 27.0\%           & 56.23\%           & 0.112\%          & 4.8118  & 1   & 5.4  & 5.4  & 2.2  & {[}1, 3, 4{]}                                   \\ \hline\hline
\multirow{4}{*}{PPO$^*$} & $\A_{sym}$ &  41.6\%          &  61.68\%          & 0.048\%          & 2.2738  & 4.8 & 91   & 15.8 & 12.2 & {[}0, 1, 2, 3, 4, 5, 6, 8, 9, 10, 11, 12, 13{]} \\ \cline{2-11} 
      & \cgray $\A_{n-0}$ & \cgray 35.4\% & \cgray 57.89\% & \cgray 0.041\% & \cgray 1.9658  & \cgray 4.6 & \cgray 80.6 & \cgray 17   & \cgray 6.8 & \cgray {[}1, 2, 3, 4, 5, 8, 12{]}                      \\ \cline{2-11} 
                         & $\A_{n-1}$ &  16.2\%          &  43.36\%          & 0.114\%          & 2.38    & 4.2 & 63.8 & 16   & 5    & {[}1, 3, 4, 5, 8{]}                             \\ \cline{2-11} 
                         & $\A_{d3qn}$&   9.0\%          &  41.94\%          & 0.117\%          & 2.1576  & 2.8 & 9    & 8.2  & 3    & {[}1, 3, 4{]}                                   \\ \hline
\end{tabular}
}
\end{adjustbox}
\caption{Validation results of different action spaces for the agents, applied to the environment case 14 sandbox without an opponent. The baseline results are highlighted in gray.}
\label{Table:ResActSp}
\end{table}
\FloatBarrier

\begin{table}[h]
\begin{adjustbox}{center}
\resizebox{1.2\textwidth}{!}
{
\scriptsize
\tabcolsep=3pt
\begin{tabular}{|l|p{0.07\linewidth}|p{0.1\linewidth}|p{0.1\linewidth}|p{0.1\linewidth}|p{0.1\linewidth}|p{0.1\linewidth}|p{0.08\linewidth}|p{0.08\linewidth}|p{0.1\linewidth}|p{0.1\linewidth}|}
\hline
\textbf{Agent} &
  \textbf{Action space} &
  \textbf{Completed episodes} &
  \textbf{Steps survived} &
  \textbf{Steps overloaded} &
  \textbf{Agent execution time {[}ms{]}} &
  \textbf{Maximum topology depth} &
  \textbf{Unique actions} &
  \textbf{Unique lines in danger} &
  \textbf{Unqique subs changed} &
  \textbf{Substations changed} \\ \hline
\multirow{4}{*}{Greedy}  & $\A_{sym}$ & 0.0\% & 9.550\%          & 1.252\%  & 479.783 & 7   & 115  & 15   & 12  & {[}0, 1, 2, 3, 4, 5, 8, 9, 10, 11, 12, 13{]}              \\ \cline{2-11} 
      & \cgray $\A_{n-0}$ & \cgray0.0\% & \cgray9.030\%          & \cgray1.217\%  & \cgray365.313 & \cgray5   & \cgray88   & \cgray14   & \cgray5   & \cgray{[}1, 3, 4, 5, 8{]}                                 \\ \cline{2-11} 
                         & $\A_{n-1}$ & 0.0\% & 9.330\%          & 1.691\%  & 208.128 & 5   & 54   & 17   & 5   & {[}1, 3, 4, 5, 8{]}                                 \\ \cline{2-11} 
                         & $\A_{d3qn}$& 0.0\% & \textbf{9.940\%} & 0.947\%  & 30.668  & 3   & 8    & 13   & 3   & {[}1, 2, 3{]}                                     \\ \hline\hline
\multirow{4}{*}{PPO}     & $\A_{sym}$ & 0.0\% & 4.738\%          & \textbf{0.681\%} & 2.3484  & 3.4 & 17.8 & 10.4 & 7.2 & {[}0, 1, 2, 3, 4, 5, 8, 9, 10, 11, 12, 13{]}    \\ \cline{2-11} 
      & \cgray $\A_{n-0}$ & \cgray0.0\% & \cgray4.690\%          & \cgray0.638\% & \cgray2.1618  & \cgray4   & \cgray 12.8 & \cgray 12.2 & \cgray 5.4 & \cgray {[}1, 2, 3, 4, 5, 8, 12{]}                      \\ \cline{2-11}
                         & $\A_{n-1}$ & 0.0\% & 4.476\%          & 1.840\% & 2.2628  & 3.4 & 33.8 & 15.4 & 4.8 & {[}1, 3, 4, 5, 8{]}                             \\ \cline{2-11} 
                         & $\A_{d3qn}$& 0.0\% & 5.376\%          & 1.639\% & 2.8682  & 1   & 4.6  & 14.8 & 1.8 & {[}1, 3, 4{]}                                   \\ \hline\hline
\multirow{4}{*}{PPO$^*$} & $\A_{sym}$ & 0.0\% & 6.220\%          & 0.692\%  & 2.4682  & 3.8 & 99.2 & 13.6 & 13  & {[}0, 1, 2, 3, 4, 5, 6, 8, 9, 10, 11, 12, 13{]} \\ \cline{2-11} 
      & \cgray  $\A_{n-0}$ & \cgray 0.0\% & \cgray 5.740\%          & \cgray 0.683\% & \cgray 2.2398  & \cgray 4.6 & \cgray 91   & \cgray 15   & \cgray 7   & \cgray {[}1, 2, 3, 4, 5, 8, 12{]}                      \\ \cline{2-11} 
                         & $\A_{n-1}$ & 0.0\% & 6.658\%          & 0.912\% & 2.9604  & 3.6 & 65.4 & 15.2 & 5   & {[}1, 3, 4, 5, 8{]}                             \\ \cline{2-11} 
                         & $\A_{d3qn}$& 0.0\% & \textbf{6.904\%} & 1.127\%  & 2.57    & 2.6 & 8.8  & 14   & 3   & {[}1, 3, 4{]}                                   \\ \hline
\end{tabular}
}
\end{adjustbox}
\caption{Validation results of different action spaces for the agents, applied to the environment case 14 sandbox \textit{with} an opponent. The baseline results are highlighted in gray.}
\label{Table:ResActSpaceOpp}
\end{table}
\FloatBarrier

From \cref{Table:ResActSpaceOpp}, we can conclude that more robust action spaces, such as $\A_{d3qn}$, help the agent perform on a network where lines can be disrupted and, similar to the results in \cref{Table:ResBaselineOpp}, the agent trained with an opponent performs better compared to the agent trained without an opponent. 

\subsection{Observation spaces}\label{Sec:ResObs}
\begin{figure}[!tbh]
\begin{adjustbox}{center}
    \resizebox{1.2\textwidth}{!}
    {
    \begin{subfigure}[b]{0.45\textwidth}
        \centering
        \includegraphics[scale=1.0]{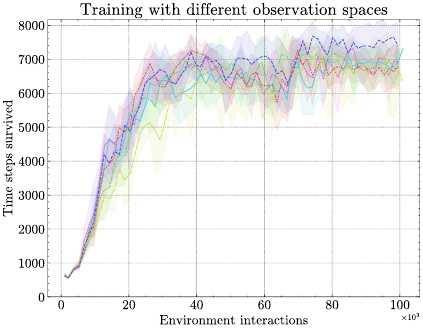}
        \caption{Without an opponent (PPO).}
        \label{Subfig:Obs}
    \end{subfigure}
     \hfill
     \begin{subfigure}[b]{0.57\textwidth}
        \centering
        \includegraphics[scale=1.0]{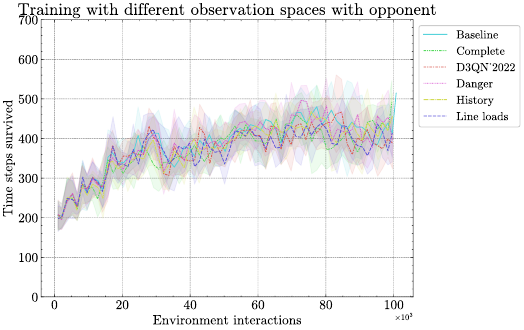}
        \caption{With an opponent (PPO$^*$).}
        \label{Subfig:ObsOpp}
    \end{subfigure}
    }
\end{adjustbox}
    \caption{Training curves of PPO agents trained with different observation spaces.} 
    \label{Fig:ObsSpacesTrainCurve}
\end{figure}

In contrast to the action space selection, the choice of observation space has a limited impact on performance.
\cref{Subfig:Obs} shows that the smallest observation space, “Line loads”, achieves the highest survival time near the end of training, suggesting that additional input features do not enhance learning progress. In the environment with an opponent (\cref{Subfig:ObsOpp}), training curves for different observation spaces largely overlap, indicating that the choice of observation space has little impact in this setting. These results raise the question of whether the additional input features beyond line loads are irrelevant for decision-making or if adjustments, such as changes to the neural network architecture, are necessary to fully utilize the extra information. Similar conclusions can be obtained by observing the validation results in \cref{Table:ResObsSpace,Table:ResObsSpace_OPP}.

In the experiments where we used alternative hyperparemeters for the PPO agent the observation space of \dddqnsecond\ gave the best results, see\cref{Table:ResPPOWrongHyperpar} in \cref{App:ResPPO_WrongHyperParam},  in contrast to the findings in \cref{Table:ResObsSpace}, highlighting the impact of hyperparameter choices.

\FloatBarrier

\begin{table}[h]
\begin{adjustbox}{center}
\resizebox{1.2\textwidth}{!}
{
\scriptsize
\tabcolsep=3pt
\begin{tabular}{|l|p{0.11\linewidth}|p{0.1\linewidth}|p{0.08\linewidth}|p{0.08\linewidth}|p{0.1\linewidth}|p{0.1\linewidth}|p{0.08\linewidth}|p{0.08\linewidth}|p{0.08\linewidth}|}
\hline
\textbf{Agent}           & 
\textbf{Observation space} &
  \textbf{Completed episodes} &
  \textbf{Steps survived} &
  \textbf{Steps overloaded} &
  \textbf{Agent execution time {[}ms{]}} &
  \textbf{Maximum topology depth} &
  \textbf{Unique actions} &
  \textbf{Unique lines in danger} &
  \textbf{Unique subs changed}\\ \hline
\multirow{6}{*}{PPO}     & \cgray Baseline & \cgray 78.6\%         & \cgray 87.31\%         & \cgray 0.008\%          & \cgray 1.9668 & \cgray 5   & \cgray 27.8 & \cgray 11.8 & \cgray 6   \\ \cline{2-10} 
                         & Complete                    & 78.0\%         & 87.71\%         & 0.010\%          & 2.5374 & 4.4 & 32.8 & 11   & 6.2 \\ \cline{2-10} 
                         & D3QN\_2022                  & 72.0\%         & 83.52\%         & 0.012\%          & 2.2846 & 4.4 & 35.4 & 12.2 & 6.4 \\ \cline{2-10} 
                         & Danger                      & 77.2\%         & 87.78\%         & 0.013\%          & 2.2118 & 4.6 & 33   & 12.8 & 6.8 \\ \cline{2-10} 
                         & History                     & 72.6\%         & 84.42\%         & 0.009\%          & 3.0548 & 4.8 & 30.4 & 11.2 & 6   \\ \cline{2-10} 
                         & Line loads                  & \textbf{92.8\%}& \textbf{96.01\%}& \textbf{0.005\%} & 1.9328 & 4.6 & 12.2 & 7.2  & 4.8 \\ \hline\hline
\multirow{6}{*}{PPO$^*$} & \cgray  Baseline & \cgray 35.4\%         & \cgray 57.89\%         & \cgray 0.041\%          & \cgray 1.9658  & \cgray 4.6 & \cgray 80.6 & \cgray 17   & \cgray 6.8  \\ \cline{2-10} 
                         & Complete                    & 25.2\%         & 52.59\%         & 0.067\%          & 3.0234 & 5   & 95.2 & 17.8 & 7     \\ \cline{2-10} 
                         & D3QN\_2022                  & 38.4\%         & 61.58\%         & 0.040\%          & 2.9538 & 4.6 & 78.6 & 16.4 & 7     \\ \cline{2-10} 
                         & Danger                      & 43.4\%         & 67.73\%         & 0.037\%          & 2.7828 & 5   & 74.4 & 15.6 & 7     \\ \cline{2-10} 
                         & History                     & 32.0\%         & 54.57\%         & 0.052\%          & 4.2958 & 4.8 & 89.4 & 17.6 & 7     \\ \cline{2-10} 
                         & Line loads                  & 52.0\%         & 68.02\%         & 0.036\%          & 2.5792 & 4.4 & 61   & 13.6 & 7     \\ \hline
\end{tabular}
}
\end{adjustbox}
\caption{Validation results of different observation spaces for the agents, applied to the environment case 14 sandbox without an opponent.}
\label{Table:ResObsSpace}
\end{table}
\FloatBarrier

\begin{table}[h]
\begin{adjustbox}{center}
\resizebox{1.2\textwidth}{!}
{
\scriptsize
\tabcolsep=3pt
\begin{tabular}{|l|p{0.11\linewidth}|p{0.1\linewidth}|p{0.08\linewidth}|p{0.08\linewidth}|p{0.1\linewidth}|p{0.1\linewidth}|p{0.08\linewidth}|p{0.08\linewidth}|p{0.08\linewidth}|}
\hline
\textbf{Agent}           & 
  \textbf{Observation space} &
  \textbf{Completed episodes} &
  \textbf{Steps survived} &
  \textbf{Steps overloaded} &
  \textbf{Agent execution time {[}ms{]}} &
  \textbf{Maximum topology depth} &
  \textbf{Unique actions} &
  \textbf{Unique lines in danger} &
  \textbf{Unique subs changed}  \\ \hline
\multirow{6}{*}{PPO}     & \cgray  Baseline & \cgray 0.0\%  & \cgray 4.690\%         & \cgray \textbf{0.638\%}& \cgray 2.1618 & \cgray 4   & \cgray 12.8 & \cgray 12.2 & \cgray 5.4 \\ \cline{2-10} 
                         & Complete                    & 0.0\%  & 4.718\%         & 0.804\%         & 3.9826 & 3.4 & 66.4 & 14.8 & 7   \\ \cline{2-10} 
                         & D3QN\_2022                  & 0.0\%  & 4.304\%         & 0.770\%         & 3.7278 & 3.6 & 64.8 & 16.8 & 6.8 \\ \cline{2-10} 
                         & Danger                      & 0.0\%  & 4.758\%         & 0.652\%         & 5.712  & 3.4 & 36   & 12.4 & 6.6 \\ \cline{2-10} 
                         & History                     & 0.0\%  & 4.400\%         & 0.766\%         & 4.7906 & 3.2 & 26.6 & 12.8 & 5.8 \\ \cline{2-10} 
                         & Line loads                  & 0.0\%  & 4.184\%         & 0.814\%         & 3.4236 & 3.2 & 45.8 & 14.6 & 6.4 \\ \hline \hline
\multirow{6}{*}{PPO$^*$} & \cgray  Baseline & \cgray 0.0\%  & \cgray \textbf{5.740\%}& \cgray 0.683\%         & \cgray 2.2398 & \cgray 4.6 & \cgray 91   & \cgray 15   & \cgray 7  \\ \cline{2-10} 
                         & Complete                    & 0.0\%  & 5.296\%         & 0.674\%         & 5.0958 & 3.4 & 97.8 & 15.6 & 7   \\ \cline{2-10} 
                         & D3QN\_2022                  & 0.0\%  & 5.468\%         & 0.672\%         & 3.9996 & 4   & 89.2 & 14.8 & 7   \\ \cline{2-10} 
                         & Danger                      & 0.0\%  & 5.586\%         & 0.718\%         & 3.96   & 4.4 & 93.4 & 15.8 & 6.8 \\ \cline{2-10} 
                         & History                     & 0.0\%  & 5.288\%         & 0.744\%         & 5.2616 & 3.4 & 91   & 16   & 7   \\ \cline{2-10} 
                         & Line loads                  & 0.0\%  & 5.250\%         & 0.762\%         & 3.5642 & 3.6 & 77.2 & 15   & 7   \\ \hline
\end{tabular}
}
\end{adjustbox}
\caption{Validation results of different observation spaces for the agents, applied to the environment case 14 sandbox \textit{with} an opponent.}
\label{Table:ResObsSpace_OPP}
\end{table}
\FloatBarrier

\subsection{Reward functions}\label{Sec:ResRw}
\begin{figure}[!tbh]
\begin{adjustbox}{center}
\resizebox{1.2\textwidth}{!}
    {
    \begin{subfigure}[b]{0.45\textwidth}
        \centering
        \includegraphics[scale=1.0]{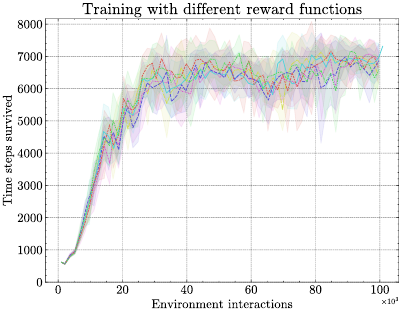}
        \caption{Without an opponent (PPO).}
    \end{subfigure}
     \hfill
     \begin{subfigure}[b]{0.55\textwidth}
        \centering
        \includegraphics[scale=1.0]{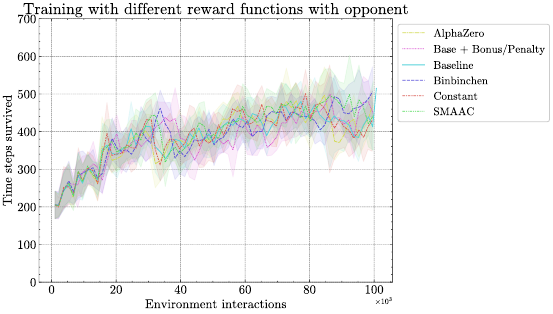}
        \caption{With an opponent (PPO$^*$).}
    \end{subfigure}
    }
\end{adjustbox}
    \caption{Training curves of PPO agents trained with different reward functions.} 
    \label{Fig:RewardsTrainCurve}
\end{figure}

\begin{figure}[!tbh]
\begin{adjustbox}{center}
\resizebox{1.2\textwidth}{!}
{
    \begin{subfigure}[b]{0.42\textwidth}
        \centering
        \includegraphics[scale=1.0]{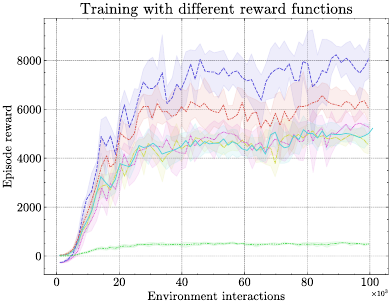}
        \caption{Without an opponent (PPO).}
    \end{subfigure}
     \hfill
     \begin{subfigure}[b]{0.55\textwidth}
        \centering
        \includegraphics[scale=1.0]{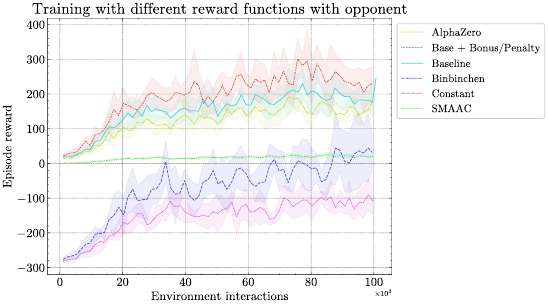}
        \caption{With an opponent (PPO$^*$).}
    \end{subfigure}
}
\end{adjustbox}
    \caption{Curves of the mean episode reward of the PPO trained with different reward functions.} 
    \label{Fig:RewardsRewardCurve}
\end{figure}

Training curves for different reward functions show little difference in the number of time steps survived (see \cref{Fig:RewardsTrainCurve}). The plots in \cref{Fig:RewardsRewardCurve} show the mean episode reward to highlight the difference between the experiments. Although episode rewards differ significantly, their impact on survival time is negligible. In the opponent-based training, the Binbinchen reward agent appeared to still be learning at the end of the 100.000 interactions. A similar trend is observed for the agent using the SMAAC reward when we zoom in on the mean episode reward for this agent. These results suggest that extended training could further improve the agents' performance.

\FloatBarrier
\FloatBarrier

The validation results in \cref{Table:ResRw,Table:ResRW_OPP} confirm that the impact of the compared reward functions on the agent's final performance is minimal. Notably, the AlphaZero reward agent trained \textit{with} an opponent (PPO$^*$) performs relatively well in the environment \textit{without} an opponent. 

\begin{table}[h]
\begin{adjustbox}{center}
\resizebox{1.2\textwidth}{!}
{
\scriptsize
\tabcolsep=3pt
\begin{tabular}{|l|p{0.11\linewidth}|p{0.1\linewidth}|p{0.08\linewidth}|p{0.08\linewidth}|p{0.1\linewidth}|p{0.1\linewidth}|p{0.08\linewidth}|p{0.08\linewidth}|p{0.08\linewidth}|}
\hline
\textbf{Agent} &
  \textbf{Reward function} &
  \textbf{Completed episodes} &
  \textbf{Steps survived} &
  \textbf{Steps overloaded} &
  \textbf{Agent execution time {[}ms{]}} &
  \textbf{Maximum topology depth} &
  \textbf{Unique actions} &
  \textbf{Unique lines in danger} &
  \textbf{Unique subs changed} \\ \hline
\multirow{6}{*}{PPO}     & \cgray  Baseline & \cgray 78.6\%          & \cgray 87.31\%          & \cgray \textbf{0.008\%} & \cgray 1.9668 & \cgray 5   & \cgray 27.8 & \cgray 11.8 & \cgray 6    \\ \cline{2-10} 
                         & AlphaZero                   & 75.6\%          & 86.46\%          & 0.015\%          & 2.2298 & 4.8 & 33   & 12.8 & 6.2  \\ \cline{2-10} 
                         & Constant                    & 78.4\%          & 87.83\%          & 0.009\%          & 2.3478 & 5   & 31.8 & 12   & 6    \\ \cline{2-10} 
                         & SMAAC                       & 71.6\%          & 84.49\%          & 0.012\%          & 2.2804 & 5   & 39.8 & 13   & 6.6  \\ \cline{2-10} 
                         & Binbinchen                  & 74.8\%          & 86.09\%          & 0.016\%          & 2.2614 & 4.2 & 34.6 & 13.2 & 6    \\ \cline{2-10} 
                         & Base + Bonus/Penalty        & \textbf{81.0\%} & \textbf{89.09\%} & 0.010\%          & 2.2558 & 5   & 30   & 12.4 & 6    \\ \hline \hline
\multirow{6}{*}{PPO$^*$} & \cgray  Baseline & \cgray 35.4\%          & \cgray 57.89\%          & \cgray 0.041\%          & \cgray 1.9658  & \cgray 4.6 & \cgray 80.6 & \cgray 17   & \cgray 6.8  \\ \cline{2-10} 
                         & AlphaZero                   & 58.2\%          & 74.19\%          & 0.023\%          & 2.0352 & 4.8 & 66.6 & 15.4 & 6.8  \\ \cline{2-10} 
                         & Constant                    & 48.0\%          & 68.69\%          & 0.033\%          & 2.084  & 5.2 & 74.8 & 15.8 & 7    \\ \cline{2-10} 
                         & SMAAC                       & 35.8\%          & 57.21\%          & 0.060\%          & 2.1178 & 4.6 & 81.4 & 16.2 & 7    \\ \cline{2-10} 
                         & Binbinchen                  & 26.2\%          & 46.07\%          & 0.091\%          & 2.072  & 4.8 & 94.8 & 16.8 & 7    \\ \cline{2-10} 
                         & Base + Bonus/Penalty        & 21.1\%          & 44.93\%          & 0.111\%          & 2.0162 & 4.6 & 91.6 & 17.4 & 7    \\ \hline
\end{tabular}
}
\end{adjustbox}
\caption{Validation results of different reward functions for the agents, applied to the environment case 14 sandbox without an opponent.}
\label{Table:ResRw}
\end{table}
\FloatBarrier

\begin{table}[h]
\begin{adjustbox}{center}
\resizebox{1.2\textwidth}{!}
{
\scriptsize
\tabcolsep=3pt
\begin{tabular}{|l|p{0.11\linewidth}|p{0.1\linewidth}|p{0.08\linewidth}|p{0.08\linewidth}|p{0.1\linewidth}|p{0.1\linewidth}|p{0.08\linewidth}|p{0.08\linewidth}|p{0.08\linewidth}|}
\hline
\textbf{Agent} &
  \textbf{Reward function} &
  \textbf{Completed episodes} &
  \textbf{Steps survived} &
  \textbf{Steps overloaded} &
  \textbf{Agent execution time {[}ms{]}} &
  \textbf{Maximum topology depth} &
  \textbf{Unique actions} &
  \textbf{Unique lines in danger} &
  \textbf{Unqique subs changed} \\ \hline
\multirow{6}{*}{PPO}     & \cgray  Baseline  & \cgray 0.0\% & \cgray 4.690\%          & \cgray 0.638\%          & \cgray 2.1618 & \cgray 4   & \cgray 12.8  & \cgray 12.2 & \cgray 5.4  \\ \cline{2-10} 
                         & AlphaZero                    & 0.0\% & 4.928\%          & \textbf{0.518\%} & 2.5898 & 3.4 & 12.2  & 11.4 & 4.6  \\ \cline{2-10} 
                         & Constant                     & 0.0\% & 4.816\%          & 0.607\%          & 2.4422 & 3.6 & 15.6  & 11.6 & 5.4  \\ \cline{2-10} 
                         & SMAAC                        & 0.0\% & 4.338\%          & 0.817\%          & 2.3208 & 3.4 & 17    & 13.2 & 6    \\ \cline{2-10} 
                         & Binbinchen                   & 0.0\% & 4.534\%          & 0.785\%          & 2.4766 & 3.8 & 19.4  & 13   & 5.4  \\ \cline{2-10} 
                         & Base + Bonus/Penalty         & 0.0\% & 4.608\%          & 0.673\%          & 2.301  & 3.4 & 12.6  & 11.2 & 5.8  \\ \hline \hline
\multirow{6}{*}{PPO$^*$} & \cgray  Baseline  & \cgray 0.0\% & \cgray 5.740\%          & \cgray 0.683\%          & \cgray 2.2398  & \cgray 4.6 & \cgray 91   & \cgray 15   & \cgray 7     \\ \cline{2-10} 
                         & AlphaZero                    & 0.0\% & \textbf{5.970\%} & 0.638\%          & 2.4438 & 3.6 & 85.8  & 15.2 & 7    \\ \cline{2-10} 
                         & Constant                     & 0.0\% & 5.484\%          & 0.669\%          & 2.3988 & 3.8 & 90.4  & 14.4 & 6.8  \\ \cline{2-10} 
                         & SMAAC                        & 0.0\% & 5.508\%          & 0.790\%          & 2.5324 & 4   & 94    & 15.4 & 7    \\ \cline{2-10} 
                         & Binbinchen                   & 0.0\% & 5.878\%          & 0.758\%          & 2.545  & 4   & 100.6 & 16.4 & 7    \\ \cline{2-10} 
                         & Base + Bonus/Penalty         & 0.0\% & 5.676\%          & 0.750\%          & 2.5292 & 3.8 & 98.2  & 15.6 & 6.8  \\ \hline
\end{tabular}
}
\end{adjustbox}
\caption{Validation results of different reward functions for the agents, applied to the environment case 14 sandbox \textit{with} an opponent.}
\label{Table:ResRW_OPP}
\end{table}
\FloatBarrier

\subsection{Curriculum training}\label{Sec:ResCurr}
Curriculum training, which starts with relatively easy environment settings, initially provides a clear performance benefit. However, once the difficulty increases, the agent struggles to maintain its high score or recover its performance.

\begin{figure}[!tbh]
\begin{adjustbox}{center}
\resizebox{1.2\textwidth}{!}
{
    \begin{subfigure}[b]{0.45\textwidth}
        \centering
        \includegraphics[scale=1.0]{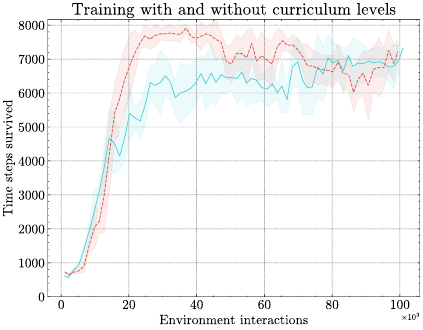}
        \caption{Without an opponent (PPO).}
    \end{subfigure}
     \hfill
     \begin{subfigure}[b]{0.55\textwidth}
        \centering
        \includegraphics[scale=1.0]{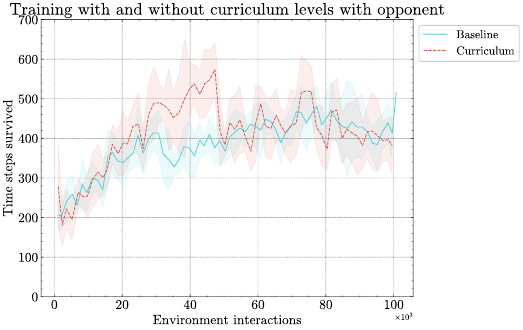}
        \caption{With an opponent (PPO$^*$).}
    \end{subfigure}
}
\end{adjustbox}
    \caption{Comparison of the training curves of PPO agents trained with and without curriculum levels setup as described in \cref{Subsec:ExpCurriculumTraining}. } 
    \label{Fig:CurrTrainCurve}
\end{figure}

As shown in \cref{Fig:CurrTrainCurve}, the training progress of the curriculum agent begins promising, but its performance rapidly decreases when the final difficulty level starts around 46.667 time steps, reaching a level similar to the baseline agent. 

While the final performance does not surpass the baseline, curriculum training offers a clear advantage in training efficiency. Specifically, training without an opponent for the baseline agent takes 350-460 minutes, whereas training for the curriculum version takes 301-360 minutes, reducing the training time by $\sim$15-20\%. Looking at the training process with an opponent, we see that the baseline agent takes 87-124 minutes and the curriculum agent takes 46-60 minutes of training, reducing the training time by $\sim$50\%. 

\FloatBarrier

\cref{Table:ResCurr,Table:ResCurr_Opp} show that the curriculum scores slightly, but not significantly, higher than the baseline agent in the environment without an opponent. However, no improvement is visible for the curriculum agents validated in the environment with an opponent. 

\begin{table}[h]
\begin{adjustbox}{center}
\resizebox{1.2\textwidth}{!}
{
\scriptsize
\tabcolsep=3pt
\begin{tabular}{|l|p{0.1\linewidth}|p{0.1\linewidth}|p{0.08\linewidth}|p{0.08\linewidth}|p{0.1\linewidth}|p{0.1\linewidth}|p{0.08\linewidth}|p{0.08\linewidth}|p{0.08\linewidth}|p{0.1\linewidth}|}
\hline
\textbf{Agent} &
  \textbf{Curriculum Training} &
  \textbf{Completed episodes} &
  \textbf{Steps survived} &
  \textbf{Steps overloaded} &
  \textbf{Agent execution time {[}ms{]}} &
  \textbf{Maximum topology depth} &
  \textbf{Unique actions} &
  \textbf{Unique lines in danger} &
  \textbf{Unique subs changed} \\ \hline
\multirow{2}{*}{PPO}     & \cgray  & \cgray 78.6\%          & \cgray 87.31\%          & \cgray \textbf{0.008\% }& \cgray 1.967 & \cgray 5.0 & \cgray 27.8 & \cgray 11.8 & \cgray 6.0 \\ \cline{2-10} 
                         & \cmark             & \textbf{81.6\%} & \textbf{90.07\%} & 0.012\%          & 2.108 & 4.8 & 31.8 & 12.2 & 6.4 \\ \hline\hline
\multirow{2}{*}{PPO$^*$} & \cgray  & \cgray 35.4\%          & \cgray 57.89\%          & \cgray 0.041\%          & \cgray 1.966 & \cgray 4.6 & \cgray 80.6 & \cgray 17   & \cgray 6.8 \\ \cline{2-10} 
                         & \cmark             & 20.2\%          & 38.39\%          & 0.100\%          & 2.241 & 4.4 & 93.2 & 16.8 & 7.0 \\ \hline
\end{tabular}
}
\end{adjustbox}
\caption{Validation results of agents trained with curriculum levels compared with the baseline agents, applied to the environment case 14 sandbox without an opponent.}
\label{Table:ResCurr}
\end{table}
\FloatBarrier

\begin{table}[h]
\begin{adjustbox}{center}
\resizebox{1.2\textwidth}{!}
{
\scriptsize
\tabcolsep=3pt
\begin{tabular}{|l|p{0.1\linewidth}|p{0.1\linewidth}|p{0.08\linewidth}|p{0.08\linewidth}|p{0.1\linewidth}|p{0.1\linewidth}|p{0.08\linewidth}|p{0.08\linewidth}|p{0.08\linewidth}|p{0.1\linewidth}|}
\hline
\textbf{Agent} &
  \textbf{Curriculum Training} &
  \textbf{Completed episodes} &
  \textbf{Steps survived} &
  \textbf{Steps overloaded} &
  \textbf{Agent execution time {[}ms{]}} &
  \textbf{Maximum topology depth} &
  \textbf{Unique actions} &
  \textbf{Unique lines in danger} &
  \textbf{Unique subs changed} \\ \hline
\multirow{2}{*}{PPO}     & \cgray  & \cgray 0.0\% & \cgray 4.690\%         & \cgray \textbf{0.638\%}  & \cgray 2.1618  & \cgray 4   & \cgray 12.8 & \cgray 12.2 & \cgray 5.4 \\ \cline{2-10} 
                         & \cmark             & 0.0\% & 4.50\%          & 0.664\%           & 2.171   & 3.8 & 17.4 & 11.2 & 5.4 \\ \hline\hline
\multirow{2}{*}{PPO$^*$} & \cgray  & \cgray 0.0\% & \cgray \textbf{5.740\%}& \cgray 0.683\%           & \cgray 2.2398  & \cgray 4.6 & \cgray 91   & \cgray 15   & \cgray 7  \\ \cline{2-10} 
                         & \cmark             & 0.0\% & 5.69\%          & 0.948\%           & 2.318   & 4.2 & 82.0 & 16.6 & 6.8 \\ \hline
\end{tabular}
}
\end{adjustbox}
\caption{Validation results of agents trained with curriculum levels compared with the baseline agents, applied to the environment case 14 sandbox with an opponent.}
\label{Table:ResCurr_Opp}
\end{table}
\FloatBarrier

\subsection{Activation threshold}\label{Sec:ResAT}
\begin{figure}[!tbh]
\begin{adjustbox}{center}
\resizebox{1.2\textwidth}{!}
{
    \begin{subfigure}[b]{0.45\textwidth}
        \centering
        \includegraphics[scale=1.0]{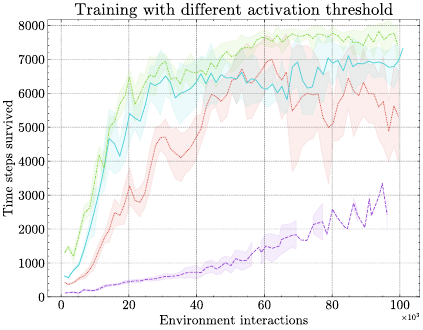}
        \caption{Without an opponent (PPO).}
    \end{subfigure}
     \hfill
     \begin{subfigure}[b]{0.57\textwidth}
        \centering
        \includegraphics[scale=1.0]{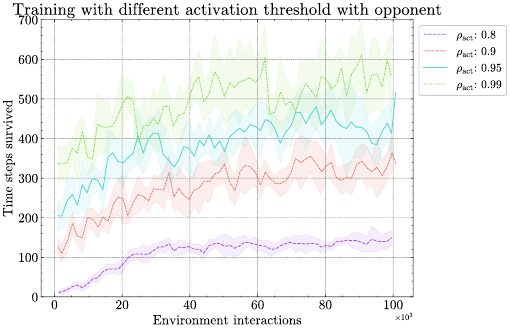}
        \caption{With an opponent (PPO$^*$).}
    \end{subfigure}
}
\end{adjustbox}
    \caption{Training curves of PPO agents trained with different activation thresholds $\AT$.} 
    \label{Fig:AT_TrainCurve}
\end{figure}

\cref{Fig:AT_TrainCurve} illustrates the significant impact of the activation threshold (AT) on training progress. The plot suggests that increasing the activation threshold improves performance up to a certain point. However, beyond a certain threshold $\AT > 1.0$, performance is likely to degrade, as the agent may act too late, allowing lines to break before intervention.

\FloatBarrier

\begin{figure}[!tbh]
\begin{adjustbox}{center}
\resizebox{1.2\textwidth}{!}
{
    \begin{subfigure}[b]{0.45\textwidth}
        \centering
        \includegraphics[scale=1.0]{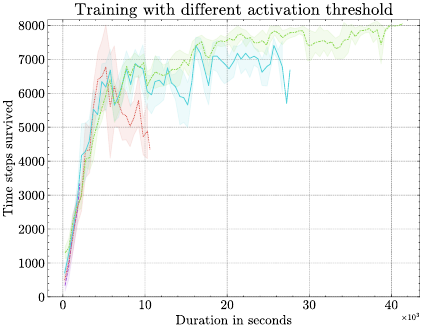}
        \caption{Without an opponent (PPO).}
    \end{subfigure}
     \hfill
     \begin{subfigure}[b]{0.57\textwidth}
        \centering
        \includegraphics[scale=1.0]{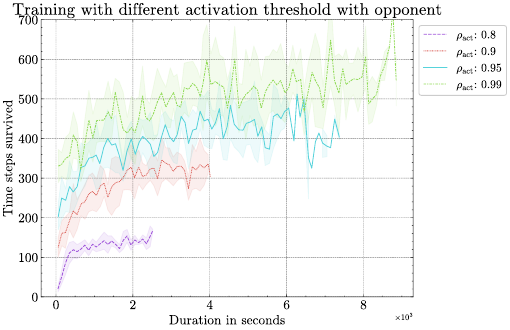}
        \caption{With an opponent (PPO$^*$).}
    \end{subfigure}
}
\end{adjustbox}
    \caption{Training curves plotting the survival time against the duration of the training process. All agents are trained for 100.000 time steps. The training curves show the PPO trained with different activation thresholds $\AT$.} 
    \label{Fig:AT_TrainCurve_time}
\end{figure}

\begin{figure}[!tbh]
\begin{adjustbox}{center}
\resizebox{1.2\textwidth}{!}
{
    \begin{subfigure}[b]{0.45\textwidth}
        \centering
        \includegraphics[scale=1.0]{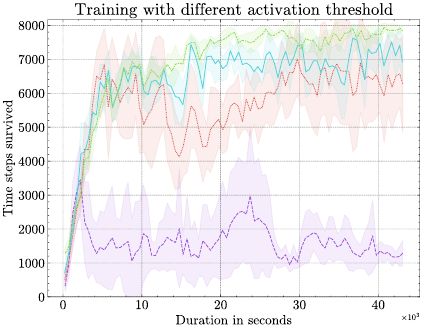}
        \caption{Without an opponent (PPO).}
    \end{subfigure}
     \hfill
     \begin{subfigure}[b]{0.57\textwidth}
        \centering
        \includegraphics[scale=1.0]{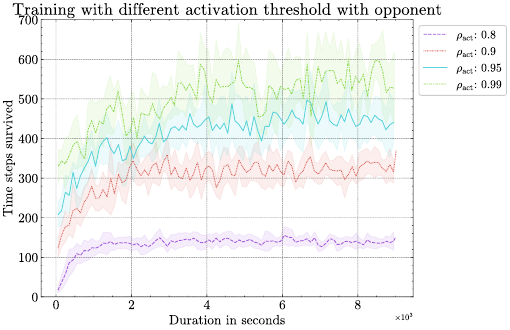}
        \caption{With an opponent (PPO$^*$).}
    \end{subfigure}
}
\end{adjustbox}
    \caption{Training curves plotting the survival time against the duration of the training process. All agents are trained for the same duration. The training curves show the PPO trained with different activation thresholds $\AT$.} 
    \label{Fig:ATLong_TrainCurve_time}
\end{figure}

We remark that when the threshold is higher, the agent interacts less with the environment per episode. Therefore, as illustrated in \cref{Fig:AT_TrainCurve_time}, training with a higher threshold also takes longer. \cref{Fig:AT_TrainCurve,Fig:AT_TrainCurve_time}, suggest that the training curve for $\AT=0.8$ may continue to improve in the no-opponent setting. To verify whether longer training improves performance, we conducted an additional experiment in which all agents were trained for the same duration, see \cref{Fig:ATLong_TrainCurve_time}. However, the upward trend for $\AT=0.8$ did not persist in the extended training experiment.

\FloatBarrier
\FloatBarrier

Intuitively, one might expect that in the presence of an opponent, anticipating earlier (i.e., using a lower AT) would be beneficial. However, \cref{Fig:AT_TrainCurve} shows that, regardless of the opponent's presence, the highest AT ($\AT=0.99$) achieves the longest survival time.

\cref{Table:ResAT,Table:ResAT_Opp} show that the best performing agents are trained and evaluated with the highest AT ($\AT=0.99$). The agent trained with $\AT=0.95$ but using $\AT=0.99$ during validation performs competitively, indicating some flexibility in selecting the AT.

Lastly, we observe that a higher AT leads to fewer redundant actions that do not alter the grid configuration. This fact indicates that a higher AT encourages more deliberate and impactful interventions, potentially improving overall stability. 

Our results suggest that a relatively high AT ($\AT\in [0.95, 0.99]$) is optimal for both training and validation, leading to longer survival times and reducing redundant interventions.

\begin{table}[h]
\begin{adjustbox}{center}
\resizebox{1.2\textwidth}{!}
{
\scriptsize
\tabcolsep=3pt
\begin{tabular}{|l|p{0.05\linewidth}|p{0.05\linewidth}|p{0.1\linewidth}|p{0.08\linewidth}|p{0.08\linewidth}|p{0.1\linewidth}|p{0.1\linewidth}|p{0.08\linewidth}|p{0.08\linewidth}|p{0.08\linewidth}|}
\hline
\textbf{Agent} &
  \textbf{AT} &
  \textbf{Train AT} &
  \textbf{Completed episodes} &
  \textbf{Steps survived} &
  \textbf{Steps overloaded} &
  \textbf{Agent execution time {[}ms{]}} &
  \textbf{Maximum topology depth} &
  \textbf{Unique actions} &
  \textbf{Unique lines in danger} &
  \textbf{Unique subs changed} \\ \hline
\multirow{16}{*}{PPO}     & \multirow{4}{*}{0.8}  & 0.8  & 13.0\%           & 44.95\%          & 0.149\%           & 1.8564 & 5     & 76.8   & 18.2   & 7     \\ \cline{3-11} 
                          &                       & 0.9  & 10.8\%           & 26.80\%          & 0.271\%           & 2.0542 & 5     & 67     & 17.4   & 7     \\ \cline{3-11} 
                          &                       & 0.95 & 1.8\%            & 19.05\%          & 0.289\%           & 1.7026 & 5     & 63.8   & 18.2   & 6.8   \\ \cline{3-11} 
                          &                       & 0.99 & 2.4\%            & 17.82\%          & 0.389\%           & 1.9152 & 5.2   & 46.8   & 17     & 7     \\ \cline{2-11} 
                          & \multirow{4}{*}{0.9}  & 0.8  & 56.8\%           & 74.66\%          & 0.036\%           & 2.1218 & 4.8   & 32.6   & 12.6   & 6.4   \\ \cline{3-11} 
                          &                       & 0.9  & 46.6\%           & 68.44\%          & 0.035\%           & 2.1662 & 4.8   & 43     & 14     & 6.6   \\ \cline{3-11} 
                          &                       & 0.95 & 42.4\%           & 66.69\%          & 0.035\%           & 1.783  & 5.2   & 46.8   & 15     & 6.6   \\ \cline{3-11} 
                          &                       & 0.99 & 32.0\%           & 57.98\%          & 0.056\%           & 2.1034 & 4.8   & 34.8   & 13.6   & 6.6   \\ \cline{2-11} 
                          & \multirow{4}{*}{0.95} & 0.8  & 64.6\%           & 78.62\%          & 0.025\%           & 2.2682 & 4     & 20.2   & 8.4    & 6.4   \\ \cline{3-11} 
                          &                       & 0.9  & 74.6\%           & 85.21\%          & 0.014\%           & 2.4084 & 4.2   & 30.2   & 11     & 6.4   \\ \cline{3-11}
                          &    & \cgray  0.95 & \cgray 78.6\%           & \cgray 87.31\%          & \cgray \textbf{0.008\% } & \cgray 1.967  & \cgray 5.0   & \cgray 27.8   & \cgray 11.8   & \cgray 6.0   \\ \cline{3-11} 
                          &                       & 0.99 & 73.8\%           & 87.33\%          & 0.013\%           & 2.3442 & 4.6   & 24.2   & 12     & 6.4   \\ \cline{2-11} 
                          & \multirow{4}{*}{0.99} & 0.8  & 60.6\%           & 75.55\%          & 0.032\%           & 2.3064 & 3.6   & 13.8   & 5.4    & 5.8   \\ \cline{3-11} 
                          &                       & 0.9  & 75.6\%           & 84.86\%          & 0.021\%           & 2.349  & 3.8   & 17.8   & 10     & 6.2   \\ \cline{3-11} 
                          &                       & 0.95 & 91.6\%           & 95.80\%          & 0.010\%           & 2.2824 & 4     & 16.6   & 7.8    & 6     \\ \cline{3-11} 
                          &                       & 0.99 & \textbf{92.6\% } & \textbf{97.07\%} & 0.009\%           & 2.4236 & 3.8   & 14.2   & 6.8    & 5     \\ \hline \hline
\multirow{16}{*}{PPO$^*$} & \multirow{4}{*}{0.8}  & 0.8  & 0.0\%            &  3.85\%          & 0.975\%            & 2.0288 & 5   & 112.4 & 18.8 & 7   \\ \cline{3-11} 
                          &                       & 0.9  & 0.0\%            &  5.37\%          & 0.746\%             & 2.0548 & 5   & 103.6 & 19.6 & 7   \\ \cline{3-11} 
                          &                       & 0.95 & 0.0\%            &  6.58\%          & 0.491\%            & 2.0866 & 5   & 97.4  & 19.2 & 7   \\ \cline{3-11} 
                          &                       & 0.99 & 0.0\%            &  9.95\%          & 0.442\%            & 1.9956 & 5.2 & 82    & 19.2 & 6.8 \\ \cline{2-11} 
                          & \multirow{4}{*}{0.9}  & 0.8  & 0.4\%            &  9.35\%          & 0.341\%            & 1.9694 & 4.2 & 106.8 & 18.4 & 7   \\ \cline{3-11} 
                          &                       & 0.9  & 2.2\%            & 22.78\%          & 0.173\%            & 2.0276 & 4.6 & 105.6 & 18.6 & 7   \\ \cline{3-11} 
                          &                       & 0.95 & 5.6\%            & 32.63\%          & 0.113\%             & 1.9982 & 5.2 & 94.6  & 17.8 & 6.6 \\ \cline{3-11} 
                          &                       & 0.99 & 24.4\%           & 48.98\%          & 0.070\%            & 2.111  & 5   & 67.4  & 16   & 6.8 \\ \cline{2-11} 
                          & \multirow{4}{*}{0.95} & 0.8  &  4.8\%           & 19.80\%          & 0.160\%              & 2.0542 & 3.8 & 103.4 & 18   & 7   \\ \cline{3-11} 
                          &                       & 0.9  & 25.0\%           & 50.13\%          & 0.063\%            & 2.2582 & 4.8 & 96.6  & 17.8 & 7   \\ \cline{3-11} 
                          &   & \cgray   0.95 & \cgray 35.4\%           & \cgray 57.89\%          & \cgray 0.041\%          & \cgray 1.9658  & \cgray 4.6 & \cgray 80.6 & \cgray 17   & \cgray 6.8\\ \cline{3-11} 
                          &                       & 0.99 & 64.8\%           & 79.36\%          & 0.018\%             & 2.0402 & 4.4 & 41.6  & 12.8 & 6.2 \\ \cline{2-11} 
                          & \multirow{4}{*}{0.99} & 0.8  &  7.2\%           & 24.92\%          & 0.145\%            & 2.2372 & 3.6 & 100.8 & 17.4 & 7   \\ \cline{3-11} 
                          &                       & 0.9  & 39.0\%           & 57.58\%          & 0.052\%            & 2.143  & 3.8 & 80.6  & 15.6 & 6.8 \\ \cline{3-11} 
                          &                       & 0.95 & 46.4\%           & 66.50\%          & 0.043\%            & 2.0842 & 4.2 & 72.4  & 16.2 & 6.4 \\ \cline{3-11} 
                          &                       & 0.99 & 68.0\%           & 81.01\%          & 0.029\%            & 2.1568 & 3.8 & 39.8  & 11.2 & 6   \\ \hline
\end{tabular}
}
\end{adjustbox}
\caption{Validation results of different activation thresholds used for the final agent agents (the column `AT') combined with different activation thresholds used for the training of the agent (column `Train AT'), applied to the environment case 14 sandbox without an opponent. The baseline results are highlighted in grey.}
\label{Table:ResAT}
\end{table}
\FloatBarrier

\begin{table}[h]
\begin{adjustbox}{center}
\resizebox{1.2\textwidth}{!}
{
\scriptsize
\tabcolsep=3pt
\begin{tabular}{|l|p{0.05\linewidth}|p{0.05\linewidth}|p{0.1\linewidth}|p{0.08\linewidth}|p{0.08\linewidth}|p{0.1\linewidth}|p{0.1\linewidth}|p{0.08\linewidth}|p{0.08\linewidth}|p{0.08\linewidth}|}
\hline
\textbf{Agent} &
  \textbf{AT} &
  \textbf{Train AT} &
  \textbf{Completed episodes} &
  \textbf{Steps survived} &
  \textbf{Steps overloaded} &
  \textbf{Agent execution time {[}ms{]}} &
  \textbf{Maximum topology depth} &
  \textbf{Unique actions} &
  \textbf{Unique lines in danger} &
  \textbf{Unique subs changed} \\ \hline
\multirow{16}{*}{PPO}     & \multirow{4}{*}{0.8}  & 0.8  & 0.0\% & 2.022\%          & 1.156\%           & 2.359  & 4   & 21.2 & 14.4 & 6.2 \\ \cline{3-11} 
                          &                       & 0.9  & 0.0\% & 1.952\%          & 1.126\%           & 2.3764 & 4.4 & 29   & 14.4 & 6   \\ \cline{3-11} 
                          &                       & 0.95 & 0.0\% & 1.972\%          & 0.554\%           & 1.8766 & 4.6 & 26.8 & 14   & 6   \\ \cline{3-11} 
                          &                       & 0.99 & 0.0\% & 2.024\%          & 0.549\%           & 2.2802 & 4.6 & 24.4 & 14.6 & 6   \\ \cline{2-11} 
                          & \multirow{4}{*}{0.9}  & 0.8  & 0.0\% & 3.498\%          & 1.094\%           & 2.373  & 3.8 & 17.2 & 13   & 6   \\ \cline{3-11} 
                          &                       & 0.9  & 0.0\% & 3.488\%          & 0.810\%           & 2.4138 & 4.2 & 16.6 & 12   & 6   \\ \cline{3-11} 
                          &                       & 0.95 & 0.0\% & 3.690\%          & \textbf{0.514\%}  & 1.9192 & 4   & 14.6 & 11.6 & 6   \\ \cline{3-11} 
                          &                       & 0.99 & 0.0\% & 3.634\%          & 0.571\%           & 2.4456 & 4.4 & 17   & 11.8 & 5.4 \\ \cline{2-11} 
                          & \multirow{4}{*}{0.95} & 0.8  & 0.0\% & 4.424\%          & 0.940\%           & 2.3154 & 3.4 & 15.6 & 13.4 & 5.4 \\ \cline{3-11} 
                          &                       & 0.9  & 0.0\% & 4.312\%          & 0.863\%           & 2.253  & 3.2 & 12   & 12.2 & 5.6 \\ \cline{3-11}  
                          &    & \cgray  0.95 & \cgray 0.0\% & \cgray 4.690\%          & \cgray 0.638\%           & \cgray 2.1618  & \cgray 4   & \cgray 12.8 & \cgray 12.2 & \cgray 5.4 \\ \cline{3-11} 
                          &                       & 0.99 & 0.0\% & 4.682\%          & 0.712\%           & 2.5526 & 3.2 & 12.8 & 12   & 5.6 \\ \cline{2-11} 
                          & \multirow{4}{*}{0.99} & 0.8  & 0.0\% & 5.488\%          & 0.928\%           & 2.2912 & 3   & 16.4 & 13.8 & 5.8 \\ \cline{3-11} 
                          &                       & 0.9  & 0.0\% & 4.982\%          & 0.824\%           & 2.3132 & 2.6 & 13.2 & 13   & 5.6 \\ \cline{3-11} 
                          &                       & 0.95 & 0.0\% & 5.450\%          & 0.652\%           & 2.7746 & 3.2 & 10.2 & 10.8 & 5.2 \\ \cline{3-11} 
                          &                       & 0.99 & 0.0\% & 4.752\%          & 0.841\%           & 2.227  & 2.8 & 6.6  & 10.2 & 4.4 \\ \hline \hline
\multirow{16}{*}{PPO$^*$} & \multirow{4}{*}{0.8}  & 0.8  & 0.0\% & 1.704\%          & 1.575\%           & 2.3988 & 4.6 & 93.8 & 18.2 & 7   \\ \cline{3-11} 
                          &                       & 0.9  & 0.0\% & 2.150\%          & 1.345\%           & 2.2218 & 4.6 & 97   & 18.4 & 7   \\ \cline{3-11} 
                          &                       & 0.95 & 0.0\% & 2.670\%          & 1.092\%           & 2.364  & 5.2 & 94.2 & 18.2 & 7   \\ \cline{3-11} 
                          &                       & 0.99 & 0.0\% & 2.808\%          & 1.062\%           & 2.2186 & 5   & 82.2 & 18.2 & 7   \\ \cline{2-11} 
                          & \multirow{4}{*}{0.9}  & 0.8  & 0.0\% & 2.872\%          & 1.124\%           & 2.582  & 4.2 & 94   & 16.2 & 7   \\ \cline{3-11} 
                          &                       & 0.9  & 0.0\% & 3.662\%          & 0.907\%           & 2.534  & 4.2 & 95.8 & 16.6 & 6.8 \\ \cline{3-11} 
                          &                       & 0.95 & 0.0\% & 4.714\%          & 0.685\%           & 2.6614 & 5   & 96.6 & 16.4 & 7   \\ \cline{3-11} 
                          &                       & 0.99 & 0.0\% & 4.834\%          & 0.682\%           & 2.301  & 4.2 & 79   & 16   & 7   \\ \cline{2-11} 
                          & \multirow{4}{*}{0.95} & 0.8  & 0.0\% & 3.784\%          & 1.002\%           & 2.3184 & 3.6 & 86.4 & 15.4 & 6.8 \\ \cline{3-11} 
                          &                       & 0.9  & 0.0\% & 5.352\%          & 0.776\%           & 3.1196 & 4.2 & 91.4 & 16.2 & 7   \\ \cline{3-11}
                          &    & \cgray  0.95 & \cgray 0.0\% & \cgray 5.740\%          & \cgray 0.683\%           & \cgray 2.2398 & \cgray 4.6 & \cgray 91   & \cgray 15   & \cgray 7     \\ \cline{3-11} 
                          &                       & 0.99 & 0.0\% & 5.852\%          & 0.692\%           & 2.2794 & 4   & 79.4 & 14.6 & 6.8 \\ \cline{2-11} 
                          & \multirow{4}{*}{0.99} & 0.8  & 0.0\% & 5.318\%          & 0.840\%           & 2.3642 & 3.2 & 89.8 & 15.4 & 6.8 \\ \cline{3-11} 
                          &                       & 0.9  & 0.0\% & 5.934\%          & 0.788\%           & 2.4672 & 3.8 & 92   & 13.4 & 6.8 \\ \cline{3-11} 
                          &                       & 0.95 & 0.0\% & 6.540\%          & 0.684\%           & 2.5556 & 3.8 & 90.8 & 14.6 & 7   \\ \cline{3-11} 
                          &                       & 0.99 & 0.0\% & \textbf{7.042\%} & 0.672\%           & 2.5936 & 4   & 80.6 & 15.6 & 6.8 \\ \hline
\end{tabular}   
}   
\end{adjustbox}
\caption{Validation results of different activation thresholds used for the final agent agents (the column `AT') combined with different activation thresholds used for the training of the agent (column `Train AT'), applied to the environment case 14 sandbox \textit{with} an opponent. The baseline results are highlighted in grey.}
\label{Table:ResAT_Opp}
\end{table}
\FloatBarrier

\subsection{Line switching: Reconnection and disconnection}\label{Sec:Res_LineSwitch}
Line-switching rules, such as reconnecting or disconnecting lines, can improve the agents performance as we avoid lines being disconnect or overloaded for an unnecessary amount of time. However, as expected, the impact depends heavily on the presence of an opponent.

\begin{figure}[!tbh]
\begin{adjustbox}{center}
\resizebox{1.2\textwidth}{!}
{
    \begin{subfigure}[b]{0.5\textwidth}
        \centering
        \includegraphics[scale=1.0]{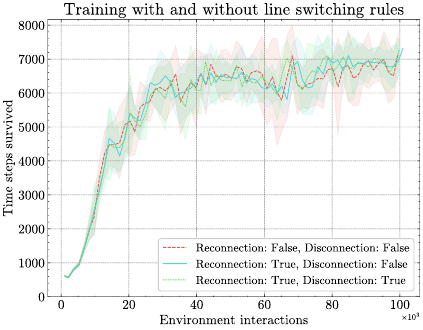}
        \caption{Without an opponent (PPO).}
        \label{Subfig:Line}
    \end{subfigure}
     \hfill
     \begin{subfigure}[b]{0.5\textwidth}
        \centering
        \includegraphics[scale=1.0]{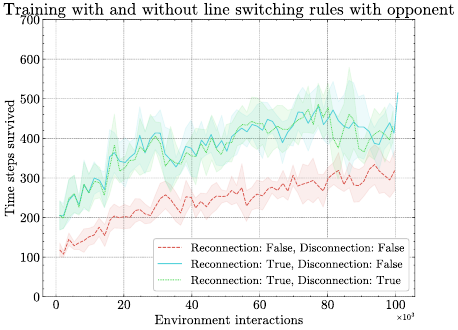}
        \caption{With an opponent (PPO$^*$).}
        \label{Subfig:LineOpp}
    \end{subfigure}
}
\end{adjustbox}
    \caption{Training curves of PPO agents trained with and without line-switching rules. The baseline corresponds to the setting where the line reconnection is applied, but the line disconnection is not. } 
    \label{Fig:Line_TrainCurve}
\end{figure}

In an environment without an opponent (\cref{Subfig:Line,Table:ResLine}), incorporating line-switching rules does not significantly affect the agent’s survival time or overall performance. Tracking the frequency of line reconnections and disconnections during training (where applicable) reveals that these actions are rarely used. At the start of training, they are activated fewer than 0.05 times per episode on average, decreasing further to below 0.005 times per episode once the agent is fully trained. This low activation frequency explains their negligible effect on performance.

\FloatBarrier

In the environment with an opponent (\cref{Fig:Line_TrainCurve,Table:ResLine_Opp}), line reconnections provide a clear performance benefit. However, including the line disconnection rule again has no very significant impact on the agents' performance. The frequency of line reconnections increases significantly in the opponent environment, with agents using them approximately 3–4 times per episode on average. The line disconnection is only used around 0.01 times on average per episode for a trained agent, which is slightly more than in an environment without an opponent, but this still does not reflect in the agent's performance.

\begin{table}[h]
\begin{adjustbox}{center}
\resizebox{1.2\textwidth}{!}
{
\scriptsize
\tabcolsep=3pt
\begin{tabular}{|l|p{0.06\linewidth}|p{0.06\linewidth}|p{0.07\linewidth}|p{0.07\linewidth}|p{0.1\linewidth}|p{0.08\linewidth}|p{0.08\linewidth}|p{0.1\linewidth}|p{0.1\linewidth}|p{0.08\linewidth}|p{0.08\linewidth}|p{0.08\linewidth}|}
\hline
\textbf{Agent} &
  \textbf{Recon-nect} &
  \textbf{Discon-nect} &
  \textbf{Train Reconnect} &
  \textbf{Train Disconnect} &
  \textbf{Completed episodes} &
  \textbf{Steps survived} &
  \textbf{Steps overloaded} &
  \textbf{Agent execution time {[}ms{]}} &
  \textbf{Maximum topology depth} &
  \textbf{Unique actions} &
  \textbf{Unique lines in danger} &
  \textbf{Unique subs changed} \\ \hline
\multirow{9}{*}{PPO}     & \multirow{3}{*}{}      & \multirow{3}{*}{}      &                        &           & 79.7\%            & 89.49\%           & 0.009\%    & 2.057  & 4.7 & 28.33 & 12.2   & 5.8            \\ \cline{4-13} 
                         &                        &                        & \multirow{2}{*}{\cmark}&           & 80.8\%            & \textbf{90.42\%}  & 0.009\%    & 2.051  & 4.4 & 25.8  & 11.8   & 6.4                 \\ \cline{5-13} 
                         &                        &                        &                        & \cmark    & 64.6\%            & 82.21\%           & 0.013\%    & 2.212  & 5.0 & 58.6  & 14     & 6.8                 \\ \cline{2-13} 
                         & \multirow{6}{*}{\cmark}& \multirow{3}{*}{}      &                        &           & 80.3\%            & 90.12\%           & 0.009\%    & 2.055  & 4.7 & 27.33 & 11.5   & 6                   \\ \cline{4-13} 
                         &                        &                        & \multirow{2}{*}{\cmark}& \cgray  & \cgray 78.6\%   & \cgray 87.31\%           & \cgray \textbf{0.008\%}    & \cgray 1.967  & \cgray 5.0 & \cgray 27.8  & \cgray 11.8   & \cgray 6                   \\ \cline{5-13} 
                         &                        &                        &                        & \cmark    &\textbf{ 81.4\%}   & 90.41\%           & 0.009\%    & 2.089  & 4.6 & 27.0  & 11.4   & 6.6                 \\ \cline{3-13} 
                         &                        & \multirow{3}{*}{\cmark}&                        &           & 78.7\%            & 89.08\%           & 0.009\%    & 2.411  & 4.5 & 30.7  & 11.3   & 5.8            \\ \cline{4-13} 
                         &                        &                        & \multirow{2}{*}{\cmark}&           & 76.2\%            & 86.46\%           & 0.008\%    & 2.150  & 5.0 & 29.4  & 10.6   & 6                   \\ \cline{5-13} 
                         &                        &                        &                        & \cmark    & \textbf{81.4\%}   & 89.24\%           & 0.011\%    & 2.587  & 4.4 & 26.8  & 11.6   & 6.4                 \\ \hline\hline
\multirow{9}{*}{PPO$^*$} & \multirow{3}{*}{}      & \multirow{3}{*}{}      &                        &           & 17.8\%            & 42.03\%           & 0.093\%    & 1.998  & 4.6 & 88.8  & 17.6   & 7                   \\ \cline{4-13} 
                         &                        &                        & \multirow{2}{*}{\cmark}&           & 36.8\%            & 58.19\%           & 0.049\%    & 2.150  & 4.8 & 80.4  & 17.8   & 7                   \\ \cline{5-13} 
                         &                        &                        &                        & \cmark    & 41.0\%            & 63.21\%           & 0.039\%    & 2.138  & 4.6 & 78.2  & 16.0   & 7                   \\ \cline{2-13} 
                         & \multirow{6}{*}{\cmark}& \multirow{3}{*}{}      &                        &           & 20.4\%            & 44.37\%           & 0.093\%    & 1.937  & 4.6 & 89.2  & 17.2   & 7                   \\ \cline{4-13} 
                         &                        &                        & \multirow{2}{*}{\cmark}& \cgray  & \cgray 35.4\%   & \cgray 57.89\%           & \cgray 0.041\%    & \cgray 1.966  & \cgray 4.6 & \cgray 80.6 & \cgray 17   & \cgray 6.8                \\ \cline{5-13} 
                         &                        &                        &                        & \cmark    & 41.8\%            & 62.00\%           & 0.046\%    & 2.060  & 4.6 & 76.4  & 16.6   & 7                   \\ \cline{3-13} 
                         &                        & \multirow{3}{*}{\cmark}&                        &           & 17.8\%            & 41.41\%           & 0.090\%    & 2.910  & 4.6 & 89.6  & 17.8   & 7                   \\ \cline{4-13} 
                         &                        &                        & \multirow{2}{*}{\cmark}&           & 35.4\%            & 57.72\%           & 0.046\%    & 2.349  & 4.4 & 75.8  & 16.8   & 7                   \\ \cline{5-13} 
                         &                        &                        &                        & \cmark    & 44.2\%            & 63.93\%           & 0.041\%    & 2.531  & 4.8 & 78.4  & 16.8   & 6.8                 \\ \hline
\end{tabular}
}   
\end{adjustbox}
\caption{Validation results of final agents with and without reconnection and disconnection rule (columns `Reconnect' and `Disconnect'), and trained with and without reconnection and disconnection rule (columns `Train Reconnect' and `Train Disconnect') applied to the environment case 14 sandbox without an opponent. The baseline results are highlighted in grey.}
\label{Table:ResLine}
\end{table}
\FloatBarrier

\begin{table}[h]
\begin{adjustbox}{center}
\resizebox{1.2\textwidth}{!}
{
\scriptsize
\tabcolsep=3pt
\begin{tabular}{|l|p{0.06\linewidth}|p{0.06\linewidth}|p{0.07\linewidth}|p{0.07\linewidth}|p{0.1\linewidth}|p{0.08\linewidth}|p{0.08\linewidth}|p{0.1\linewidth}|p{0.1\linewidth}|p{0.08\linewidth}|p{0.08\linewidth}|p{0.08\linewidth}|}
\hline
\textbf{Agent} &
  \textbf{Recon-nect} &
  \textbf{Discon-nect} &
  \textbf{Train Reconnect} &
  \textbf{Train Disconnect} &
  \textbf{Completed episodes} &
  \textbf{Steps survived} &
  \textbf{Steps overloaded} &
  \textbf{Agent execution time {[}ms{]}} &
  \textbf{Maximum topology depth} &
  \textbf{Unique actions} &
  \textbf{Unique lines in danger} &
  \textbf{Unique subs changed} \\ \hline
\multirow{9}{*}{PPO}     & \multirow{3}{*}{     } & \multirow{3}{*}{     } &                          &                 & 0.0\%     & 2.442\%           & 1.671\%           & 2.170  & 4.0  & 17.8    & 12.6   & 6.2  \\ \cline{4-13} 
                         &                        &                        & \multirow{2}{*}{\cmark}  &                 & 0.0\%     & 2.396\%           & 1.386\%           & 1.985  & 3.6  & 13.8    & 12.2   & 6.0  \\ \cline{5-13} 
                         &                        &                        &                          & \cmark          & 0.0\%     & 2.354\%           & 1.489\%           & 2.507  & 3.6  & 17.0    & 13.4   & 5.8  \\ \cline{2-13} 
                         & \multirow{6}{*}{\cmark}& \multirow{3}{*}{     } &                          &                 & 0.0\%     & 4.814\%           & 0.658\%           & 2.707  & 3.6  & 12.0    & 10.8   & 4.6  \\ \cline{4-13} 
                         &                        &                        & \multirow{2}{*}{\cmark}  & \cgray & \cgray 0.0\%     & \cgray 4.690\%          & \cgray 0.638\%           & \cgray 2.162  & \cgray 4.0  & \cgray 12.8    & \cgray 12.2   & \cgray 5.4  \\ \cline{5-13} 
                         &                        &                        &                          & \cmark          & 0.0\%     & 4.452\%           & 0.777\%           & 2.485  & 3.2  & 11.0    & 12.0   & 5.2  \\ \cline{3-13} 
                         &                        & \multirow{3}{*}{\cmark}&                          &                 & 0.0\%     & 4.832\%           & 0.654\%           & 3.535  & 3.8  & 12.6    & 11.8   & 5.4  \\ \cline{4-13} 
                         &                        &                        & \multirow{2}{*}{\cmark}  &                 & 0.0\%     & 4.604\%           & \textbf{0.626\%}  & 3.315  & 3.4  & 11.8    & 11.8   & 5.2  \\ \cline{5-13} 
                         &                        &                        &                          & \cmark          & 0.0\%     & 4.816\%           & 0.631\%           & 3.536  & 3.6  & 12.6    & 13.4   & 6.0  \\ \hline\hline
\multirow{9}{*}{PPO$^*$} & \multirow{3}{*}{     } & \multirow{3}{*}{     } &                          &                 & 0.0\%     & 3.754\%           & 1.567\%           & 2.071  & 4.0  & 93.6    & 16.4   & 7.0  \\ \cline{4-13} 
                         &                        &                        & \multirow{2}{*}{\cmark}  &                 & 0.0\%     & 2.542\%           & 1.995\%           & 1.951  & 4.0  & 86.6    & 13.8   & 7.0  \\ \cline{5-13} 
                         &                        &                        &                          & \cmark          & 0.0\%     & 2.622\%           & 2.149\%           & 2.331  & 3.6  & 88.4    & 14.4   & 7.0  \\ \cline{2-13} 
                         & \multirow{6}{*}{\cmark}& \multirow{3}{*}{     } &                          &                 & 0.0\%     & 5.478\%           & 0.867\%           & 2.244  & 4.0  & 95.2    & 15.8   & 7.0  \\ \cline{4-13} 
                         &                        &                        & \multirow{2}{*}{\cmark}  & \cgray & \cgray 0.0\%    & \cgray 5.740\%           & \cgray 0.683\%           & \cgray 2.2398  & \cgray 4.6 & \cgray 91   & \cgray 15   & \cgray 7    \\ \cline{5-13} 
                         &                        &                        &                          & \cmark          & 0.0\%     & 5.644\%           & 0.640\%           & 2.372  & 3.6  & 90.6    & 15.4   & 6.6  \\ \cline{3-13} 
                         &                        & \multirow{3}{*}{\cmark}&                          &                 & 0.0\%     & 5.658\%           & 0.808\%           & 3.265  & 3.8  & 94.6    & 16.4   & 7.0  \\ \cline{4-13} 
                         &                        &                        & \multirow{2}{*}{\cmark}  &                 & 0.0\%     & 5.834\%           & 0.631\%           & 3.329  & 3.6  & 89.0    & 15.4   & 7.0  \\ \cline{5-13} 
                         &                        &                        &                          & \cmark          & 0.0\%     & \textbf{5.870\%}  & 0.671\%           & 3.732  & 4.2  & 96.0    & 16.8   & 7.0  \\ \hline
        
\end{tabular}           
}   
\end{adjustbox}
\caption{Validation results of final agents tested with and without reconnection and disconnection rule (columns `Reconnect' and `Disconnect'), and trained with and without reconnection and disconnection rule (columns `Train Reconnect' and `Train Disconnect') applied to the environment case 14 sandbox \textit{with} an opponent. The baseline results are highlighted in grey.}
\label{Table:ResLine_Opp}
\end{table}
\FloatBarrier

\subsection{Revert to reference topology}\label{Sec:ResRT}
\begin{figure}[!tbh]
\begin{adjustbox}{center}
\resizebox{1.2\textwidth}{!}
{
    \begin{subfigure}[b]{0.45\textwidth}
        \centering
        \includegraphics[scale=1.0]{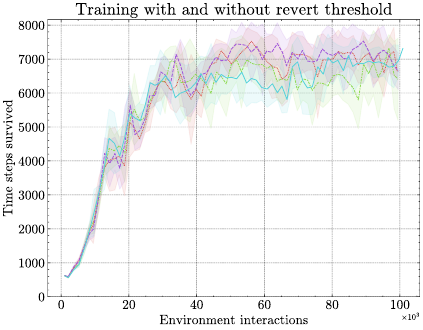}
        \caption{Without an opponent (PPO).}
    \end{subfigure}
     \hfill
     \begin{subfigure}[b]{0.55\textwidth}
        \centering
        \includegraphics[scale=1.0]{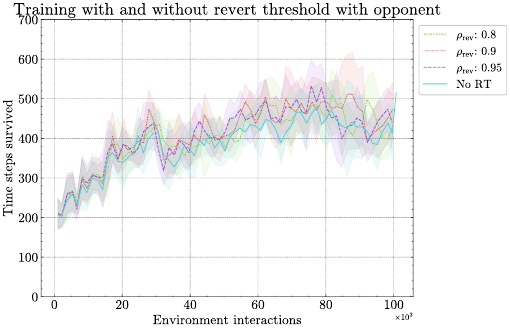}
        \caption{With an opponent (PPO$^*$).}
    \end{subfigure}
}
\end{adjustbox}
    \caption{Training curves of PPO agents trained with and without the revert to reference topology rule. The baseline is without reverting to the reference topology.} 
    \label{Fig:RT_TrainCurve}
\end{figure}

Training with or without the revert-to-reference-topology rule has a minor effect on the agent's survival time during training. For agents trained without an opponent, the training curve with the highest revert thresholds (RTs) is slightly higher near the end, as shown in \cref{Fig:RT_TrainCurve} where the training curves in exhibit similar trends across all configurations. 

By tracking the frequency of the revert actions, we see that in the environment without an opponent, trained agents revert substations to the reference topology approximately twice per episode for $\RT=0.8$, and around 3-4 times per episode for $\RT=0.9$ or $\RT=0.95$. With a survival time of around 7.000 time steps, this is not very frequent and, therefore, explains the minimal effect on the training curves.
In the environment with an opponent, substations are reverted to the reference topology between 0.2 and 0.4 times per episode, depending on the RT value.

\FloatBarrier

During validation in the environment without an opponent, the best-performing configuration uses an $\RT= 0.9$ during training and an $\RT= 0.95$ for the final agent, see \cref{Table:ResRT}. 

\cref{Table:ResRT_Opp}  shows that the PPO$^*$ agent trained with $\RT = 0.8$ and evaluated with $\RT = 0.9$ achieves the highest performance among all the configurations tested on the environment with an opponent.

In general, training with an RT but omitting it for the final agent decreases performance. This is expected as the agent is trained with a rule that it does not ultimately use.
Furthermore, the results suggest that performance improves when the RT during training is slightly lower (making the rule less strict), leading to less frequent activation compared to the RT used in the final agent. 

\begin{table}[h]
\begin{adjustbox}{center}
\resizebox{1.2\textwidth}{!}
{
\scriptsize
\tabcolsep=3pt
\begin{tabular}{|l|p{0.05\linewidth}|p{0.05\linewidth}|p{0.1\linewidth}|p{0.08\linewidth}|p{0.08\linewidth}|p{0.1\linewidth}|p{0.1\linewidth}|p{0.08\linewidth}|p{0.08\linewidth}|p{0.08\linewidth}|}
\hline
\textbf{Agent} &
  \textbf{RT} &
  \textbf{Train RT} &
  \textbf{Completed episodes} &
  \textbf{Steps survived} &
  \textbf{Steps overloaded} &
  \textbf{Agent execution time {[}ms{]}} &
  \textbf{Maximum topology depth} &
  \textbf{Unique actions} &
  \textbf{Unique lines in danger} &
  \textbf{Unique subs changed} \\ \hline
\multirow{16}{*}{PPO}     & \multirow{4}{*}{0.0}  & \cgray  0.0  & \cgray 78.6\%          & \cgray 87.31\%          & \cgray 0.008\%          & \cgray 1.967 & \cgray 5.0 & \cgray 27.8 & \cgray 11.8 & \cgray 6.0 \\ \cline{3-11} 
                          &                       & 0.8                     & 54.4\%          & 72.90\%          & 0.040\%          & 2.286 & 4.6 & 23.4 & 11.4 & 5.8 \\ \cline{3-11} 
                          &                       & 0.9                     & 34.0\%          & 58.87\%          & 0.069\%          & 2.221 & 4.8 & 18.2 & 8.4  & 6.2 \\ \cline{3-11} 
                          &                       & 0.95                    & 38.2\%          & 62.72\%          & 0.064\%          & 2.195 & 5.2 & 17.2 & 8.6  & 5.6 \\ \cline{2-11} 
                          & \multirow{4}{*}{0.8}  & 0.0                     & 77.6\%          & 87.64\%          & 0.008\%          & 2.097 & 4.6 & 28.8 & 12.2 & 6.2 \\ \cline{3-11} 
                          &                       & 0.8                     & 77.2\%          & 87.50\%          & 0.011\%          & 2.205 & 4.6 & 24.4 & 11.4 & 6.0 \\ \cline{3-11} 
                          &                       & 0.9                     & 81.0\%          & 89.88\%          & 0.020\%          & 2.174 & 4.6 & 24.6 & 12.0 & 6.6 \\ \cline{3-11} 
                          &                       & 0.95                    & 82.2\%          & 91.02\%          & 0.017\%          & 2.164 & 5.0 & 25.0 & 9.6  & 6.4 \\ \cline{2-11} 
                          & \multirow{4}{*}{0.9}  & 0.0                     & 78.4\%          & 87.04\%          & \textbf{0.007\%} & 2.143 & 4.6 & 28.2 & 11.4 & 6.2 \\ \cline{3-11} 
                          &                       & 0.8                     & 80.6\%          & 89.49\%          & 0.011\%          & 2.510 & 4.6 & 25.8 & 11.6 & 5.6 \\ \cline{3-11} 
                          &                       & 0.9                     & 87.8\%          & 93.45\%          & 0.016\%          & 2.376 & 5.0 & 24.4 & 10.8 & 6.4 \\ \cline{3-11} 
                          &                       & 0.95                    & 88.2\%          & 93.90\%          & 0.013\%          & 2.433 & 5.0 & 25.8 & 10.2 & 6.4 \\ \cline{2-11} 
                          & \multirow{4}{*}{0.95} & 0                       & 78.0\%          & 87.47\%          & 0.009\%          & 2.190 & 4.8 & 31.0 & 10.6 & 6.0 \\ \cline{3-11} 
                          &                       & 0.8                     & 79.2\%          & 88.42\%          & 0.011\%          & 2.415 & 4.6 & 25.8 & 12.2 & 5.8 \\ \cline{3-11} 
                          &                       & 0.9                     & \textbf{88.8\%} & \textbf{94.24\%} & 0.016\%          & 2.297 & 5.0 & 26.8 & 10.6 & 6.4 \\ \cline{3-11} 
                          &                       & 0.95                    & 87.6\%          & 92.97\%          & 0.014\%          & 2.410 & 5.0 & 25.6 & 10.6 & 6.4 \\ \hline \hline
\multirow{16}{*}{PPO$^*$} & \multirow{4}{*}{0.0}  & \cgray  0.0  & \cgray 35.4\%          & \cgray 57.89\%          & \cgray 0.041\%          & \cgray 1.9658 & \cgray 4.6 & \cgray 80.6 & \cgray 17   & \cgray 6.8 \\ \cline{3-11} 
                          &                       & 0.8                     & 33.4\%          & 55.64\%          & 0.045\%          & 2.1132 & 4.6 & 80.4 & 17.2 & 6.8 \\ \cline{3-11} 
                          &                       & 0.9                     & 31.6\%          & 51.95\%          & 0.060\%          & 2.2036 & 4.8 & 80.4 & 16.6 & 6.8 \\ \cline{3-11} 
                          &                       & 0.95                    & 59.6\%          & 74.64\%          & 0.019\%          & 2.0416 & 5   & 56.6 & 14.8 & 6.4 \\ \cline{2-11} 
                          & \multirow{4}{*}{0.8}  & 0.0                     & 32.0\%          & 55.43\%          & 0.050\%          & 2.132  & 4.6 & 86   & 17.2 & 7   \\ \cline{3-11} 
                          &                       & 0.8                     & 36.0\%          & 57.72\%          & 0.043\%          & 2.0768 & 4.6 & 79.8 & 16.4 & 6.8 \\ \cline{3-11} 
                          &                       & 0.9                     & 29.4\%          & 54.00\%          & 0.051\%          & 2.1914 & 4.4 & 80.4 & 16.2 & 6.8 \\ \cline{3-11} 
                          &                       & 0.95                    & 54.6\%          & 73.19\%          & 0.023\%          & 2.6348 & 4.8 & 60.6 & 14.8 & 6.6 \\ \cline{2-11} 
                          & \multirow{4}{*}{0.9}  & 0.0                     & 31.6\%          & 54.92\%          & 0.055\%          & 1.9902 & 4.8 & 82.8 & 17   & 7   \\ \cline{3-11} 
                          &                       & 0.8                     & 35.8\%          & 57.04\%          & 0.041\%          & 2.047  & 4.2 & 80.8 & 16.6 & 6.6 \\ \cline{3-11} 
                          &                       & 0.9                     & 29.2\%          & 55.33\%          & 0.049\%          & 2.1774 & 4.2 & 83.6 & 17.4 & 6.6 \\ \cline{3-11} 
                          &                       & 0.95                    & 57.2\%          & 76.04\%          & 0.024\%          & 2.1238 & 4.6 & 59.8 & 16.2 & 6.6 \\ \cline{2-11} 
                          & \multirow{4}{*}{0.95} & 0.0                     & 31.0\%          & 56.75\%          & 0.050\%          & 2.6844 & 4.4 & 88.4 & 16.6 & 6.8 \\ \cline{3-11} 
                          &                       & 0.8                     & 36.0\%          & 57.37\%          & 0.045\%          & 2.1208 & 4.2 & 80.6 & 16   & 6.8 \\ \cline{3-11} 
                          &                       & 0.9                     & 31.2\%          & 56.11\%          & 0.049\%          & 2.2034 & 4.2 & 80.2 & 16.8 & 6.8 \\ \cline{3-11} 
                          &                       & 0.95                    & 56.8\%          & 77.12\%          & 0.024\%          & 2.1824 & 5   & 63.4 & 15.6 & 6.6 \\ \hline
\end{tabular}
}   
\end{adjustbox}
\caption{Validation results of different revert thresholds used for the final agent agents (the column `RT') combined with different revert thresholds used for the training of the agent (column `Train RT'), applied to the environment case 14 sandbox without an opponent. RT=0.0 means that no reverting to topology is applied. The baseline results are highlighted in gray.}
\label{Table:ResRT}
\end{table}
\FloatBarrier

\begin{table}[h]
\begin{adjustbox}{center}
\resizebox{1.2\textwidth}{!}
{
\scriptsize
\tabcolsep=3pt
\begin{tabular}{|l|p{0.05\linewidth}|p{0.05\linewidth}|p{0.1\linewidth}|p{0.08\linewidth}|p{0.08\linewidth}|p{0.1\linewidth}|p{0.1\linewidth}|p{0.08\linewidth}|p{0.08\linewidth}|p{0.08\linewidth}|}
\hline
\textbf{Agent} &
  \textbf{RT} &
  \textbf{Train RT} &
  \textbf{Completed episodes} &
  \textbf{Steps survived} &
  \textbf{Steps overloaded} &
  \textbf{Agent execution time {[}ms{]}} &
  \textbf{Maximum topology depth} &
  \textbf{Unique actions} &
  \textbf{Unique lines in danger} &
  \textbf{Unique subs changed} \\ \hline
\multirow{16}{*}{PPO}     & \multirow{4}{*}{0.0}  & \cgray  0.0  & \cgray 0.0\% & \cgray 4.690\%          & \cgray 0.638\%            & \cgray 2.162 & \cgray 4.0 & \cgray 12.8 & \cgray 12.2 & \cgray 5.4 \\ \cline{3-11} 
                          &                       & 0.8                     & 0.0\% & 4.854\%          & 0.610\%            & 2.733 & 3.8 & 9.4  & 11.4 & 4.6 \\ \cline{3-11} 
                          &                       & 0.9                     & 0.0\% & 4.966\%          & 0.703\%            & 2.297 & 4.4 & 10.2 & 11.0 & 5.4 \\ \cline{3-11} 
                          &                       & 0.95                    & 0.0\% & 4.594\%          & 0.755\%            & 2.413 & 4.0 & 10.8 & 12.0 & 5.0 \\ \cline{2-11} 
                          & \multirow{4}{*}{0.8}  & 0.0                     & 0.0\% & 4.650\%          & 0.712\%            & 2.212 & 3.6 & 12.6 & 11.6 & 5.6 \\ \cline{3-11} 
                          &                       & 0.8                     & 0.0\% & 4.590\%          & 0.662\%            & 2.425 & 3.2 & 8.8  & 11.0 & 4.6 \\ \cline{3-11} 
                          &                       & 0.9                     & 0.0\% & 4.670\%          & 0.650\%            & 2.420 & 4.0 & 9.6  & 11.4 & 5.4 \\ \cline{3-11} 
                          &                       & 0.95                    & 0.0\% & 4.676\%          & 0.658\%            & 2.367 & 4.0 & 12.4 & 11.6 & 5.4 \\ \cline{2-11} 
                          & \multirow{4}{*}{0.9}  & 0.0                     & 0.0\% & 4.650\%          & \textbf{0.581\%}   & 2.064 & 3.4 & 11.2 & 12.4 & 5.6 \\ \cline{3-11} 
                          &                       & 0.8                     & 0.0\% & 4.726\%          & 0.668\%            & 2.628 & 3.6 & 9.0  & 11.6 & 5.0 \\ \cline{3-11} 
                          &                       & 0.9                     & 0.0\% & 4.722\%          & 0.652\%            & 2.426 & 3.6 & 8.2  & 11.4 & 4.8 \\ \cline{3-11} 
                          &                       & 0.95                    & 0.0\% & 4.694\%          & 0.688\%            & 2.489 & 4.2 & 11.8 & 11.2 & 5.4 \\ \cline{2-11} 
                          & \multirow{4}{*}{0.95} & 0.0                     & 0.0\% & 4.620\%          & 0.658\%            & 2.082 & 3.6 & 12.8 & 11.4 & 5.4 \\ \cline{3-11} 
                          &                       & 0.8                     & 0.0\% & 4.918\%          & 0.651\%            & 2.523 & 3.0 & 9.2  & 11.8 & 4.6 \\ \cline{3-11} 
                          &                       & 0.9                     & 0.0\% & 4.588\%          & 0.673\%            & 2.411 & 3.8 & 9.2  & 11.2 & 4.4 \\ \cline{3-11} 
                          &                       & 0.95                    & 0.0\% & 4.810\%          & 0.632\%            & 2.349 & 3.8 & 9.0  & 12.2 & 5.4 \\ \hline \hline
\multirow{16}{*}{PPO$^*$} & \multirow{4}{*}{0.0}  & \cgray  0.0  & \cgray 0.0\% & \cgray 5.740\%          & \cgray 0.683\%            & \cgray 2.2398  & \cgray 4.6 & \cgray 91   & \cgray 15   & \cgray 7    \\ \cline{3-11} 
                          &                       & 0.8                     & 0.0\% & 2.644\%          & 2.285\%            & 2.1626 & 4.2 & 86   & 15.6 & 7   \\ \cline{3-11} 
                          &                       & 0.9                     & 0.0\% & 2.692\%          & 2.019\%            & 2.262  & 4.4 & 87.2 & 15.6 & 7   \\ \cline{3-11} 
                          &                       & 0.95                    & 0.0\% & 2.564\%          & 2.141\%            & 2.379  & 4   & 88   & 15.8 & 6.8 \\ \cline{2-11} 
                          & \multirow{4}{*}{0.8}  & 0.0                     & 0.0\% & 5.190\%          & 0.813\%            & 1.9972 & 3.6 & 89.2 & 16.2 & 6.8 \\ \cline{3-11} 
                          &                       & 0.8                     & 0.0\% & 5.620\%          & 0.692\%            & 2.8124 & 3.6 & 89.8 & 15.6 & 7   \\ \cline{3-11} 
                          &                       & 0.9                     & 0.0\% & 6.066\%          & 0.633\%            & 2.497  & 3.4 & 93   & 15.2 & 7   \\ \cline{3-11} 
                          &                       & 0.95                    & 0.0\% & 6.180\%          & 0.619\%            & 2.3248 & 4   & 89.6 & 15.6 & 7   \\ \cline{2-11} 
                          & \multirow{4}{*}{0.9}  & 0.0                     & 0.0\% & 5.586\%          & 0.688\%            & 1.9358 & 3.6 & 88.6 & 14.4 & 7   \\ \cline{3-11} 
                          &                       & 0.8                     & 0.0\% & \textbf{6.460\%} & 0.596\%            & 2.8822 & 3.6 & 91.6 & 15.4 & 7   \\ \cline{3-11} 
                          &                       & 0.9                     & 0.0\% & 6.032\%          & 0.689\%            & 2.3828 & 3.2 & 94.6 & 15.2 & 7   \\ \cline{3-11} 
                          &                       & 0.95                    & 0.0\% & 6.088\%          & 0.639\%            & 2.467  & 3.6 & 89.8 & 14.2 & 7   \\ \cline{2-11} 
                          & \multirow{4}{*}{0.95} & 0.0                     & 0.0\% & 5.716\%          & 0.652\%            & 2.073  & 3.8 & 88.6 & 14.6 & 6.8 \\ \cline{3-11} 
                          &                       & 0.8                     & 0.0\% & 6.146\%          & 0.639\%            & 2.3474 & 3.2 & 87.4 & 14.2 & 6.8 \\ \cline{3-11} 
                          &                       & 0.9                     & 0.0\% & 5.858\%          & 0.675\%            & 2.3884 & 3.6 & 90.4 & 14.8 & 7   \\ \cline{3-11} 
                          &                       & 0.95                    & 0.0\% & 6.256\%          & 0.629\%            & 2.4976 & 3   & 91   & 14.2 & 7   \\ \hline
\end{tabular}
}   
\end{adjustbox}
\caption{Validation results of different revert thresholds used for the final agent agents (the column `RT') combined with different revert thresholds used for the training of the agent (column `Train RT'), applied to the environment case 14 sandbox \textit{with} an opponent. RT=0.0 means that no reverting to topology is applied. The baseline results are highlighted in grey.}
\label{Table:ResRT_Opp}
\end{table}
\FloatBarrier

\subsection{Rainbow agent}\label{Sec:ResRainbow}
This section showcases the performance of RL agents that combine the best modeling choices in view of the results described in previous sections. Inspired by \cite{hessel2018rainbow}, we refer to such an agent as `\textit{Rainbow agent}'.

\subsection{Agent configuration}
To configure the Rainbow agent, we compare on the validation results in \cref{Table:ResActSp,Table:ResActSpaceOpp,Table:ResObsSpace,Table:ResObsSpace_OPP,Table:ResRw,Table:ResRW_OPP,Table:ResCurr,Table:ResCurr_Opp,Table:ResAT,Table:ResAT_Opp,Table:ResLine,Table:ResLine_Opp,Table:ResRT,Table:ResRT_Opp}. For the agents trained and tested in an environment without an opponent (PPO), we compared the percentages of completed episodes and steps that survived to select the most promising configuration. The setup selected for the PPO Rainbow agent is reported in \cref{Table:SetupRainbow}. 

\begin{table}[h]
\centering
% \scriptsize
\tabcolsep=3pt
\begin{tabular}{|l|p{0.16\linewidth}|p{0.16\linewidth}|p{0.16\linewidth}|p{0.13\linewidth}|}
\hline
                              & {\textbf{Train}}         & \textbf{Validate}       & \textbf{Completed episodes} & \textbf{Steps survived} \\ \hline
\textbf{Action space}         & \multicolumn{2}{c|}{$\A_{n-0}$ (Baseline)}                                   & 78.6\%                      & 87.31\%                 \\ \hline
\textbf{Observation space}    & \multicolumn{2}{c|}{Line loads}                                       & 92.8\%                      & 96.01\%                 \\ \hline
\textbf{Reward function}      & \multicolumn{2}{c|}{Base + Bonus/Penalty}                             & 81.0\%                      & 89.09\%                 \\ \hline
\textbf{Currriculum}          & \multicolumn{2}{c|}{Curriculum training (yes)}                              & 81.6\%                      & 90.07\%                 \\ \hline
\textbf{Activation threshold ($\AT$)} & {0.99}                   & 0.99                    & 92.6\%                      & 97.07\%                 \\ \hline
\textbf{Line switching}       & {Reconnect + Disconnect} & Reconnect + Disconnect\tablefootnote{Although validation results without the disconnection rule slightly outperformed those with it (\cref{Sec:Res_LineSwitch}), we chose to use the rule during both training and validation to maintain consistency, as the performance difference was negligible.} & 81.4\%                      & 90.41\%                 \\ \hline
\textbf{Revert threshold ($\RT$)}      & {0.9}                    & 0.95                    & 88.8\%                      & 94.24\%                 \\ \hline
\end{tabular}
\caption{Setup for the PPO Rainbow agent trained and evaluated on an environment without an opponent. The column `Train' shows the setup used during training and the column `Validate' shows the setup used when running the validation of the final trained agents. The columns `Completed episodes' and `Steps survived' show the scores obtained in the previous experiments.}
\label{Table:SetupRainbow}
\end{table}
\FloatBarrier

For agents trained and tested on an environment \textit{with} an opponent (PPO$^*$), we compare the values in the columns “Steps survived” and “Steps overloaded”, since the percentage of completed episodes is 0.0\% for all agents in an environment with an opponent. The setup selected for the PPO$^*$ Rainbow agent is provided in \cref{Table:SetupRainbow_Opp}. 

\begin{table}[h]
\centering
% \scriptsize
\tabcolsep=3pt
\begin{tabular}{|l|p{0.16\linewidth}|p{0.16\linewidth}|p{0.13\linewidth}|p{0.16\linewidth}|}
\hline
\textbf{}               & {\textbf{Train}}         & \textbf{Validate}      & \textbf{Steps survived} & \textbf{Steps overloaded} \\ \hline
\textbf{Action space}         & \multicolumn{2}{c|}{$\A_{d3qn}$}        & 6.904\% & 1.127\% \\ \hline
\textbf{Observation space}    & \multicolumn{2}{c|}{Baseline}    & 5.740\% & 0.683\% \\ \hline
\textbf{Reward function}      & \multicolumn{2}{c|}{AlphaZero}   & 5.970\% & 0.638\% \\ \hline
\textbf{Currriculum}          & \multicolumn{2}{c|}{Baseline (no)}    & 5.740\% & 0.683\% \\ \hline
\textbf{Activation threshold ($\AT$)} & {0.99} & 0.99 & 7.042\% & 0.672\% \\ \hline
\textbf{Line switching} & {Reconnect + Disconnect} & Reconnect + Disconnect & 5.870\%                 & 0.671\%                   \\ \hline
\textbf{Revert threshold ($\RT$)}      & {0.8}  & 0.9  & 6.460\% & 0.596\% \\ \hline
\end{tabular}
\caption{Setup for the PPO$^*$ Rainbow agent trained and evaluated on an environment \textit{with} an opponent. The column "Train" shows the setup used during training and the column "Validate" shows the setup used when running the validation of the final trained agents. The columns "Completed episodes" and "Steps survived" show the scores obtained in the previous experiments.}
\label{Table:SetupRainbow_Opp}
\end{table}
\FloatBarrier

For each of the selected configurations, we test its contribution to the Rainbow agents by reverting it to its default configuration from the baseline while maintaining all other configurations for the Rainbow agent, referred to as \textit{ablation studies} in \cite{hessel2018rainbow}.

\subsection{Evaluating the Rainbow Agent}
This subsection shows the results of the PPO Rainbow agent trained and validated in an environment without opponent and the PPO$^*$ agent trained and validated in an environment with an opponent.

\begin{figure}[!tbh]
    \centering
    \includegraphics[width=\linewidth]{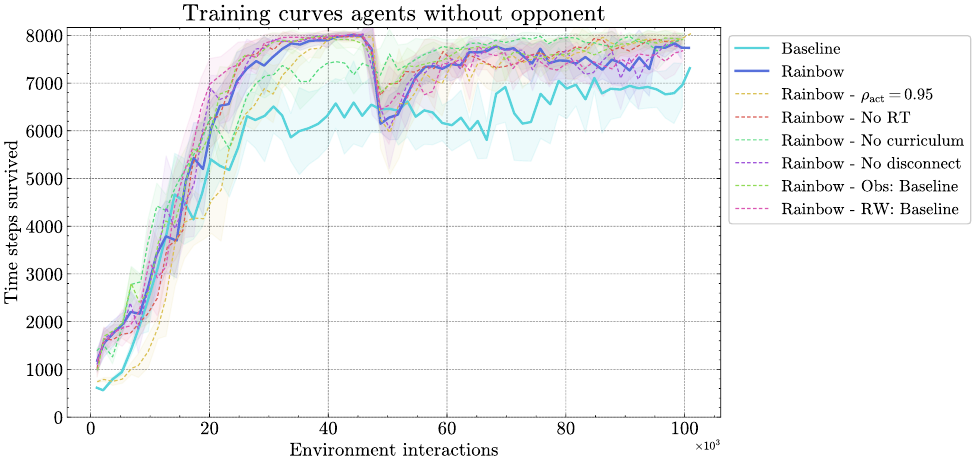}
    \caption{Training curves of the PPO Rainbow agent (trained on an environment without an opponent) compared to the Baseline and six different ablations (dashed, thinner lines).} 
    \label{Fig:Rainbow_TrainCurve}
\end{figure}

\begin{figure}[!tbh]
    \centering
    \includegraphics[width=\linewidth]{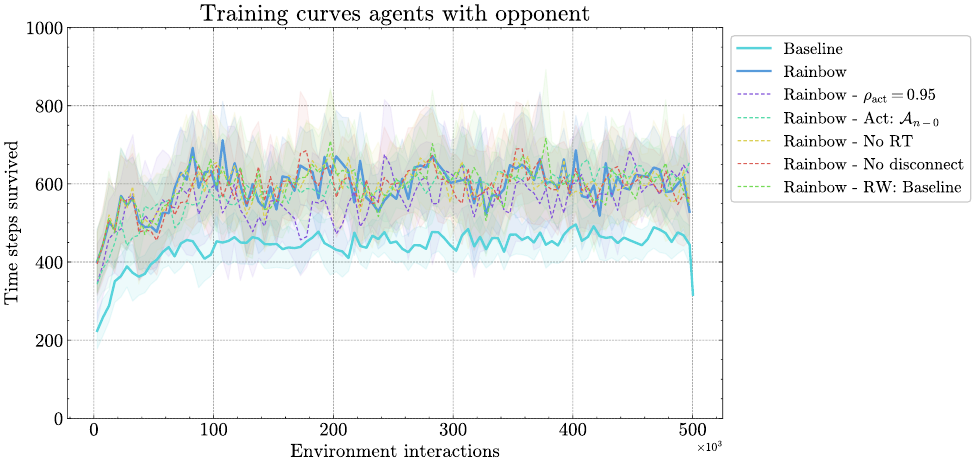}
    \caption{Training curves of the PPO$^*$ Rainbow agent (trained on an environment \textit{with} an opponent) compared to the Baseline and five different ablations (dashed, thinner lines). These curves are smoothed with a moving average over 5.000 environment interactions to make the visualization clearer (instead of 1500).} 
    \label{Fig:RainbowOPP_TrainCurve}
\end{figure}

From \cref{Fig:Rainbow_TrainCurve,Fig:RainbowOPP_TrainCurve} we see that the Rainbow agent, configured with all best found configurations, indeed improves the mean episode survival time during training for both the PPO and the PPO$^*$ agent, when compared to the Baseline agent. 
The comparable training curves of the ablation studies reveal that the performance of reverting to a single configuration closely matches that of the Rainbow agent. For example, we see that the impact of not including the change in activation threshold (Rainbow - $\AT = 0.95$) is not as big as one might expect based on the previous results for the activation threshold. 

\FloatBarrier

\FloatBarrier

In \cref{Table:ResRainbow}, the Rainbow agent achieves a notable improvement over the Baseline, completing 94.0\% of episodes and surviving 96.85\% of steps on average, corresponding to relative increases of $15.6\%$ and $9.54\%$, respectively. 
However, the ablation variants have comparable or even better performance. The agent using the original observation space and the agent without curriculum training are specifically performing well. This outcome suggests that the improvements observed with the Rainbow agent may be primarily driven by a subset of enhancements, or that interactions between the enhancements produce unexpected effects. A more in-depth analysis is needed to disentangle the contributions of individual components and understand their combined impact on performance.

\begin{table}[h]
\begin{adjustbox}{center}
\resizebox{1.2\textwidth}{!}
{
\scriptsize
\tabcolsep=3pt
\begin{tabular}{|l|p{0.1\linewidth}|p{0.08\linewidth}|p{0.08\linewidth}|p{0.1\linewidth}|p{0.1\linewidth}|p{0.08\linewidth}|p{0.08\linewidth}|p{0.08\linewidth}|}
\hline
\textbf{Agent} &
  \textbf{Completed episodes} &
  \textbf{Steps survived} &
  \textbf{Steps overloaded} &
  \textbf{Agent execution time {[}ms{]}} &
  \textbf{Maximum topology depth} &
  \textbf{Unique actions} &
  \textbf{Unique lines in danger} &
  \textbf{Unique subs changed} \\ \hline
Baseline                  & 78.6\%          & 87.31\%          & 0.008\%          & 1.967 & 5.0 & 27.8 & 11.8 & 6.0 \\ \hline
Rainbow                   & 94.0\%          & 96.85\%          & 0.013\%          & 2.728 & 4.0 & 7.0  & 5.8  & 4.4 \\ \hline
Rainbow - Obs: Baseline   & \textbf{97.8\%} & \textbf{98.68\%} & 0.010\%          & 2.805 & 4.0 & 9.6  & 7.8  & 4.8 \\ \hline
Rainbow - RW: Baseline    & 92.6\%          & 96.25\%          & 0.012\%          & 2.477 & 4.0 & 6.2  & 5.0  & 4.2 \\ \hline
Rainbow - No curriculum   & 97.2\%          & 98.50\%          & 0.010\%          & 2.439 & 3.8 & 6.2  & 4.4  & 4.0 \\ \hline
Rainbow - AT: 0.95        & 96.3\%          & 98.22\%          & \textbf{0.003\%} & 2.129 & 4.0 & 9.3  & 6.3  & 5.3 \\ \hline
Rainbow - No disconnect   & 95.4\%          & 97.86\%          & 0.014\%          & 2.190 & 4.0 & 8.8  & 6.6  & 5.2 \\ \hline
Rainbow - No RT           & 91.2\%          & 95.15\%          & 0.009\%          & 2.745 & 3.6 & 6.6  & 5.4  & 4.0 \\ \hline

\end{tabular}
}
\end{adjustbox}
\caption{Validation results of the PPO Rainbow agent and the six ablations compared to the baseline agent.}
\label{Table:ResRainbow}
\end{table}
\FloatBarrier

\cref{Table:ResRainbowOpp} shows more convincingly that combining all the best-found modeling choices results in the best performance for the final agent. With a mean survival percentage of $8.65\%$, the Rainbow agent improves on the Baseline agent by $2.63\%$. Furthermore, compared to the best-scoring agent including only one enhancement, the agent with AT=$0.99$ (\cref{Table:SetupRainbow_Opp}), the Rainbow agent improves by $1.60\%$.
Using the original larger action space $\A_{n-0}$ (agent \textit{Rainbow - Act: $\A_{n-0}$} in \cref{Table:ResRainbowOpp}), performed relatively well on the mean steps survived while keeping the percentage of steps overloaded low, making this configuration of modeling choices an interesting option. However, we observed that the training duration of this agent took twice as long compared to the Rainbow agent with the reduced action space ($\A_{d3qn}$), underlining the importance of action space reduction for larger grids for computational efficiency.

\begin{table}[h]
\begin{adjustbox}{center}
\resizebox{1.2\textwidth}{!}
{
\scriptsize
\tabcolsep=3pt
\begin{tabular}{|l|p{0.1\linewidth}|p{0.08\linewidth}|p{0.08\linewidth}|p{0.1\linewidth}|p{0.1\linewidth}|p{0.08\linewidth}|p{0.08\linewidth}|p{0.08\linewidth}|}
\hline
\textbf{Agent} &
  \textbf{Completed episodes} &
  \textbf{Steps survived} &
  \textbf{Steps overloaded} &
  \textbf{Agent execution time {[}ms{]}} &
  \textbf{Maximum topology depth} &
  \textbf{Unique actions} &
  \textbf{Unique lines in danger} &
  \textbf{Unique subs changed} \\ \hline
Baseline (long)                     & 0.0\% & 6.018\%          & \textbf{0.684\%} & 2.2808 & 4.2 & 98.6 & 15.8 & 7 \\ \hline
Rainbow                             & 0.0\% & \textbf{8.646\%} & 1.294\% & 3.7762 & 2.6 & 9    & 14.4 & 3 \\ \hline
Rainbow - Act: $\A_{n-0}$ & 0.0\% & 8.368\%          & 0.694\% & 4.4156 & 3.2 & 91.4 & 15.2 & 7 \\ \hline
Rainbow - RW: Baseline              & 0.0\% & 7.882\%          & 1.294\% & 3.8726 & 2.4 & 9    & 13.2 & 3 \\ \hline
Rainbow - AT: 0.95           & 0.0\% & 7.674\%          & 1.343\% & 3.036  & 3   & 9    & 14.8 & 3 \\ \hline
Rainbow - No disconnect      & 0.0\% & 8.270\%          & 1.238\% & 2.2876 & 2.6 & 9    & 13.6 & 3 \\ \hline
Rainbow - No RT              & 0.0\% & 7.978\%          & 1.302\% & 4.2582 & 2.6 & 9    & 14.6 & 3 \\ \hline
\end{tabular}
}
\end{adjustbox}
\caption{Validation results of the PPO$^*$ Rainbow agent and the five ablations compared to the baseline agent. All agents are trained for 500.000 environment interactions, explaining the slightly higher score for the baseline agent compared to previous results.}
\label{Table:ResRainbowOpp}
\end{table}
\FloatBarrier

While the Rainbow agents show a notable improvement over the RL baseline, they still fall short of outperforming the Greedy baseline. Further research should explore advanced techniques and fine-tuning hyperparameters to bridge this performance gap.

%%%%%%%%%%%%%%%%%%%%%%%%%%%%%%%%%%%%%%%%%%%%%%%%%%%%%%%%%%%%%%%%%%%%%%%%%%%%%%%
%%%%%%%%%%%%%%%%%%%%%%%%%%%%%%%%%%%%%%%%%%%%%%%%%%%%%%%%%%%%%%%%%%%%%%%%%%%%%%%
\section{Discussion and Future Work}\label{sec:discussion}
The application of reinforcement learning (RL) to power network control (PNC) has evolved rapidly since the introduction of the Learning to Run a Power Network (L2RPN) challenge in 2019. This survey has highlighted a significant body of research in this field, demonstrating the potential of RL and other machine learning techniques, such as imitation learning (IL). However, several challenges remain open. In this chapter, we discuss key issues and propose directions for future research.

\paragraph{Scalability of RL methods for large-scale grids}
One of the main challenges in applying RL to PNC is the exponential growth of the action and the state spaces as the grid size increases. Currently, researchers have explored two main approaches.
\begin{itemize}
    \item \textbf{Action space reduction:} 
    As shown in \cref{Table:Case36ActReduction,Table:Case118ActReduction} most researchers employ a greedy reduction method for larger grids (cases 36 and 118). While effective and relatively simple to implement, this approach may exclude beneficial actions that could be valuable when combined with others. Future research should explore more sophisticated reduction techniques.
    \item \textbf{Factorization of the state and action space:} 
    Most of the solutions that propose decomposition of the PNC problem define a (hierarchical) multi-agent architecture where each agent is responsible for a subset of the grid. These solutions define a factorization of the action space. However, as noted in \cite{losapio2024state}, these agents still observe the entire grid, leading to excessive information processing. 
    % Due too the highly interconnected nature of the grid it is not easy to find a proper state space factorization. 
    \cite{losapio2024state,henka2022power} propose potentially interesting state and action factorization that could enhance scalability. Both have yet to be tested in MARL settings.
\end{itemize}
% Lastly, some researchers refer to expert knowledge to select actions. It would be interesting to hear the reasoning of the experts on why they selected these specific actions as being good configurations for the grid.

Another challenge with MARL-based approaches is the need for agents to accumulate sufficient experience. Transfer learning, which enables agents to reuse knowledge across tasks \cite{taylor2019parallel,da2019survey}, could significantly accelerate training in this multi-agent setup. While promising, transfer learning has not yet been explored in RL for PNC. Future studies could investigate its applicability in MARL setups and for adapting policies to unseen grid configurations.

Given the complexity of power grids, additional approaches beyond action reduction and factorization could be considered \cite{heeswijk2023five}. The scalability of RL remains a key open challenge in this domain.

\paragraph{Advancing RL Techniques for Power Grid Optimization}
Our numerical studies in \cref{sec:experiments,sec:results} evaluated a selection of proposed modeling choices for an RL-based agent applied to case 14 in Grid2Op \cite{donnot2020grid2op}. While the Rainbow agent, which combined our best findings, shows a notable improvement over the original RL-based baseline, it still did not surpass the Greedy baseline. 
% As mentioned multiple times, we did not perform any hyperparameter tuning, which could still improve the performance of the current Rainbow agent. 
Several avenues remain open for exploration:

\begin{itemize}
    \item \textbf{Imitation Learning (IL):} 
    Prior work \cite{lan2020ai,binbinchen2020neurips,lehna2023managing,lajavaness2023l2rpn,artelys2024,lehna2024hugo,dejong2024imitation,dejong2025generalizable,hassouna2025learning} suggests that IL can significantly boost RL performance or perhaps even be used on its own. Future research should explore the benefits of IL further.
    \item \textbf{Off-policy techniques:}
    Since our experiments relied on PPO, an on-policy algorithm, we did not explore off-policy methods such as prioritized experience replay or advanced exploration strategies. These techniques could be tested using DQN-based approaches.
    \item \textbf{Curriculum Learning:} 
    Curriculum training has potential but requires careful design choices. Future research could explore different difficulty design choices and adaptive hyperparameter tuning during training.
    \item \textbf{Neural Network Architectures:} 
    While FCNNs are most commonly used, recent studies \cite{dejong2025generalizable,hassouna2025learning} indicate that graph neural networks (GNNs) can improve performance in IL settings. The benefits of GNNs in RL-based PNC warrant further investigation. 
    Additionally, recurrent architectures, such as LSTMs, could improve learning in environments with temporal dependencies.
    \item \textbf{Model-Based RL:}
    Most studies reviewed focus on model-free RL. The approach \alphazero\ in \cite{dorfer2022power}, which integrates Monte Carlo Tree Search (MCTS), demonstrated strong results in L2RPN 2022. Extending model-based planning approaches could offer new insights.
\end{itemize}

% Furthermore, when opting for the fully-connected neural network, there are many design choices to make, such as depth, width of network and which parts of the architecture the actor and the critic share. When moving towards more complicated neural networks such as GNN there are even more roads on which one can go. Research often shows the results of the final network choices, but not the effects of choosing something different. This makes it difficult to set up a strategy for designing a neural network for RL applied in a different, perhaps larger, power grid setting. 
% Lastly, there are other advanced neural network architectures, such as LSTM, that are still open for research in the area of RL applied to PNC. 

Although our numerical studies explored only a subset of possible RL techniques, we hope our methodology and our overview provide a foundation for future research.

\paragraph{Extend findings to larger and realistic grids}
Our experiments focused on Case 14 due to computational constraints. Future research should extend these findings to larger test cases (36 and 118) in Grid2Op. However, full-scale experiments with large action spaces may be infeasible. Our results indicate that environments with opponents benefit from smaller, more robust action spaces, suggesting that extensive action-space experiments may not be necessary for larger grids.

A significant challenge remains in bridging the gap between Grid2Op test cases and real-world power grids. A recent enhancement in Grid2Op provides the option to adjust the number of busbars per substation, bringing simulations closer to reality. However, to the best of our knowledge, this new setup has not yet been explored. Future studies should evaluate their impact on RL training and performance.

Lastly, in \cite{dejong2024imitation}, the lines attacked by the opponent differ from \cite{manczak2023hierarchical}. \cite{dejong2024imitation} found that by disabling some of the lines the network becomes almost entirely inoperable. Therefore, they consider a particular subset of lines for disablement. Our experiments confirm that agents struggle in the current adversarial settings, as proposed by \cite{manczak2023hierarchical}. Re-running these experiments with a more balanced opponent configuration could yield further insights.
 
% - opponent in case 14 too complex? \cite{dejong2024imitation}
% % - improve results with hyperparemeter tuning.
% - larger grids with and without opponent.
% - bridge between real world. Larger grids, more busbars at a substation

% To discuss:
% - Limit of directly applying a single agent algorithm
% - Importance of hyperparameter tuning
% - Potential solutions to explore further:
%     - MCTS
%     - GNN

%%%%%%%%%%%%%%%%%%%%%%%%%%%%%%%%%%%%%%%%%%%%%%%%%%%%%%%%%%%%%%%%%%%%%%%%%%%%%%%
%%%%%%%%%%%%%%%%%%%%%%%%%%%%%%%%%%%%%%%%%%%%%%%%%%%%%%%%%%%%%%%%%%%%%%%%%%%%%%%
\section{Conclusion}\label{sec:conclusion}
In this survey, we have explored the application of reinforcement learning (RL) to power grid topology optimization, highlighting key methods, challenges, and recent advancements in the field. Since the introduction of the L2RPN challenge in 2019, research in this area has grown significantly, with diverse approaches emerging to address the complexities of power network control (PNC). Our study has categorized and analyzed various techniques, including action space reduction, hierarchical multi-agent systems, and advanced RL algorithms, offering insights into their respective advantages and limitations. Our comparative study further provides quantitative insights into commonly used methods, offering a clearer understanding of their impact on performance.

Despite notable progress, several challenges remain, including the need for more scalable RL algorithms, improved generalization across networks, and bridging the gap between simulated test cases and existing grid operation frameworks. Addressing these challenges is important to realize the full potential of RL in power grid optimization. 

    Nevertheless, given the progress already achieved, the rapid growth of research in RL, and the many underexplored techniques with potential, such as imitation learning, off-policy techniques, curriculum learning, different neural network architectures and model-based RL mentioned in \cref{sec:discussion}, we believe RL remains a promising direction to support grid operators in PNC.

We hope that this survey provides a valuable foundation for researchers and practitioners interested in RL-based power grid topology optimization and inspires future work to address the outstanding challenges in this field.

% - developments of the problem

\paragraph{Acknowledgements}
We want to thank Jan Viebahn (TenneT) and Davide Barbiere (TenneT) for the interesting discussions. Furthermore, we would like to thank Barbera de Mol for the collaboration at the beginning of this project on a joint codebase.
Lastly, we thank RTE France for developing and providing the Grid2Op environment, which facilitates the RL-based research on power network control.

\bibliographystyle{elsarticle-num}  %{unsrt}%
\bibliography{references}  %%% Remove comment to use the external .bib file (using bibtex).

\appendix
\section{Acronyms} \label{App:Acronyms}
Throughout this paper, we use acronyms to refer to the solutions discussed. An overview of these acronyms, along with their corresponding reference papers and code if available, is provided in \cref{Table:SolAcr}. 

In addition, \cref{Table:Acronyms} provides an overview of the other acronyms used throughout this paper.

\begin{table}[h]
\centering
\footnotesize  
\tabcolsep=3pt
\begin{tabular}{l|p{0.1\linewidth}|p{0.7\linewidth}}
% \hline
\textbf{Solution acronym}     
& \textbf{Reference paper}   
& \textbf{Reference code}
\\ \hline
\aaacfirst  &  \cite{matavalam2022curriculum}  & \url{https://github.com/amar-iastate/L2RPN-using-A3C} \\
\ddqn       &  \cite{lan2020ai}                & \url{https://github.com/shidi1985/L2RPN} \\
\smaac      &  \cite{yoon2021winning}          & \url{https://github.com/sunghoonhong/SMAAC} \\
\aaacsecond &  \cite{yan2020l2rpn}             & \url{https://github.com/ZM-Learn/L2RPN_WCCI_a_Solution} \\
\sas        &  \cite{zhou2021action}           & \url{https://github.com/PaddlePaddle/PARL/tree/develop/examples/NeurIPS2020-Learning-to-Run-a-Power-Network-Challenge} \\
\binbin     &  \cite{binbinchen2020neurips}    & \url{https://github.com/AsprinChina/L2RPN_NIPS_2020_a_PPO_Solution} \\
\dddqnfirst &  \cite{zhihong2020neurips}       & \url{https://github.com/lujasone/NeurIPS_2020_L2RPN_Comp_An_Approach} \\
\cem        &  \cite{Subramanian2021}          & N.A. \\
\dddqnsecond & \cite{damjanovic2022deep}       & N.A. \\
\powrl      &  \cite{chauhan2023powrl}         & N.A. \\
\alphazero  &  \cite{dorfer2022power}          & \url{https://github.com/enlite-ai/maze-l2rpn-2022-submission}\tablefootnote{Their preceding version can be found at \url{https://github.com/enlite-ai/maze-l2rpn-2021-submission}.} \\
\brute      &  \cite{alibaba2022l2rpn}         & \url{https://github.com/AlibabaResearch/l2rpn-wcci-2022} \\
\hri        &  \cite{hri2022trial}             & \url{https://github.com/HRI-EU/l2rpn_tae_agent} \\
\curriculum &  \cite{lehna2023managing}        & \url{https://github.com/FraunhoferIEE/CurriculumAgent} \\
\hrl        &  \cite{manczak2023hierarchical}  & \url{https://github.com/bmanczak/runPowerNetworks} \\
\marl       &  \cite{vandersar2023marl}        & \url{https://gitlab.com/ericavandersar/marl4powergridtopo} \\
\ljn        &  \cite{lajavaness2023l2rpn}      & N.A. \\
\artelys    &  \cite{artelys2024}              & N.A. \\
\hugo       &  \cite{lehna2024hugo}            & \url{https://github.com/FraunhoferIEE/CurriculumAgent}  \\
\il         &  \cite{dejong2024imitation}      & \url{https://github.com/MatthijsdeJ/Imitation_Learning_Topology_Control} \\
\bdqn       &  \cite{wang2024alleviating}      & N.A. \\
\zonal      &  \cite{boguslawskiemulation}     & N.A. \\
\gnnil      &  \cite{dejong2025generalizable}  & \url{https://github.com/MatthijsdeJ/GNN_PN_Imitation_Learning} \\
\CCMA       &  \cite{demol2025centrally}       & N.A. \\      
\SoftIL     &  \cite{hassouna2025learning}     & N.A. \\
\end{tabular}
\caption{Overview of the solution acronyms used in this paper and references to the corresponding papers.}
\label{Table:SolAcr}
\end{table}
\FloatBarrier

\begin{table}[]
\centering
\begin{tabular}{l|l}
\textbf{Acronym} & \textbf{Meaning}                    \\
\hline
A3C              & Asynchronous-Advantage-Actor-Critic \\
AT               & Activation Threshold                \\
BC               & Behavioral Cloning                  \\
CE               & Cross-Entropy                       \\
D3QN             & Double Dueling DQN                  \\
DN-action        & Do-Nothing action                   \\
DQN              & Deep Q-Learning (or Deep Q-Network) \\
DRL              & Deep-RL                             \\
FCNN             & Fully Connected Neural Network      \\
GNN              & Graph Neural Network                \\
HRL              & Hierarchical Reinforcement Learning \\
IL               & Imitation Learning                  \\
IRL              & Inverse RL                          \\
L2RPN            & Learning to Run a Power Network     \\
LODF             & Line Outage Distribution Factor     \\
MARL             & Multi-Agent Reinforcement Learning  \\
MCTS             & Monte Carlo Tree Search             \\
MDP              & Markov Decision Process             \\
PER              & Prioritized Experience Replay       \\
PNC              & Power Network Control               \\
PPO              & Proximal Policy Optimization        \\
RL               & Reinforcement Learning              \\
RT               & Revert Threshold                    \\
SAC              & Soft Actor-Critic                   \\
TD               & Temporal Difference                 \\
TSO              & Transmission System Operator        \\
TT               & Target Topology                    \\
\end{tabular}
\caption{Overview of acronyms used in this paper.}
\label{Table:Acronyms}
\end{table}
\FloatBarrier

\section{Observation spaces}\label{App:ObsSpaces}
\cref{Table:SizeObsSpaceExp} shows a quick overview of the observation spaces considered in the experiments and their size.
\begin{table}[h]
\begin{adjustbox}{center}
\resizebox{1.1\textwidth}{!}
{
% \scriptsize
\tabcolsep=3pt
\begin{tabular}{|p{0.15\linewidth}|r|p{0.6\linewidth}|p{0.2\linewidth}|}
\hline
\textbf{Observation Space}    & \multicolumn{1}{l|}{\textbf{Total size}} & \textbf{Grid2Op features included}     & \textbf{Custom features included}                                                                                                                                                     \\ \hline
Baseline    & 154           & $load\_p$, $gen\_p$, $p\_or$, $p\_ex$, $\rho$, $timestep\_overflow$, $topo\_vect$                                             &                                              \\ \hline
Danger       & 174           & $load\_p$, $gen\_p$, $p\_or$, $p\_ex$, $\rho$, $timestep\_overflow$, $topo\_vect$                                             & $danger$                                     \\ \hline
History    & 1044          & $load\_p$, $gen\_p$, $p\_or$, $p\_ex$, $\rho$, $timestep\_overflow$, $topo\_vect$                                             & $danger$ and 6 time steps of observed values      \\ \hline
Complete     & 387           & $load\_p$, $gen\_p$, $p\_or$, $p\_ex$, $load\_q$, $gen\_q$, $q\_or$, $q\_ex$, $load\_v$, $gen\_v$, $v\_or$, $v\_ex$, $load\_\theta,$ 
                               $gen\_\theta$, $\theta\_or$, $\theta\_ex$, $a\_or$, $a\_ex \rho$, $timestep\_overflow$, $topo\_vect$                          & $time\_of\_day$, $day\_of\_year$               \\ \hline
\dddqnsecond         & 140           & $v\_or$, $v\_ex$, $a\_or$, $a\_ex$, $\rho$, $topo\_vect$                                                                    &                                              \\ \hline
\end{tabular}
}
\end{adjustbox}
\caption{Overview of the size of the observation space that is used as input for the PPO agent in the experiments.}
\label{Table:SizeObsSpaceExp}
\end{table}
\FloatBarrier

\section{Results PPO with different hyperparameters} 
This appendix contains the validation results of a PPO agent trained without an opponent using the hyperparameters that are presented in the column “Value with opponent (PPO$^*$)” in \cref{Table:Params}. The values in this table show how different hyperparameters can influence the performance of the agent.

\label{App:ResPPO_WrongHyperParam}
{
\scriptsize
\tabcolsep=3pt
\begin{longtable}{|ll|lll|}
\hline
    \multicolumn{2}{|l|}{\textbf{}}&
      \multicolumn{1}{l|}{\textbf{Completed episodes}} &
      \multicolumn{1}{l|}{\textbf{Steps survived}} &
      \textbf{Steps overloaded} \\ \hline
\endhead
        \hline
        \multicolumn{5}{r@{}}{Continued on next page\ldots}\\
\endfoot 
    % \bottomrule     
\endlastfoot  
\multicolumn{2}{|l|}{\textbf{Baseline}}                              & \multicolumn{1}{l|}{70.8\%} & \multicolumn{1}{l|}{84.45\%} & 0.010\% \\ 
\hline
\multicolumn{2}{|l|}{\textbf{Action Spaces}}                         & \multicolumn{3}{l|}{}                                                \\ \hline
\multicolumn{2}{|l|}{$\A_{sym}$}                                          & \multicolumn{1}{l|}{\textbf{89.0\%}} & \multicolumn{1}{l|}{\textbf{95.14\%}} & \textbf{0.009\%} \\ \hline
\multicolumn{2}{|l|}{$\A_{n-1}$}                                   & \multicolumn{1}{l|}{86.6\%} & \multicolumn{1}{l|}{93.99\%} & 0.015\% \\ \hline
\multicolumn{2}{|l|}{$\A_{d3qn}$}                                           & \multicolumn{1}{l|}{19.0\%} & \multicolumn{1}{l|}{47.37\%} & 0.125\% \\ \hline
\multicolumn{2}{|l|}{\textbf{Observation Spaces}}                   & \multicolumn{3}{l|}{}                                                \\ \hline
\multicolumn{2}{|l|}{Complete}                                       & \multicolumn{1}{l|}{69.2\%} & \multicolumn{1}{l|}{82.77\%} & 0.013\% \\ \hline
\multicolumn{2}{|l|}{D3QN\_2022}                                     & \multicolumn{1}{l|}{\textbf{72.8\%}} & \multicolumn{1}{l|}{\textbf{86.59\%}} & 0.010\% \\ \hline
\multicolumn{2}{|l|}{Danger}                                         & \multicolumn{1}{l|}{70.6\%} & \multicolumn{1}{l|}{84.40\%} & \textbf{0.009\%} \\ \hline
\multicolumn{2}{|l|}{History}                                        & \multicolumn{1}{l|}{65.2\%} & \multicolumn{1}{l|}{81.56\%} & 0.013\% \\ \hline
\multicolumn{2}{|l|}{Line loads}                                     & \multicolumn{1}{l|}{66.2\%} & \multicolumn{1}{l|}{82.05\%} & 0.013\% \\ \hline
\multicolumn{2}{|l|}{\textbf{Reward functions}}                      & \multicolumn{3}{l|}{\textbf{}}                                       \\ \hline
\multicolumn{2}{|l|}{AlphaZero}                                      & \multicolumn{1}{l|}{\textbf{71.8\%}} & \multicolumn{1}{l|}{\textbf{85.85\%}} & 0.011\% \\ \hline
\multicolumn{2}{|l|}{Constant}                                       & \multicolumn{1}{l|}{71.0\%} & \multicolumn{1}{l|}{85.07\%} & 0.011\% \\ \hline
\multicolumn{2}{|l|}{SMAAC}                                          & \multicolumn{1}{l|}{70.4\%} & \multicolumn{1}{l|}{84.00\%} & 0.014\% \\ \hline
\multicolumn{2}{|l|}{Binbinchen}                                     & \multicolumn{1}{l|}{71.4\%} & \multicolumn{1}{l|}{85.63\%} & 0.011\% \\ \hline
\multicolumn{2}{|l|}{Base + Bonus/Penalty}                           & \multicolumn{1}{l|}{69.9\%} & \multicolumn{1}{l|}{84.52\%} & 0.010\% \\ \hline
\multicolumn{2}{|l|}{\textbf{Curriculum Training}}                   & \multicolumn{3}{l|}{\textbf{}}                                       \\ \hline
\multicolumn{2}{|l|}{Level 1: 20000, level 2: 46667} &
  \multicolumn{1}{l|}{59.8\%} &
  \multicolumn{1}{l|}{80.03\%} &
  0.016\% \\ \hline
\multicolumn{2}{|l|}{\textbf{Activation Threshold}}                  & \multicolumn{3}{l|}{\textbf{}}                                       \\ \hline
\multicolumn{1}{|l|}{AT}                    & Train AT               & \multicolumn{3}{l|}{}                                                \\ \hline
\multicolumn{1}{|l|}{\multirow{4}{*}{0.8}}  & 0.8                    & \multicolumn{1}{l|}{0.0\%}  & \multicolumn{1}{l|}{6.83\%}  & 0.497\% \\ \cline{2-5} 
\multicolumn{1}{|l|}{}                      & 0.9                    & \multicolumn{1}{l|}{0.0\%}  & \multicolumn{1}{l|}{8.46\%}  & 0.411\% \\ \cline{2-5} 
\multicolumn{1}{|l|}{}                      & 0.95                   & \multicolumn{1}{l|}{0.0\%}  & \multicolumn{1}{l|}{6.48\%}  & 0.459\% \\ \cline{2-5} 
\multicolumn{1}{|l|}{}                      & 0.99                   & \multicolumn{1}{l|}{0.0\%}  & \multicolumn{1}{l|}{5.80\%}  & 0.499\% \\ \hline
\multicolumn{1}{|l|}{\multirow{4}{*}{0.9}}  & 0.8                    & \multicolumn{1}{l|}{2.8\%}  & \multicolumn{1}{l|}{26.44\%} & 0.135\% \\ \cline{2-5} 
\multicolumn{1}{|l|}{}                      & 0.9                    & \multicolumn{1}{l|}{29.6\%} & \multicolumn{1}{l|}{59.07\%} & 0.033\% \\ \cline{2-5} 
\multicolumn{1}{|l|}{}                      & 0.95                   & \multicolumn{1}{l|}{12.6\%} & \multicolumn{1}{l|}{50.47\%} & 0.052\% \\ \cline{2-5} 
\multicolumn{1}{|l|}{}                      & 0.99                   & \multicolumn{1}{l|}{3.4\%}  & \multicolumn{1}{l|}{32.91\%} & 0.091\% \\ \hline
\multicolumn{1}{|l|}{\multirow{3}{*}{0.95}} & 0.8                    & \multicolumn{1}{l|}{24.0\%} & \multicolumn{1}{l|}{51.50\%} & 0.056\% \\ \cline{2-5} 
\multicolumn{1}{|l|}{}                      & 0.9                    & \multicolumn{1}{l|}{72.4\%} & \multicolumn{1}{l|}{84.94\%} & 0.013\% \\ \cline{2-5} 
\multicolumn{1}{|l|}{}                      & 0.99                   & \multicolumn{1}{l|}{54.4\%} & \multicolumn{1}{l|}{77.49\%} & 0.019\% \\ \hline
\multicolumn{1}{|l|}{\multirow{4}{*}{0.99}} & 0.8                    & \multicolumn{1}{l|}{30.4\%} & \multicolumn{1}{l|}{52.57\%} & 0.063\% \\ \cline{2-5} 
\multicolumn{1}{|l|}{}                      & 0.9                    & \multicolumn{1}{l|}{85.0\%} & \multicolumn{1}{l|}{90.56\%} & 0.014\% \\ \cline{2-5} 
\multicolumn{1}{|l|}{}                      & 0.95                   & \multicolumn{1}{l|}{\textbf{91.0\%}} & \multicolumn{1}{l|}{\textbf{95.09\%}} & \textbf{0.008\%} \\ \cline{2-5} 
\multicolumn{1}{|l|}{}                      & 0.99                   & \multicolumn{1}{l|}{86.2\%} & \multicolumn{1}{l|}{94.92\%} & 0.010\% \\ \hline
\multicolumn{2}{|l|}{\textbf{Line switching}}                        & \multicolumn{3}{l|}{\textbf{}}                                       \\ \hline
\multicolumn{1}{|l|}{Final}                 & Train                  & \multicolumn{3}{l|}{}                                                \\ \hline
\multicolumn{1}{|l|}{No rules}              & No rules               & \multicolumn{1}{l|}{70.0\%} & \multicolumn{1}{l|}{83.87\%} & 0.013\% \\ \hline
\multicolumn{1}{|l|}{}                      & Reconnect              & \multicolumn{1}{l|}{\textbf{71.0\%}} & \multicolumn{1}{l|}{\textbf{85.19\%}} & \textbf{0.009\%} \\ \hline
\multicolumn{1}{|l|}{}                      & Reconnect + Disconnect & \multicolumn{1}{l|}{64.6\%} & \multicolumn{1}{l|}{82.21\%} & 0.013\% \\ \hline
\multicolumn{1}{|l|}{Reconnect}             & No rules               & \multicolumn{1}{l|}{70.0\%} & \multicolumn{1}{l|}{83.65\%} & 0.013\% \\ \hline
\multicolumn{1}{|l|}{}                      & Reconnect + Disconnect & \multicolumn{1}{l|}{64.4\%} & \multicolumn{1}{l|}{81.13\%} & 0.012\% \\ \hline
\multicolumn{1}{|l|}{Reconnect + Disconnect} &
  No rules &
  \multicolumn{1}{l|}{67.8\%} &
  \multicolumn{1}{l|}{83.62\%} &
  0.010\% \\ \hline
\multicolumn{1}{|l|}{}                      & Reconnect              & \multicolumn{1}{l|}{70.2\%} & \multicolumn{1}{l|}{84.58\%} & 0.009\% \\ \hline
\multicolumn{1}{|l|}{}                      & Reconnect + Disconnect & \multicolumn{1}{l|}{65.8\%} & \multicolumn{1}{l|}{82.45\%} & 0.012\% \\ \hline
\multicolumn{2}{|l|}{\textbf{Revert Topology}}                       & \multicolumn{3}{l|}{\textbf{}}                                       \\ \hline
\multicolumn{1}{|l|}{RT}                    & Train RT               & \multicolumn{3}{l|}{}                                                \\ \hline
\multicolumn{1}{|l|}{\multirow{3}{*}{0.0}}  & 0.8                    & \multicolumn{1}{l|}{55.6\%} & \multicolumn{1}{l|}{78.61\%} & 0.022\% \\ \cline{2-5} 
\multicolumn{1}{|l|}{}                      & 0.9                    & \multicolumn{1}{l|}{66.4\%} & \multicolumn{1}{l|}{82.12\%} & 0.014\% \\ \cline{2-5} 
\multicolumn{1}{|l|}{}                      & 0.95                   & \multicolumn{1}{l|}{65.6\%} & \multicolumn{1}{l|}{80.47\%} & 0.015\% \\ \hline
\multicolumn{1}{|l|}{\multirow{4}{*}{0.8}}  & 0.0                    & \multicolumn{1}{l|}{\textbf{72.8\%}} & \multicolumn{1}{l|}{\textbf{85.45\%}} & \textbf{0.009\%} \\ \cline{2-5} 
\multicolumn{1}{|l|}{}                      & 0.8                    & \multicolumn{1}{l|}{60.2\%} & \multicolumn{1}{l|}{78.41\%} & 0.014\% \\ \cline{2-5} 
\multicolumn{1}{|l|}{}                      & 0.9                    & \multicolumn{1}{l|}{66.2\%} & \multicolumn{1}{l|}{81.19\%} & 0.014\% \\ \cline{2-5} 
\multicolumn{1}{|l|}{}                      & 0.95                   & \multicolumn{1}{l|}{63.4\%} & \multicolumn{1}{l|}{79.45\%} & 0.019\% \\ \hline
\multicolumn{1}{|l|}{\multirow{4}{*}{0.9}}  & 0.0                    & \multicolumn{1}{l|}{72.4\%} & \multicolumn{1}{l|}{85.18\%} & 0.010\% \\ \cline{2-5} 
\multicolumn{1}{|l|}{}                      & 0.8                    & \multicolumn{1}{l|}{63.2\%} & \multicolumn{1}{l|}{80.07\%} & 0.013\% \\ \cline{2-5} 
\multicolumn{1}{|l|}{}                      & 0.9                    & \multicolumn{1}{l|}{67.4\%} & \multicolumn{1}{l|}{80.73\%} & 0.014\% \\ \cline{2-5} 
\multicolumn{1}{|l|}{}                      & 0.95                   & \multicolumn{1}{l|}{66.4\%} & \multicolumn{1}{l|}{80.84\%} & 0.018\% \\ \hline
\multicolumn{1}{|l|}{\multirow{4}{*}{0.95}} & 0                      & \multicolumn{1}{l|}{72.2\%} & \multicolumn{1}{l|}{85.29\%} & 0.011\% \\ \cline{2-5} 
\multicolumn{1}{|l|}{}                      & 0.8                    & \multicolumn{1}{l|}{59.8\%} & \multicolumn{1}{l|}{78.62\%} & 0.016\% \\ \cline{2-5} 
\multicolumn{1}{|l|}{}                      & 0.9                    & \multicolumn{1}{l|}{66.2\%} & \multicolumn{1}{l|}{81.46\%} & 0.012\% \\ \cline{2-5} 
\multicolumn{1}{|l|}{}                      & 0.95                   & \multicolumn{1}{l|}{63.6\%} & \multicolumn{1}{l|}{77.83\%} & 0.016\% \\ \hline

\caption{Validation results of PPO agents with different design choices, trained on an environment without an opponent, using the hyperparameters from PPO$^*$ in \cref{Table:Params}.}
\label{Table:ResPPOWrongHyperpar}
\end{longtable}
\FloatBarrier
}

\end{document}